\pdfoutput=1
\documentclass[draftclsnofoot,onecolumn,11pt]{IEEEtran}
\usepackage{amsfonts}
\usepackage{times}
\usepackage{graphicx}

\usepackage{latexsym}
\usepackage{dsfont}
\usepackage{amssymb}
\usepackage{amsmath}
\usepackage{cite}
\usepackage{verbatim}

\newcommand{\figref}[1]{{Fig.}~\ref{#1}}

\def\bb0{{\mathbb{0}}}

\def\ba{{\mathbf{a}}}
\def\bb{{\mathbf{b}}}

\def\bff{{\mathbf{f}}}

\def\bn{{\mathbf{n}}}

\def\bs{{\mathbf{s}}}

\def\bw{{\mathbf{w}}}
\def\bx{{\mathbf{x}}}
\def\by{{\mathbf{y}}}

\def\b0{{\mathbf{0}}}

\def\bA{{\mathbf{A}}}

\def\bF{{\mathbf{F}}}

\def\bH{{\mathbf{H}}}

\def\bU{{\mathbf{U}}}
\def\bV{{\mathbf{V}}}
\def\bW{{\mathbf{W}}}

\def\cA{\mathcal{A}}

\def\sf0{{\mathsf{0}}}

\def\bandwidth{{\mathrm{B}}}
\def\rate{{\mathrm{R}}}
\def\load{\Psi}
\def\ratecov{{\mathcal{R}}}
\def\sinrcov{{\mathcal{S}}}
\def\BSdensity{\lambda}
\def\UEdensity{\lambda_u}

\usepackage{amsmath, amssymb, bm, cite, epsfig, psfrag, sansmath,dsfont}
\usepackage{multirow}
\usepackage{graphicx}
\usepackage{color}
\usepackage{epstopdf} 
\usepackage{subfig}
\usepackage{tabulary}
\def\sf{\sansmath}

\def\beq{\begin{equation}}
\def\eeq{\end{equation}}
\def\beqa{\begin{eqnarray}}
\def\eeqa{\end{eqnarray}}
\def\beqan{\begin{eqnarray*}}
\def\eeqan{\end{eqnarray*}}

\def\argmin{\mathop{\mathrm{arg\,min}}}

\def\sir{\mathsf{SIR}}
\def\snr{\mathsf{SNR}}
\def\sinr{\mathsf{SINR}}
\newcommand{\PP}{\mathbb{P}}

\def\Nrx{N_{\rm rx}}
\def\Nrf{N_{\rm RF}}
\def\Ntx{N_{\rm tx}}

\def\BS{{\rm BS}}
\def\MS{{\rm MS}}

\def \Frf{\bF_{\rm RF}}
\def \Fbb{\bF_{\rm BB}}

\def \il{b}

\newtheorem{theorem}{Theorem}
\newtheorem{lemma}{Lemma}

\usepackage{acronym}
\usepackage{url}
\acrodef{bps}{bits per second}

\acrodef{UHF}{ultra high frequency}
\acrodef{mmWave}{millimeter wave}
\acrodef{LOS}{line-of-sight}
\acrodef{NLOS}{non-line-of-sight}
\acrodef{SINR}{signal-to-interference-plus-noise ratio}
\acrodef{RMS}{root mean square}
\acrodef{IID}{identically and independently distributed}
\acrodef{MIMO}{multiple-input and multiple-output}
\acrodef{FBR}{front-to-back ratio}
\acrodef{LTE}{long-term evolution}
\acrodef{PPP}{Poisson point process}
\acrodef{TDD}{time-division duplex}
\acrodef{SNR}{signal-to-noise ratio}
\acrodef{SIR}{signal-to-interference ratio}
\acrodef{MRC}{maximum ratio combining}
\acrodef{ZF}{zero-forcing}
\acrodef{MMSE}{minimum mean square error}
\acrodef{CSI}{channel state information}
\acrodef{BS}{base station}
\acrodef{i.i.d.}{independently and identically distributed}
\acrodef{PMF}{probability mass function}
\acrodef{HetNet}{heterogeneous network}

\title{Modeling and Analyzing Millimeter Wave Cellular Systems}

\author{Jeffrey~G.~Andrews, 
	Tianyang Bai, 
	Mandar Kulkarni, 
	Ahmed Alkhateeb, \\
	Abhishek Gupta, 
        Robert W. Heath, Jr. \\  
        \vspace{0.3in}
       \emph{Invited Paper}
\thanks{J. G. Andrews (jandrews@ece.utexas.edu) is the contact author.  The authors are all with the University of Texas at Austin, USA. This research has been supported by the National Science Foundation, CIF-1514275.   Article last revised: \today}}

\begin{document}

\setcounter{page}{1}

\maketitle
        \vspace{-0.5in}

\begin{abstract}
We provide a comprehensive overview of mathematical models and analytical techniques for millimeter wave (mmWave) cellular systems.  The two fundamental physical differences from conventional Sub-6GHz cellular systems are (i) vulnerability to blocking, and (ii) the need for significant directionality at the transmitter and/or receiver, which is achieved through the use of large antenna arrays of small individual elements.   We overview and compare models for both of these factors, and present a baseline analytical approach based on stochastic geometry that allows the computation of the statistical distributions of the downlink signal-to-interference-plus-noise ratio (SINR) and also the per link data rate, which depends on the SINR as well as the average load.   There are many implications of the models and analysis: (a) mmWave systems are significantly more noise-limited than at Sub-6GHz for most parameter configurations; (b) initial access is much more difficult in mmWave; (c) self-backhauling is more viable than in Sub-6GHz systems which makes ultra-dense deployments more viable, but this leads to increasingly interference-limited behavior; and (d) in sharp contrast to Sub-6GHz systems cellular operators can mutually benefit by sharing their spectrum licenses despite the uncontrolled interference that results from doing so.  We conclude by outlining several important extensions of the baseline model, many of which are promising avenues for future research.
\end{abstract}

\section{Introduction}
\label{sec:intro}

Until  recently, \ac{mmWave} frequencies -- spanning from 30-300 GHz -- were not considered useful for dynamic communication environments such as cellular systems.  Millimeter waves have been used extensively for long-distance point-to-point communication in satellite and terrestrial applications, now they are being investigated and developed for commercial cellular systems.  This new application is much more challenging due to unpredictable propagation environments and strict constraints on size, cost, and power consumption (particularly in the mobile handset).  Given the extreme shortage of available spectrum at traditional cellular frequencies -- often referred to in the industry as Sub-6GHz -- along with a booming demand for broadband and other wireless data services, the possibility of using mmWaves for cellular has generated intense interest starting about five years ago \cite{PiKha11}. 

\subsection{Millimeter Wave: What's New?}

The misconception that \ac{mmWave} frequencies do not propagate well in free space stems from the $\lambda_\mathrm{c}^2 = (c/f_\mathrm{c})^2$ dependence in the well-known Friis equation, where $\lambda_\mathrm{c}$ is the carrier wavelength, $f_\mathrm{c}$ is the carrier frequency, and $c$ the speed of light.   The baseline Friis equation, however, applies to omnidirectional transmission and reception with a specific type of antenna where the effective antenna area is $\lambda_\mathrm{c}^2/4\pi$, which implies that a great deal of energy is lost simply because the antennas have a small effective area and cannot radiate or capture much energy.  The key observation is that for a \emph{fixed} two dimensional antenna area, the \emph{number} of antenna elements -- each proportional in length and/or width to $\lambda_\mathrm{c}$ -- increases as $\lambda_\mathrm{c}^2$.  Thus, the small effective area of each antenna can be overcome by a moderately sized 2-D array of small antenna elements.  With such 2-D arrays at both the transmitter and receiver, this aggregate loss of $\lambda_\mathrm{c}^2$ turns into a theoretical aggregate gain of $\lambda_\mathrm{c}^2$ due to the gain of $\lambda_\mathrm{c}^2$ at each end.

This simple observation has been known long before the recent excitement about \ac{mmWave} cellular.  For example, a paper \cite{Fel56} in 1956 on ``Millimeter waves and their applications" makes many of the same points.  Its abstract reads ``Investigations in the vast 30,000- to 300,000-mc [MHz] frequency range is proving that it can accommodate many of the communications services, especially where there is need for high-gain, high-directional antennas, and large bandwidth.''   This one sixty year old sentence summarizes the basic idea even today: that with sufficient directionality, millimeter waves can be used in cellular communications as well, although such environments are usually very different than free space.   This required directionality stemming from large antenna arrays is the key distinguishing feature of \ac{mmWave} cellular systems, and it has far-reaching implications on how to model, analyze, design, and implement them.  

Another important trait of \ac{mmWave} cellular systems is their vulnerability to \emph{blocking}.  Although Sub-6GHz cellular systems also suffer from blocking, the effects are much more severe for \ac{mmWave}.  Millimeter waves are particularly sensitive to blocking for four main reasons.  First, they suffer much higher penetration losses when passing through many common materials (including concrete, tinted glass, and water \cite{Rappaport2013a}), owing to their smaller wavelength.  Second, \ac{mmWave} frequencies do not diffract well in terrestrial environments because the wavelength is much smaller than the objects it would preferably bend around.  Quantitatively, the Fresnel zone is proportional to $\sqrt{\lambda_c}$, and this determines the size of the LOS region between a transmitter and receiver.  Therefore, an environment that would be effectively line-of-sight (LOS)\footnote{``Effectively LOS'' means that there is a strong path between the transmitter and receiver that attenuates similarly to free space -- e.g. inside the Fresnel zone -- it need not be literally line of light (completely unobstructed). } for Sub-6GHz is Non-line-of-sight (NLOS) for \ac{mmWave} and thus attenuates more rapidly.  Third, because of the aforementioned required directionality, both the transmitter and receiver beam patterns are focused over a more narrow beamwidth, which affords millimeter wave signals fewer chances to avoid strong blocking than in a nearly omnidirectional transmit/receive scenario where energy is radiated and collected over much wider angles.  Fourth, because \ac{mmWave} systems have large bandwidths and relatively low transmit powers, as well as various other hardware constraints, the received signal-to-noise ratio ($\snr$) is generally quite low even with nontrivial beamforming gain, and so additional power loss from blocking cannot be tolerated.  

Along with the strong required directionality, \ac{mmWave} cellular's susceptibility to blocking requires important changes to the cellular network architecture and deployment.  This in turn requires nontrivial changes to their modeling and analysis. Providing a comprehensive overview of how to adapt to these changes from a theoretical perspective is the main focus of this paper.

\subsection{Scope and Organization}
This tutorial article focuses on the communication theory aspects of \ac{mmWave} cellular systems, unify. As such we focus on the modeling and analysis at the physical layer, with implications on the network architecture and higher layer protocols.   Specifically, this tutorial covers the following topics, which also correspond to the sections of the paper.

\begin{itemize}
\item \textbf{History and state of the art}.  We provide a brief survey of the history of \ac{mmWave} cellular, including recent developments in both theory and practice.
\item \textbf{Blocking}.  We overview several proposed blocking models, and discuss their relative merits and accuracy.
\item \textbf{Directionality via large antenna arrays}.  We discuss different antenna architectures and their tradeoffs, in particular analog and hybrid beamforming, single user spatial multiplexing, and multiuser MIMO.
\item \textbf{Analytical tools and approaches.}  In this, the main technical section, we show how to analyze key metrics like the signal-to-interference-plus-noise ratio (SINR) and per link data rate in a \ac{mmWave} cellular system.  Specific contents are:
	\begin{itemize}
	\item Define metrics of SINR and rate
	\item	Define and describe a baseline mmWave cellular system model
	\item Derivation of the SINR distribution
	\item Approximation of the per link rate distribution
	\end{itemize}
\item \textbf{Design Implications.} Based on the analysis, we discuss key considerations that are distinct in mmWave systems including the need for novel initial access techniques; noise vs. interference limited behavior in the context of densification; the viability of self-backhauling; and novel spectrum licensing paradigms.
\item \textbf{Extensions of Baseline Model.}  A great many extensions are possible, and we overview a few key ones.  These include the uplink; joint coverage with the more robust Sub-6GHz network including outdoor-to-indoor coverage; and MIMO techniques beyond directional analog beamforming.\end{itemize}

\section{A Brief History on Millimeter Wave Systems}
\label{sec:background}

Millimeter wave frequencies have been in use for various applications for a long time; it is only recently that they have been seriously considered for use in commercial cellular systems, notably 5G.  In this section, we provide a concise chronology of how and why \ac{mmWave} is now viable for cellular.

\subsection{Pre-cellular \ac{mmWave}}
Millimeter wave frequencies have been considered and studied for cellular-like systems as early as 1985 \cite{Walke1985}, wherein the use of ``fan antennas'' to provide directionality gains at a $60$ GHz carrier frequency were claimed to be able to reach a range of 500 meters, assuming the use of spread spectrum and targeting a very low data rate (tens of kbps).  However, with the possible exception of other obscure outliers, very little consideration was given to the use of \ac{mmWave} frequencies in cellular applications over the subsequent 25 years.

During the interim, \ac{mmWave} frequencies were leveraged in a host of non-communication applications like radar sensing \cite{Currie1987}, automotive navigation \cite{Russell1997, Clark1998}, and medical imaging \cite{Woodward2002,Taylor2011}.  As far as communication applications, \ac{mmWave} was mostly considered early for two diametrically opposed applications.  The first being medium to long-range LOS communication using large direction (e.g. dish) antennas, including backhaul over a few km and satellite communications.    The second main application was vehicular communication  \cite{Kato2001,Meinel1995,Tsugawa2005,VaShiBan:Millimeter-Wave-Vehicular:16}, allowing cars to internetwork directly or through the infrastructure. Though there was an ISO standard \cite{ISO-21216-CALM-MM}, dedicated short range communication at $5.9$ GHz has become the defacto standard for communication between cars \cite{Kenney2011}. 

The consumer revolution in mmWave came with the release of the  the large unlicensed band around 60 GHz \cite{Park2007,DanHea07}, which has now culminated in wireless personal area network (WPAN) and WLAN standards \cite{11ad,Baykas2011,WirelessHDSrandard2007}. The main target application was for very short range ``cable replacement'' type applications. The most successful products thus far has been the proprietary standard WirelessHD, though IEEE 802.11ad \cite{11ad}, also known also as ``WiGig'', is now gaining commercial traction. Both achieve data rates on the order of several Gbps over short ranges (within a room).  The commercial advance of these short-range standards is an important tangential development for \ac{mmWave} cellular, because it has produced considerable consumer grade device-side innovation, for example establishing the viability of small adaptive arrays and advancing the low power capabilities of RF and mixed-signal circuitry \cite{Rappaport2014book}. 

\subsection{Understanding the \ac{mmWave} channel}

The case for \ac{mmWave} cellular systems relies on an accurate understanding of their signal propagation and channel characteristics. While indoor \ac{mmWave} channels have been extensively studied, especially at the $60$ GHz unlicensed band \cite{Tharek1988,Davies1991,Manabe1994,Smulders1995,Smulders1997,Manabe1996,Xu2002,Moraitis2004a,Anderson2004,Collonge2004,Zwick2005a,Geng2009,WyneMolisch2011TWC,Ben-Dor2011}, thorough measurements of outdoor \ac{mmWave} channels began much more recently, after \cite{PiKha11}.

The measurement and characterization effort for indoor \ac{mmWave} channels started in the late 1980's at The University of Bristol \cite{Tharek1988}. Inside university rooms,  channel measurements for the 60 GHz band were performed in both LOS and NLOS conditions \cite{Tharek1988}. In \cite{Davies1991}, some wideband channel characteristics, such as the excess delay and RMS delay spread, of the 1.78 GHz and 60 GHz bands were compared, and the impact of some propagation characteristics like the atmospheric absorption was illustrated. Later in \cite{Manabe1994}, the multi-path propagation characteristics in a modern office building were measured at 60GHz.  Similar studies in different room settings were then conducted in \cite{Smulders1995,Smulders1997}. In \cite{Manabe1996}, the impact of antenna polarization and radiation pattern on the indoor \ac{mmWave} signal propagation was characterized. With the increased interest in defining a 60 GHz WLAN standard, more measurement work has been conducted for the sake of accurately modeling the channel and signal propagation characteristics in this band \cite{Xu2002,Moraitis2004a,Anderson2004,Collonge2004,Zwick2005a,Geng2009,WyneMolisch2011TWC,Ben-Dor2011}.

Outdoor \ac{mmWave} channel measurement data increased greatly in the last few years \cite{murdock201238,rappaport2012cellular,Rappaport2013a,Rappaport2015, Weiler14,haneda20165g}. In \cite{murdock201238,rappaport2012cellular}, 38 GHz outdoor urban cellular channels were measured across the campus of UT Austin using directional transmit antennas of 25 dBi power gain and 7$^\circ$ beamwidth, and with a transmit power of 21 dBm. These measurements showed that acceptable SNR can be achieved in outdoor \ac{mmWave} links up to a distance of approximately 200m, with a bandwidth of 800 MHz.  In \cite{Rappaport2013a},  measurements at 28 GHz and 38 GHz for outdoor urban environments around UT Austin and New York University provided data on the angles of arrival/departure, RMS delay spread, path loss, and building penetration and reflection coefficients, leading to further models for cellular \ac{mmWave} channels \cite{Rappaport2015}. Following \cite{rappaport2012cellular}, outdoor mmWave channel measurements by different groups were also conducted \cite{Weiler14,haneda20165g}.

These extensive measurements in \cite{murdock201238,rappaport2012cellular,Rappaport2013a,Rappaport2015, Weiler14,haneda20165g} have demonstrated that although \ac{mmWave} signals share basic propagation characteristics (like power law path loss) with their lower frequency counterparts, they also have some very important differences. It is also important to note that the use of directional antennas changes the effective channel seen at the receiver. For example, directional antennas reduce delay spread \cite{MacCartney2015} and Doppler spread \cite{Va2015}, but introduce other impairments such as pointing (beam misalignment) errors.  Another example is the classic two-ray ground reflection model, which results in the path loss exponent changing from $\alpha = 2$ to $\alpha = 4$ even for LOS \cite{Goldsmith2005}.  Directional antennas make such a model even more questionable, since the ground (or other) reflection will likely not occur to do the focused beam pattern.
   
We summarize the key takeaways to date from these measurement campaigns as follows:
\begin{itemize}
	\item  There is sharp difference between \ac{LOS} and \ac{NLOS} propagation for \ac{mmWave}.  	
	\item Because of poor diffraction (due to a smaller Fresnel zone, as discussed earlier), NLOS conditions in mmWave are due to reflections and scattering.
	\item There is usually more attenuation on NLOS paths when compared to Sub-6GHz, due to high penetration losses and energy losses due to scattering. 
	\item Indoor-to-outdoor (and vice versa) penetration losses are  much higher at mmWave in most materials, to the extent that it usually will not be possible to serve indoor users with outdoor base stations.
	\item Delay spread is generally much lower at \ac{mmWave}, but the symbol time is also much smaller due to the large bandwidth. Therefore, equalization requirements may even be higher at mmWave. 
	\item \ac{mmWave} channels are often sparse in the angular domain, with a few scattering clusters, each with several rays, in addition to a dominant LOS path. 
\end{itemize}

These differences are important to bring into any mathematical model for a mmWave cellular system. 

\subsection{The recent push for \ac{mmWave} cellular}

Around the start of this decade, Jerry Pi and Farooq Khan in Samsung's Dallas Technology Lab were the first to publicly make the case for \ac{mmWave} cellular, providing a detailed link budget analysis and other persuasive arguments in \cite{PiKha11}.   Their link budget showed that  with high gain antennas at both the transmitter and the receiver (about $15-30$ dB), the propagation losses can be overcome and Gbps-type data rates can be obtained in a cellular architecture; at least theoretically.  This was followed by the propagation studies in Ted Rappaport's group at UT Austin that developed extensive channel measurements for outdoor \ac{mmWave} communication, culminating in \cite{Rappaport2013a}, which triumphantly (although perhaps prematurely) declared the viability of \ac{mmWave} cellular, up to cell radii on the order of 200 m.

Notable early prototypes and feasibility studies were carried out by Nokia \cite{NSN13, Cud14, Ghosh14} and Samsung Electronics \cite{Roh2014,Hong2014} shortly thereafter.  For example, in \cite{Cud14}, Nokia presented an experimental system operating at 73.5 GHz with a 1 GHz bandwidth, with the BS having a steerable dielectric lens antenna offering 28 dB gain over a narrow 3 degree beamwidth. The mobile station (MS) had an open ended wave guide antenna with a 60 degree beamwidth.  Samsung's prototype \cite{Roh2014} instead offered transmit and receive arrays each with 32 antenna elements arranged in an $8 \times 4$ uniform planar array, in a compact area of 6 cm $\times $3 cm.  The antennas were grouped into 4 subarrays of 8 antennas each, with one RF unit per subarray, known as \emph{hybrid beamforming}. The resulting beamwidth was approximately $10^\mathrm{o}$ horizontally and $20^\mathrm{o}$ vertically with an overall beamforming gain of 18 dB.  The reported peak data rate with no mobility was about 1 Gbps, over a range up to 1.7 km (LOS) or 200 meters (NLOS).

Universal coverage and the support of mobility are arguably the key distinguishing features of cellular networks: simply supporting a link budget (especially outdoors-to-outdoors) is not sufficient.  Links need to be able to be set up quickly regardless of the mobile's location, and mobile users need to be tracked and communicated to on demand.   The mobility study in \cite{Roh2014} claims that users moving at about 8 km/hr can achieve 500 Mbps with 1$\%$ block error rates using steerable antennas. Although this is much less mobility than LTE can support, performed under specific conditions, it is hopefully a first step towards supporting the many dynamics inherent to cellular networks.  We will not model or analyze mobility in this paper either, but the difficulty in supporting mobility and dynamic on-demand connectivity should be kept in mind.

To test the feasibility of realizing large antenna arrays at mmWave mobile terminals and its biological implications, \cite{Hong2014} prepared a prototype for a mmWave 5G cellular phone equipped with a pair of 16-element antenna arrays. This study found that the electromagnetic filed absorbed by a user at 28 GHz is more localized compared to that at 1.9 GHz. The skin penetration depth, however, at 28 GHz is much less---around 3 mm compared to 45 mm at 1.9 GHz. This implies that most of the absorbed energy is limited to the epidermis at mmWave communications. The biological impact of mmWave radiation has also been further studied in \cite{Wu2015,Wu2015a}.

\subsection{Performance analysis}
The highly motivating link budget analysis in \cite{PiKha11} was followed up in parallel by several simulation and analysis efforts, e.g. \cite{Akoum2012, BaiHea14,NSN13,Rangan2014,Akdeniz2013a,Ghosh14,Kim13, Abouelseoud13}. As far as the simulation-based studies, in \cite{Rangan2014,Akdeniz2013a}, a measurement-based \ac{mmWave} channel model that incorporated blockage effects and angle spread was proposed and further used to simulate the \ac{mmWave} cellular network capacity. It was found that the achievable rate in \ac{mmWave} networks outperforms conventional cellular networks by an order-of-magnitude owing to the large available bandwidth. It was also observed that the impact of thermal noise on coverage dominates that of out of cell interference in mmWave networks. In \cite{NSN13}, a systematic ray tracing study including roads, sidewalks, and rectangular buildings with outdoor users showed that mean throughput and cell edge rates can be improved from 3 to 5.8 Gbps and 25 to 1400 Mbps (a factor of 56!), respectively, by increasing number of base stations in the $0.72$km$^2$ region under consideration from 36 to 96 (corresponds to increasing base station density from 50/km$^2$ to 133/km$^2$). This shows the importance of density in mmWave cellular networks for enhancing throughput, especially the cell edge throughput, which is strongly noise-limited. Around the same time, similar observations were also reported in \cite{Ghosh14,Abouelseoud13}. 

The impact of the number of antennas on system performance was reported in \cite{Ghosh14,Kim13}. In \cite{Kim13}, it was reported that the mean rates improve from about 500 Mbps to more than 4 Gbps when the antenna configuration is changed from (4,2) to (32,8), where the first number in the bracket indicates the number of antennas at the base station and the second number indicates number of user antennas. Similarly, cell edge rates increase from about 50 Mbps to 200 Mbps. Similar observations were reported in \cite{Ghosh14}. Hybrid analog/digital beamforming was used in \cite{Kim13} to tackle the analog to digital converter (ADC) power consumption issue in large antenna mmWave networks. The importance of enabling as many number of radio frequency (RF) chains as possible given the power constraints was highlighted in this work.

Although these simulation results appear encouraging, the claimed rates and outage probabilities are not transparently related to the many simulation parameters.  As with any communication system, a mathematical model approximating the key features of the system is desirable.  A mathematical analysis of a \ac{mmWave} cellular system can help expose key dependencies and bottlenecks in the system, and provide a mechanism for incubating and comparing new ideas and different design approaches without building and running a system-level simulation to test each hypothesis.  Although generally theorists use simulations to validate analysis, the reverse can also be helpful for system engineers: analysis can provide a way to sanity check complex simulations that could have any number of bugs.

In parallel to the excitement over \ac{mmWave} cellular systems, a new analytical approach to cellular systems was pioneered starting with \cite{AndBac11}.  This approach provided a mechanism for mathematically deriving the SINR distribution in a downlink cellular system.  This framework relies on stochastic geometry \cite{HaeAnd09,BacNOW,HaenggiBook}, which is an increasingly sophisticated subfield of applied probability, wherein the  BS locations are assumed to follow a stochastic point process, rather than taking up deterministic grid-like locations.  Such an approach has been shown by an increasing body of evidence to be quite accurate for Sub-6GHz macrocell-based cellular networks, at least as accurate as the conventional hexagonal grid model in typical circumstances, and typically being pessimistic by a nearly fixed horizontal SINR shift (i.e. independent of the actual SINR or coverage probability) of 1-3 dB \cite{Hae14,GuoHae15}.

Stochastic geometry was first applied to analyze the SINR and rate in \ac{mmWave} cellular in 2012 in \cite{Akoum2012}, where the results indicated that when the link budget is satisfied using large arrays in \ac{mmWave} systems, \ac{mmWave} could provide comparable SINR coverage and much higher rate compared with conventional cellular networks. The critical effect of blockages was first incorporated in \cite{Bai2014b}, and then extended to \ac{mmWave} specifically in \cite{BaiHea14}, and subsequently \cite{Singh2015}.  We will provide a more detailed description of these and related contributions in the next several sections. We now begin our attempt to mathematically model and analyze mmWave cellular systems  with an in depth discussion of one of their key differentiating traits: its susceptibility to blocking.

\begin{table}[t!]\label{notation}
\caption{Summary of Notation}
\begin{tabulary}{\columnwidth}{ |l | L | }\hline
{\bf Notation} &{\bf Description}\\ \hline
$\lambda_\mathrm{c}, f_\mathrm{c}$ & Carrier wavelength and frequency. \\ \hline
$\Phi, \lambda,X_\il$ & PPP for BS locations, BS density, location of $\il$th BS. \\ \hline
$\Phi_\mathrm{u}, \lambda_\mathrm{u}$ & PPP for user locations, user density. \\ \hline
$C_p$  & Path loss at 1 m where $p\in\{\mathrm{LOS,NLOS}\}$. For reference distance path loss model, it equals $\left(\frac{\lambda_c}{4\pi}\right)^2$ irrespective of $p$ and are just curve-fit parameters for the floating intercept model\cite{Rappaport2015}.  \\ \hline
$\alpha_p$ & Slope of power-law path loss where $p\in\{\mathrm{LOS,NLOS}\}$. Interpreted as path loss exponents for reference distance model and are curve-fitting parameters in the floating intercept model\cite{Rappaport2015}.\\\hline
$\ell(d)$ & Path loss at distance $d\in\mathbb{R}^+\cup\{0\}$. Equals $C_p d^{\alpha_p}$ where $p\in\{\mathrm{LOS,NLOS}\}$. \\\hline
$P_\mathrm{LOS}(d)$& Probability that a link of length $d$ is LOS.\\ \hline
$\mathbf{H}_\il$ & The downlink channel from the $\il$th BS to the typical user.\\\hline
$h_{\il,p}$ & The small scale fading of the $p$th path from BS $\il$.\\\hline
$N_p$  & Nakagami fading parameter where $p\in\{\mathrm{LOS,NLOS}\}$.\\ \hline
$\mathrm{B}$ & Total bandwidth.\\\hline
$\sigma^2$ & Noise power \\\hline
$\lambda_\mathrm{bldg}; \mathbb{E}[L_\mathrm{bldg}],\mathbb{E}[A_\mathrm{bldg}]$ & density of buildings; average building perimeter, area \\\hline
$R_\mathrm{B},p_l$ & Size of the LOS ball, average fraction of  LOS links in the LOS ball.\\\hline
$N_\mathrm{tx},N_\mathrm{rx}, N_\mathrm{RF}$ & Number of transmit antennas, receive antennas, RF chains, .\\\hline
$\mathbf{f}_\mathrm{RF}$ & The analog beamforming vector \\\hline
$\mathbf{F}_\mathrm{BB},\mathbf{F}_\mathrm{RF}$ & The baseband precoder, RF precoder for hybrid precoding.\\\hline
$\mathbf{W}_\mathrm{BB},\mathbf{W}_\mathrm{RF}$ & The baseband combiner, RF combiner for hybrid precoding.\\\hline
$\mathbf{a}_\mathrm{BS},\mathbf{a}_\mathrm{MS}$ & Array response vector at the BS, MS\\\hline
$\theta_{\il,p},\phi_{\il,p}$ & The $p$th path spatial angles of arrival and departure at MS, BS, from $\il$th BS.\\\hline
$G_\il$ & Total directivity gain in the link from the $\il$th BS.\\\hline
$(a_k,b_k)$ & PMF parameters of the random variable $G_\il$: $b_k$ is the probability that $G_\il=a_k$ for $k\in\{1,2,3,4\}$.\\\hline
$M_s,m_s,\theta_s$ &The main lobe gain, the side lobe gain and the main lobe beamwidth where $s\in\{\mathrm{MS,BS}\}$.\\\hline
$\mathcal{S}(T)$ & The coverage probability at SINR $T$,  $\mathcal{S}(T) = \mathbb{P}[\mathrm{SINR}>T]$.\\\hline
$\mathcal{A}_p$ & The association probability of the typical user for $p\in\{\mathrm{LOS,NLOS}\}$.\\\hline
$\Psi_\il$ & Number of users connected to the $\il^\text{th}$ BS.\\\hline$\mathcal{R}(\rho)$& The rate coverage probability at rate $\rho$, $\mathcal{R}(\rho) = \mathbb{P}[R>\rho]$.\\\hline
\end{tabulary}
\end{table}

\section{Novel Modeling Aspects: Blocking}
\label{sec:blocking}

Obstacles in the environment affect wireless communication channels owing to reflection, diffraction, scattering, absorption, and refraction. These effects are complicated and environment-specific, and so the received signal power from a transmitter is often modeled statistically, as a function of distance.  The traditional first-order approach to incorporate randomness is to introduce a shadowing random variable, most often log-normal distributed, on top of the average received signal, which is modeled as a function of the distance, e.g. the deterministic power law path loss model $\ell(d) \propto d^{-\alpha}$. The log-normal distribution for shadowing has a classical interpretation in terms of the central limit theorem in view of many independent obstructions \cite{Goldsmith2005}. Shadowing, however, does not accurately capture blocking in dense networks. For instance, blockage not only adds randomness to the average path loss, but also can dramatically change the effective path loss exponent \cite{Rappaport2014book}.  Though the 3GPP standards\cite{3GPPTR36.8142010,3GPP3D} suggest different channel statistics for \ac{LOS} and \ac{NLOS} links in simulations, blockage seems to be a secondary effect in macrocell Sub-6GHz networks mostly due to the fact that the links are long and thus mostly NLOS anyway. Besides, in Sub-6 GHz bands, the path loss exponent $\alpha$ (typically $\alpha > 3$), fitted from measurements using omni-directional antennas, already takes account for the blocking effects, as well as other effects, including diffractions and ground reflections.

Recent experimental investigations have shown a high sensitivity of the \ac{mmWave} channel to blockage effects. To begin with, penetration losses through buildings can be as high as $40-80$ dB\cite{5GChannel}, which is usually insurmountable, and so indoor and outdoor mmWave systems can be considered to be isolated from one another. Moreover, even focusing on the scenario of outdoor-to-outdoor communication, measurements show that static blockages like buildings lead to a large difference in the path loss laws, usually modeled via different path loss exponents, between LOS and NLOS mmWave links\cite{5GChannel,Rappaport2014book}. In the presence of blocking, the path loss in the NLOS links can be much higher, as diffractions are weak \cite{Rappaport2014book,ZhaRyu15}, and a larger fraction of signal energy is scattered in the mmWave bands \cite{Langen1994}. On the positive side, it should be noted that blocking also applies to interfering signals but even more so, since interferers are typically farther than the desired transmitter and thus more likely to be blocked. Besides buildings and other static objects, \ac{mmWave} signals are also attenuated by smaller objects of smaller sizes, e.g. the human body and trees.  At \ac{mmWave} frequencies, the penetration loss through the human body is as high as 20-40 dB \cite{Lu2012,Rajagopal2012}.  Given that most use cases for mmWave involve human users interacting with the device (as well as other humans frequently being nearby), this is a particularly important type of intermittent and severe blocking that changes on a much smaller time scale.

Recent theoretical work in \cite{BaiHea14,Ding2015,Bai2014c,Singh2015,ZhaAnd15} has shown that the coverage and rate trends with blockage switching the path loss exponents can be substantially different from the prior results assuming a conventional power law path loss with a single $\alpha$ value.  This will be discussed further in Section~\ref{sec:implications}. The experimental investigations as well as system capacity results together signify the importance of accurate yet tractable blockage models for analysis of \ac{mmWave} cellular networks. In this section, we first describe the empirical 3GPP blockage model in Section~\ref{sec:blockage:3GPP}. Then, we introduce the analytical blockage models: the random shape theory model in Section~\ref{sec:blockage:RST}, LOS ball model in Section~\ref{sec:blockage:LOS}, and Poisson line model in Section~\ref{sec:blockage:manhattan}. We discussed the model for body and foliage blocking in Section~\ref{sec:blockage:human}. In the end, we present some comparisons between them using real geographic data in Section~\ref{sec:blockage:summary}.

\subsection{3GPP model for incorporating blockages}\label{sec:blockage:3GPP}
The 3GPP standards\cite{3GPPTR36.8142010,3GPP3D}, suggest modeling building blockages by differentiating the \ac{LOS} and \ac{NLOS} links using a stochastic model.  A function $P_{\mathrm{LOS}}(d)$ is a deterministic non-increasing function of $d$ that takes values in $[0,1]$ and is interpreted as the probability that an arbitrary link of length $d$ is LOS. Although 3GPP refers to $P_{\mathrm{LOS}}(d)$ as the ``LOS probabilty function'', it should be understood that it is \emph{not} a traditional probability function (such as a PDF, CDF, or CCDF) but rather just a mapping from a positive distance $d$ to a probability of being LOS in $[0,1]$.   The function $P_{\mathrm{LOS}}(d)$ is modeled differently for varying environments, e.g. urban, suburban and rural areas. For instance, in urban areas with regular street layouts,
\begin{align}\label{eqn:urban_LOS}
P_{\mathrm{LOS}}(d)=\min\left(\frac{A}{d},1\right)\left(1-\mathrm{e}^{-\frac{d}{B}}\right)+\mathrm{e}^{-\frac{d}{B}},
\end{align}
where $A=18$ m, and $B=63$ m\cite{3GPPTR36.8142010}. In suburban areas,
\begin{align}\label{eqn:sub_LOS}
P_{\mathrm{LOS}}(d)=\mathrm{e}^{-d/C},
\end{align}
where $C=200$ m. Note that when $d$ is large, the urban LOS probability in (\ref{eqn:urban_LOS}) has a heavier tail than in (\ref{eqn:sub_LOS}). Intuitively, in a regular urban street grid, users are fairly likely to receive \ac{LOS} signals from far-away base stations on the same street.

The specific values taken for $A, B, C$ in the 3GPP blockage model are based on a relatively limited number of measurements from the WINNER II 2007 document, which pre-dates the deployment of LTE \cite{WINNERII}. In \cite{3GPP3D}, the parameter values were modified to incorporate building heights in the 3D channel model; in \cite{Sun2015,Akdeniz2013a}, the urban LOS probabilities were re-fitted using measurement data in the New York city.  For areas with irregular building deployments, one analytical approach is to fit the parameters based on a few first-order statistics of buildings, e.g the average size and perimeter \cite{Bai2014b}. This last approach will be discussed in the next section.
 
It is essential to classify the links into the LOS and NLOS type, where different path loss laws are applied. Clearance of blockages from the first Fresnel zone of a link has been known to be a good indicator for LOS links \cite{Xia93,Rappaport2014book,Jarvelainen2016}. Fresnel zones are frequency dependent and thus, links that are LOS at 73 GHz need not be LOS at 3 GHz, which has a larger Fresnel clearance zone. A recent white paper\cite{5GChannel} written jointly by Nokia, Qualcomm, Docomo, Huawei, Samsung, Intel, Ericsson and others proposes a 3GPP UMi-like LOS probability function for static blockages which is frequency \emph{independent} for all bands up to 100 GHz. Evaluations of LOS probability incorporating the Fresnel effects in \cite{Jarvelainen2016}, alternatively, suggested significant variations between Sub-6GHz and $f_c >$ 15 GHz networks, but smaller variations in the 16-63 GHz range. This suggests that across the \ac{mmWave} bands it should be possible to use a frequency independent building blockage model, since the Fresnel zone above 15 GHz is narrow. A different model for Sub-6GHz, though, with a significantly wider Fresnel zone, is likely needed.

The blocking models we will now present, in addition to the baseline 3GPP model in \eqref{eqn:urban_LOS}, are all frequency independent.  
Studies to develop frequency-dependent blockage models are still in a nascent stage \cite{Jarvelainen2016}. 

\subsection{Random shape theory model}\label{sec:blockage:RST}
\begin{figure}[t]
	\centering
    \subfloat[Random shape theory model for blockages.]{
		\includegraphics[height=170pt,width=.48\columnwidth]{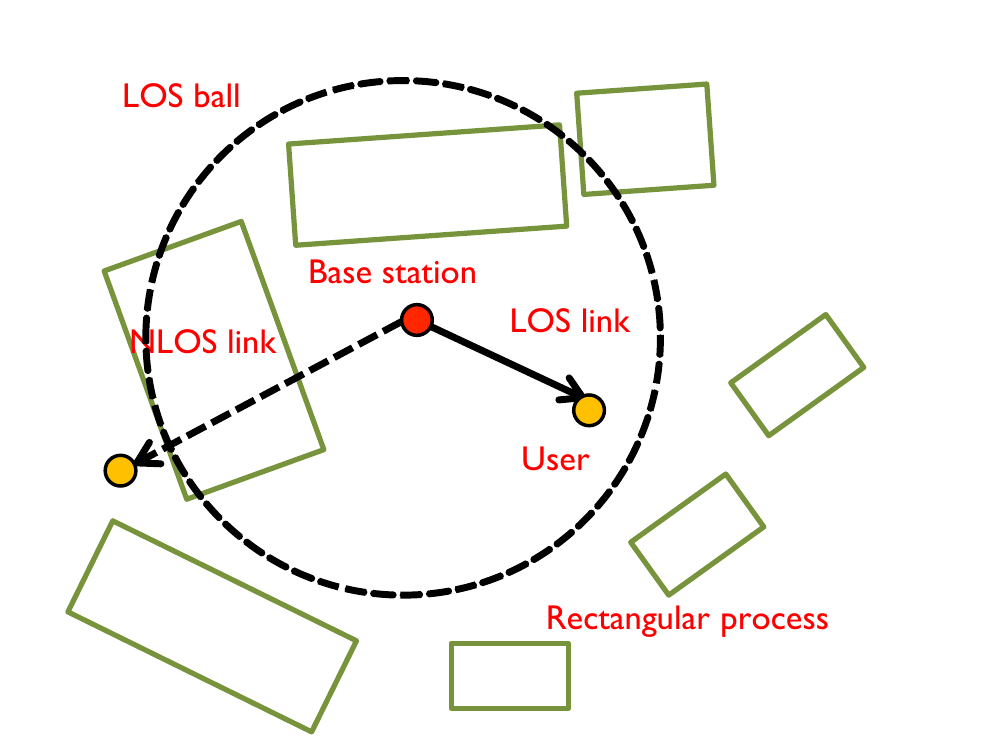}
		}
    \subfloat[Poisson line model for Manhatten-type blockages]{
		\includegraphics[height=170pt,width=.48\columnwidth]{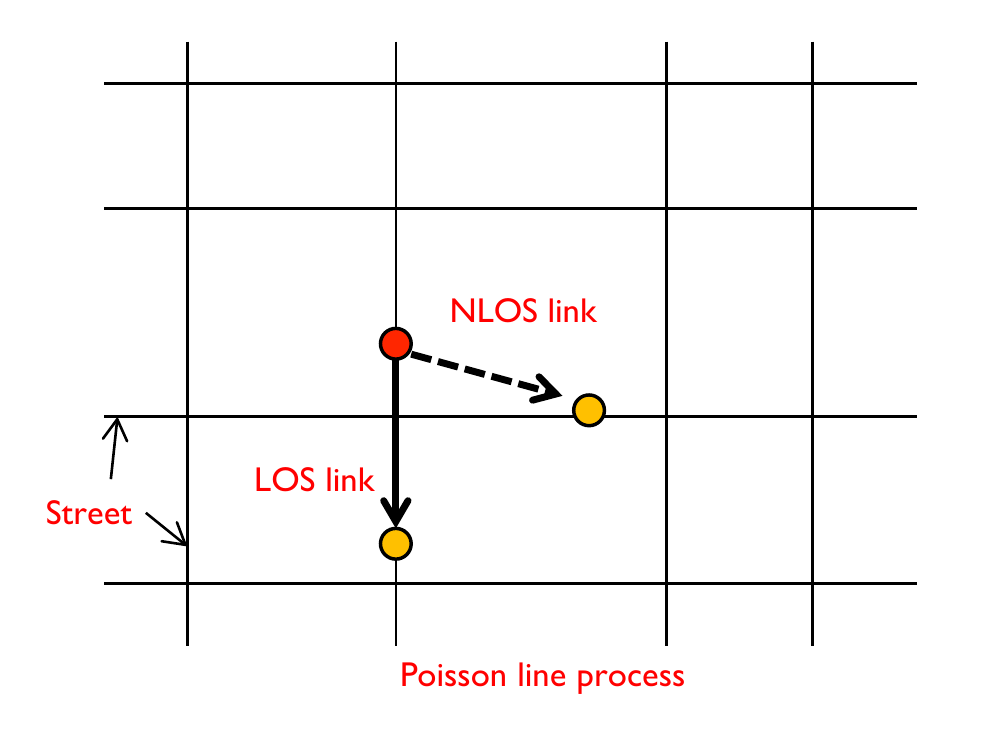}
		}
\caption{Analytical models for building blockages. In (a), the irregular LOS region to the typical user determined by nearby buildings is approximated by a ball in the LOS ball model.}\label{fig:blockage_model}
\end{figure}

To model irregular building deployments, one stochastic blockage model was proposed in \cite{Bai2014b}, based on random shape theory. The Boolean germ grain model is the simplest process of objects in random shape theory \cite{Cowan1989}, where the centers of objects form a \ac{PPP}, and each object is allowed to have independent shape, size, and orientation according to certain distributions. As shown in Fig. \ref{fig:blockage_model}(a), the randomly located buildings are modeled as a Boolean model of rectangles. Interestingly, the analysis in \cite{Bai2014b} showed that the derived LOS probability function has the same form as the 3GPP suburban function in (\ref{eqn:sub_LOS}), which is a negative exponential function of the link length $d$. More importantly, based on the random shape model, the parameter $C$ in (\ref{eqn:sub_LOS}) can be analytically computed using the statistics of the buildings in the area. For example, assuming the orientation of the buildings are uniformly distributed in space,
\begin{equation}
\label{eqn:rst1}
C=\frac{\pi}{\lambda_{\mathrm{bldg}}\mathbb{E}\left[L_\mathrm{bldg}\right]},
\end{equation}
where $\lambda_{\mathrm{bldg}}$ is the average number of buildings in a unit area, and $\mathbb{E}\left[L_\mathrm{bldg}\right]$  is the average perimeter of the buildings in the investigated region. Another way to obtain $C$ is to choose
\begin{equation}\label{eqn:rst2}
C=-\frac{\pi \mathbb{E}\left[A_\mathrm{bldg}\right]}{\ln(1-\kappa)\mathbb{E}\left[L_\mathrm{bldg}\right]},
\end{equation}
where $\mathbb{E}\left[A_\mathrm{bldg}\right]$ is the average area of the buildings in the investigated region and $\kappa$ is the fraction of area under buildings.
The results in (\ref{eqn:rst1}) and (\ref{eqn:rst2}) provide a quick way to approximate the parameters of the LOS probability function without performing extensive simulations and measurements. Since the buildings in a geographical region are not necessarily rectangles, (\ref{eqn:rst1}) and (\ref{eqn:rst2}) lead to slightly different estimates in general. For example, the UT Austin building topology in Fig.~\ref{fig:buildings} corresponds to $C = 100$m with (\ref{eqn:rst1}) and $C = 85$m with (\ref{eqn:rst2}). 

\subsection{LOS ball model}\label{sec:blockage:LOS}
To simplify the mathematical derivation in the system-level analysis, a {\it LOS ball model} was proposed \cite{bai2013e,BaiHea14}, where the LOS probability function is modeled as a simple step function
\begin{align}\label{eqn:LOSball}
P_{\mathrm{LOS}}(d)=\mathds{1}(d<R_\mathrm{B}),
\end{align}
$\mathds{1}(\cdot)$ is the indicator function, and $R_\mathrm{B}$ is the maximum length of a LOS link. As shown in Fig. \ref{fig:blockage_model} (a), in the LOS ball model, {\it the LOS area}, defined as the area that is LOS to a typical user, is characterized by a ball of radius $R_\mathrm{B}$. Consequently, the maximum LOS length $R_\mathrm{B}$ can be determined by fitting the average size of the \ac{LOS} area, from either the \ac{LOS} probability functions derived from other models or geographic datasets. For instance, to have the same average \ac{LOS} area with the random shape theory model,
\begin{align}\label{eqn:rst}
R_\mathrm{B}=\frac{\sqrt{2}\lambda_{\mathrm{bldg}}\mathbb{E}(L_\mathrm{bldg})}{\pi}.
\end{align}
When the base station density is high, only a minor gap in terms of the SINR distributions is observed using a fitted \ac{LOS} ball model versus the random shape theory model, which is impressive given its simplicity.

In \cite{Singh2015}, a \emph{generalized LOS ball model} was proposed and validated using 2-D real building locations in Manhattan and Chicago downtown regions. The probability of a link being LOS in this model is:
\begin{align}\label{eqn:LOSball2}
P_{\mathrm{LOS}}(d)=p_l\mathds{1}(d<R_\mathrm{B}),
\end{align}
where the LOS fraction constant $p_l\in[0,1]$ represents the average fraction of the LOS area in the ball of radius $R_\mathrm{B}$. Clearly, for $p_l = 1$, this reverts to the previous LOS ball model. MATLAB code to extract and process building data, and differentiate between LOS and NLOS links has been made available online\cite{KulBlockageCodes15}. 

\subsection{Poisson line model}\label{sec:blockage:manhattan}
To model a dense urban environment, a Poisson line model was proposed in \cite{Baccelli2015}. As shown in Fig. \ref{fig:blockage_model}(b), the streets are abstracted as a grid of Poisson lines; the intersections along one line are assumed to be randomly distributed as a Poisson process. The users and base stations located on the lines are considered outdoor, where the locations inside the blocks are indoor; two outdoor locations are considered to be \ac{LOS} if and only if they are on the same line. In \cite{Baccelli2015}, it was shown that the Poisson line model offers a tractable way to incorporate the correlations in the \ac{LOS} probabilities between different links, which was ignored in prior analysis and simulations. The results in \cite{Baccelli2015} show that the tail behavior of the SINR distributions can be different when incorporating the correlations in the \ac{LOS} probability.

\subsection{Human body and foliage blockage models}\label{sec:blockage:human}
The models discussed thus far are primarily motivated from macroscopic rigid obstructions like buildings. There are some recent attempts to model blockage effects due to smaller objects like trees or the human body.  In \cite{Rappaport2015a}, the foliage loss in dB is found to be a linear function of the path length through tree canopies. In \cite{ThomasVook14}, ray tracing was used to come up with a distance-based blocking probability function by other users and foliage was fitted from ray-tracing simulations as a linear function of the link length $x$. The LOS  probability was found to be of the form $\mathrm{min}(a x+ b,c)$, where the parameters $a,b,c$ are deployment dependent.
In \cite{Bai2014c}, a cone-blocking model was proposed to model the probability of self-body blocking in outdoor \ac{mmWave} cellular networks, where all the signals from a cone in the angle space are assumed to be blocked by the user's self-body, and the fraction of the blocking cone can be estimated based on the position and size of the user.  In \cite{Venugopal2016}, a human body blocking model was proposed for indoor mmWave wearable networks, where the bodies of both the self-user and other users are modeled as cylinders of certain sizes, and the blocking probability of a link was computed as a function of the relative locations. Most of the current analysis focuses on static blockages and users. In the future, it would be interesting to incorporate time dynamics to study the impact of penetration losses on coverage from mobile obstacles and the resulting impact on handover rates. For example, the dynamics of self-body blocking can be modeled as a shift of the blocking cone over time \cite{Bai2014c}.

\subsection{Comparison and conclusions on blockage models}\label{sec:blockage:summary}

\begin{figure}[t]
	\centering
     \subfloat[UT Austin neighbourhood as of 2013\cite{Austinbuildings13}, 1km by 1.3km area centered at ($30^o 17'13.2''\text{N}, 97^o 44'16.9''\text{W}$)]{
		\includegraphics[width=.48\columnwidth]{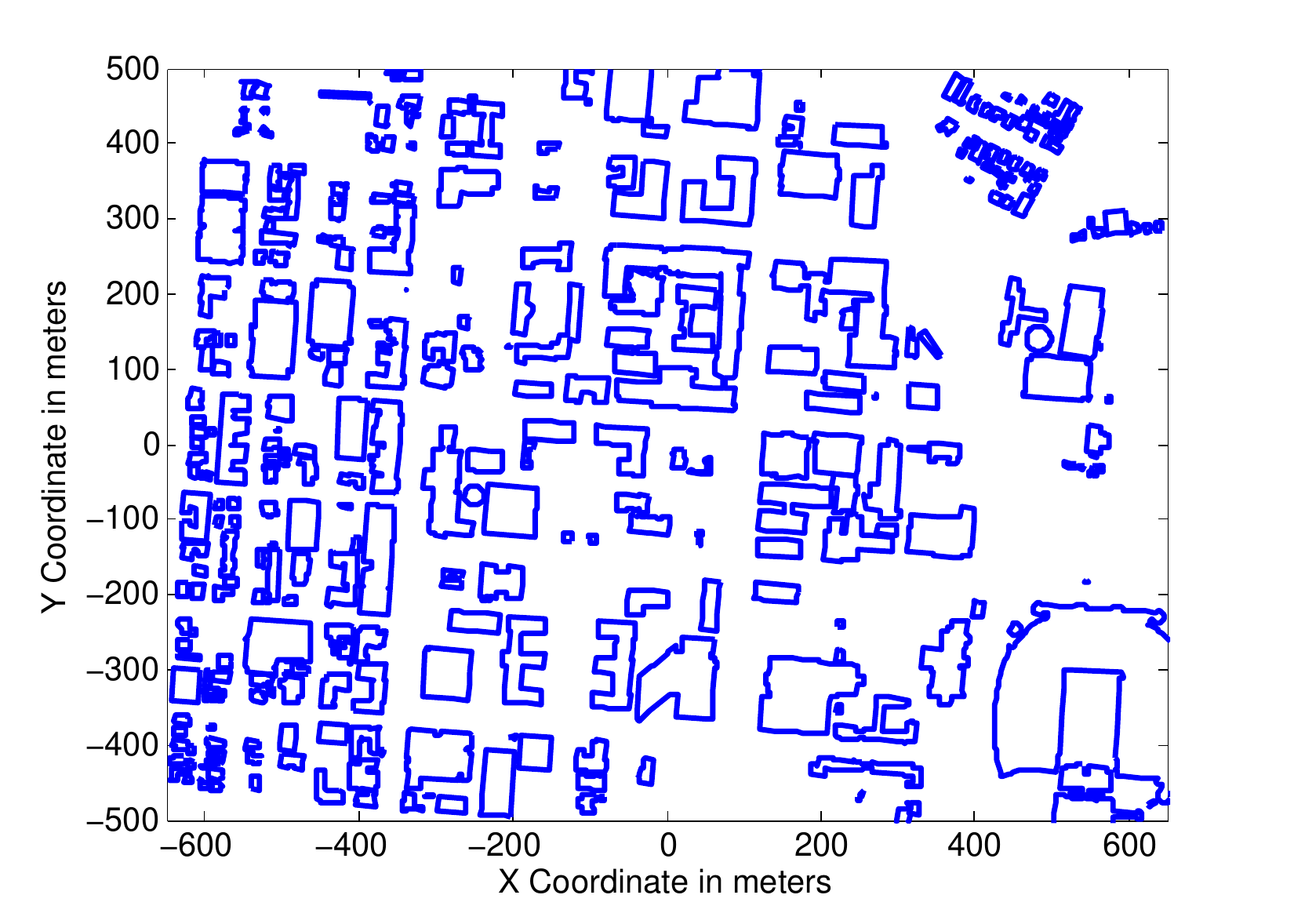}
		}
     \subfloat[LA downtown as of 2008\cite{LAbuildings08}, 2km by 2km area centered at ($34^o 02'30.8''\text{N}, 118^o 15'06.8''\text{W}$)]{
		\includegraphics[width=.48\columnwidth]{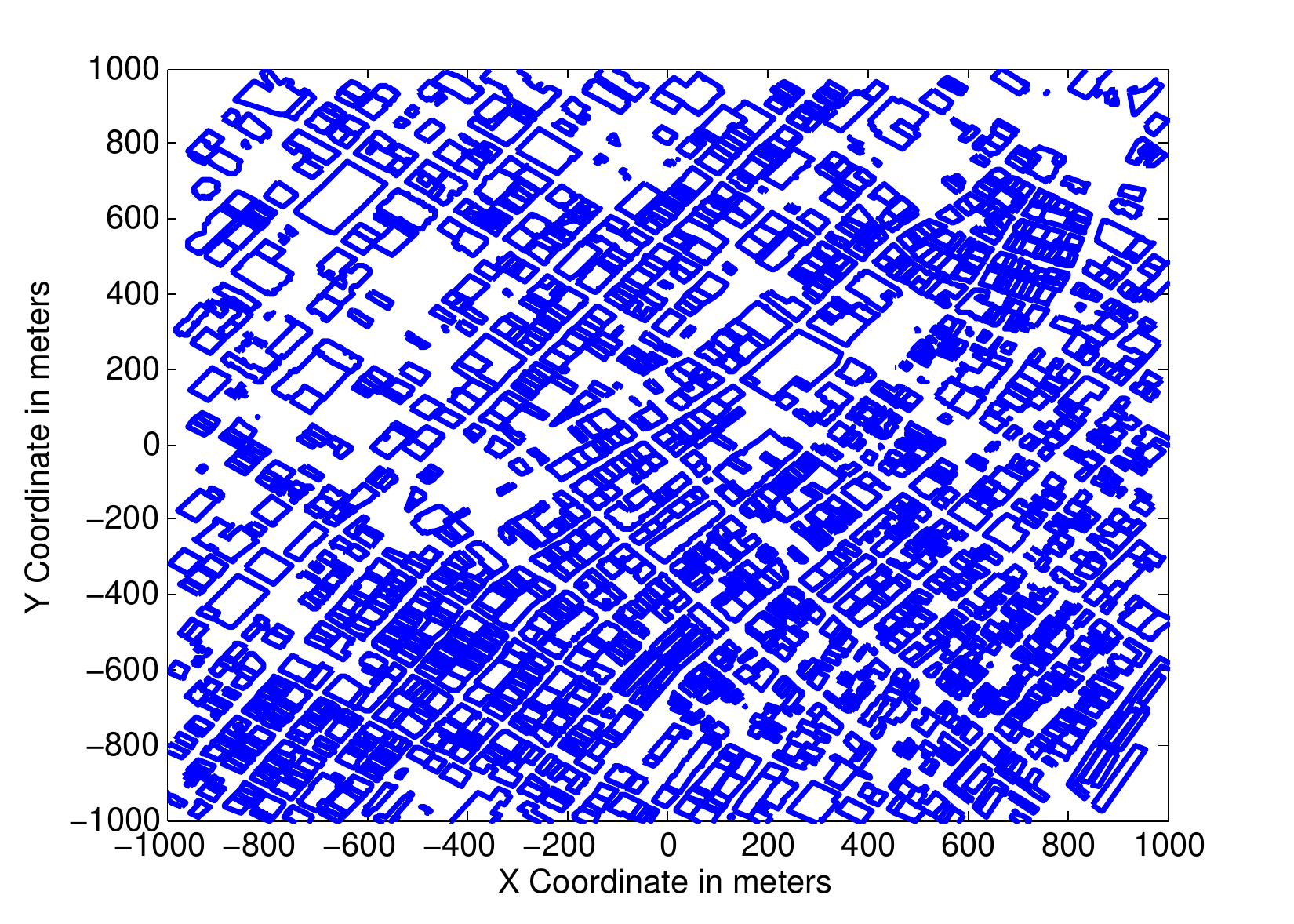}
		}
	\caption{2-D building locations used for comparing blockage models.}\label{fig:buildings}
\end{figure}

We have overviewed several blockage models, each with their own set of modeling assumptions. An obvious question is when to use which blockage model? We attempt to answer this question using simulation methodology similar to \cite{Singh2015}, based on 2-D real building data in the UT Austin and downtown LA regions as shown in Fig.~\ref{fig:buildings}. Though, every different environment will experience different blocking behavior, observations based on these two environments, along with our previous studies for NYC and Chicago, share a few common points.

We consider a 28 GHz carrier frequency with 200 MHz of bandwidth operating in the downlink. Path loss exponents are chosen to be 2 for LOS and 3.3 for NLOS, and lognormal shadow fading has standard deviation of 3.1 dB for LOS and 8.2 dB for NLOS\cite{5GChannel}. The noise figure is 10 dB and the transmit power is 30 dBm. UEs are assumed to be omnidirectional and BSs have a step beam pattern (refer Fig. \ref{fig:Beam_Pattern}) with 10 degree 3 dB beamwidth, 18 dB maximum gain and 20 dB front to back ratio. For simulations with actual buildings, we consider a dense network with an average 30 BSs/km$^2$ distributed randomly in the outdoor region. If the urban region has dimensions $\mathrm{X}\times \mathrm{Y}$, then the user location whose performance is to be evaluated is placed outdoors randomly in the central $\mathrm{X}/2\times \mathrm{Y}/2$ rectangle.  We now summarize some key observations and suggest methodology for choosing the parameters for the blockage models.
\begin{figure}[t]
\centering
\includegraphics[width=.48\columnwidth]{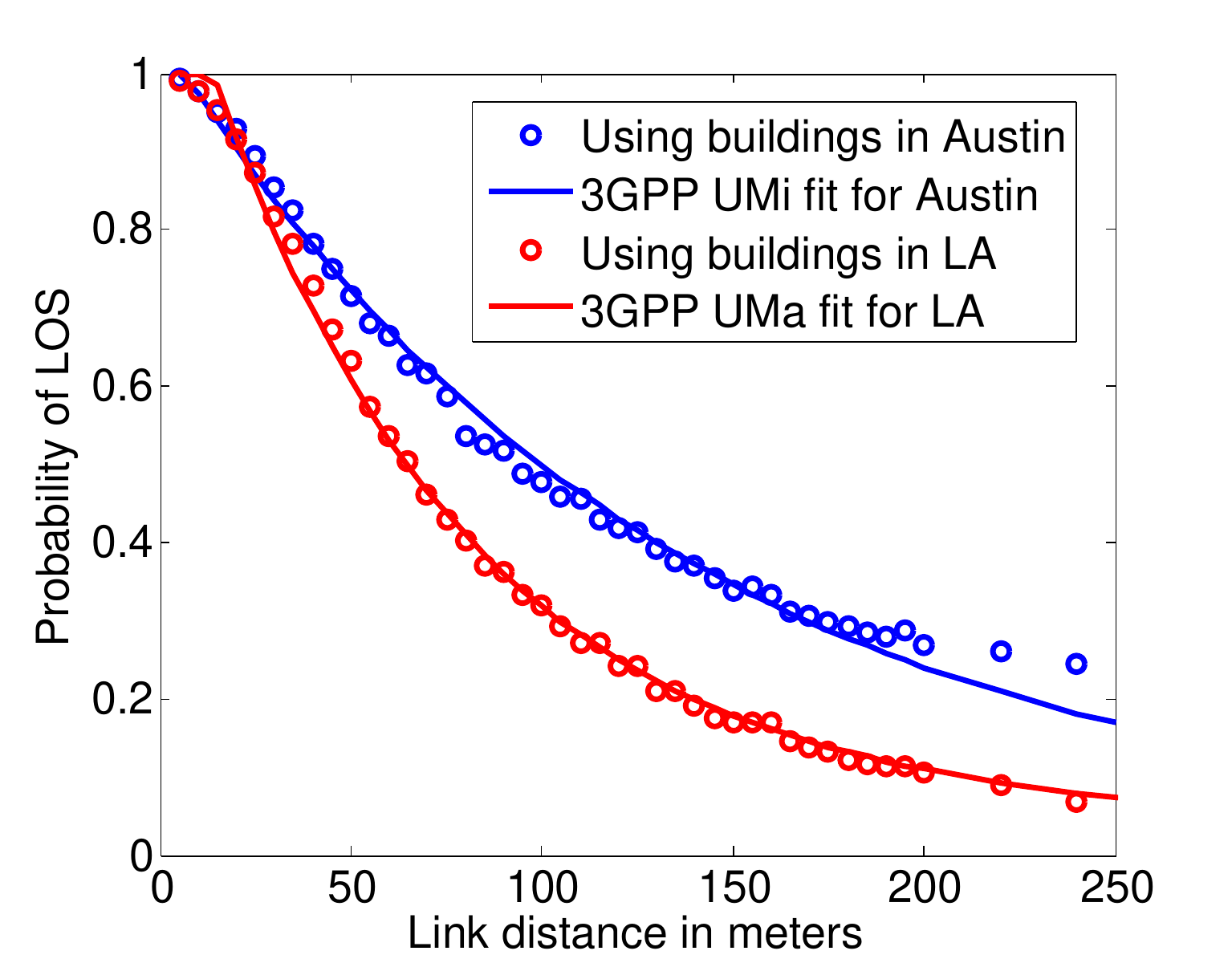}
\caption{Fitting LOS probability for 3GPP-like UMi model.}\label{fig:comparelosprobability}
\end{figure}
\begin{figure}[t]
	\centering
     \subfloat[UT Austin]{
		\includegraphics[width=.48\columnwidth]{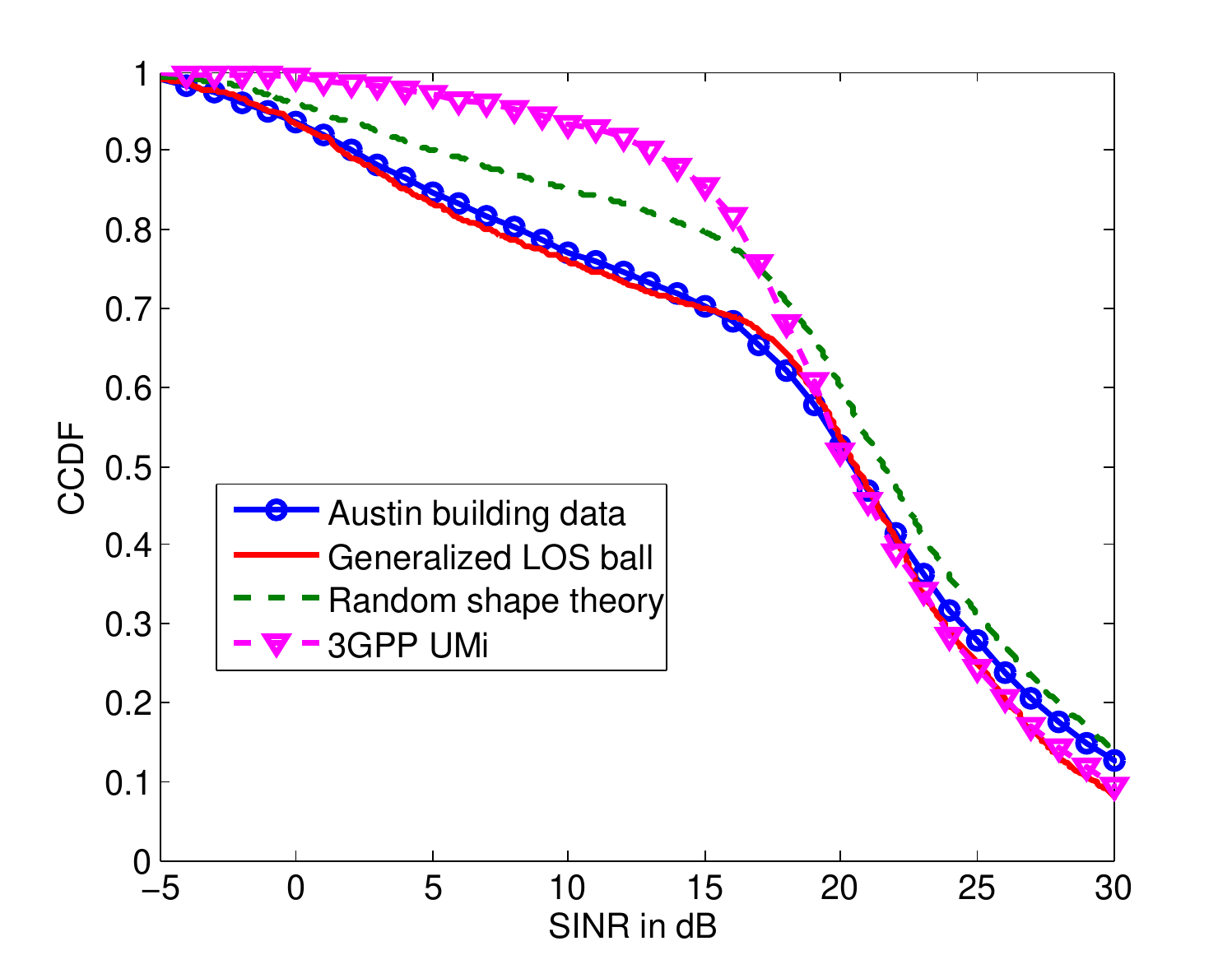}
		}
     \subfloat[LA Downtown]{
		\includegraphics[width=.48\columnwidth]{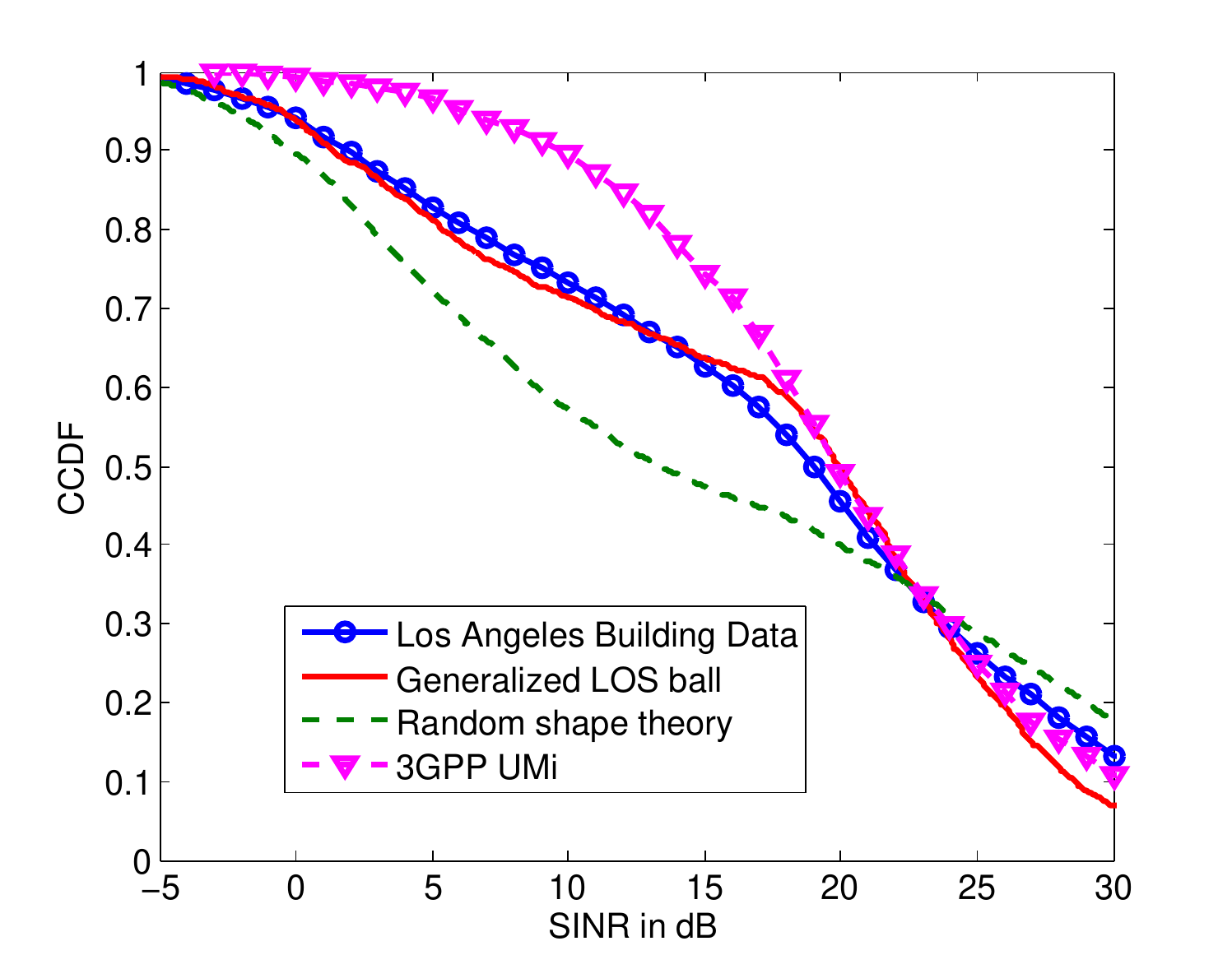}
		}
	\caption{Comparison of blocking models.}\label{fig:comparesinr}
\end{figure}

\paragraph{3GPP-like model} We curve fit the LOS probability obtained using the building locations with that in (\ref{eqn:urban_LOS}). As can be seen from Figure~\ref{fig:comparelosprobability}, the fit is good with root mean squared error $2.18\%$ for the Austin region and $1.45\%$ for LA region. This matches the insight in \cite{5GChannel}, that 3GPP-like models could be sufficient to fit the LOS probability in urban regions. The parameters obtained for Austin and LA are as follows. $A= 6.659$m and $B = 129.9$m for Austin and $A = 13.89$m and $B = 63.76$m for LA. However, as shown in Figure~\ref{fig:comparesinr}, fitting the LOS probability but neglecting the correlation does not necessarily mean a good fit to metrics of interest, like \ac{SINR} or rate coverage.

\paragraph{Random shape theory model} 
Conditioned on the users and BSs being outdoors, a simple upper bound on the LOS probability is $P_{\rm LOS}(d) = \min\left(\exp(-d/\mathrm{C}+\delta),1\right)$, where here $\delta = -\ln(1-\kappa)$, where $\kappa$ is the fraction of area under buildings. Using (\ref{eqn:rst2}), $\mathrm{C}=85$m for Austin and $\mathrm{C} = 42$m for Los Angeles. Also, $\kappa = 0.27$ for Austin and $\kappa = 0.42$ for LA.
From Fig.~\ref{fig:comparesinr} it can be seen that this LOS probability gives a reasonably tight upper bound to the SINR coverage obtained using real building locations near UT Austin, similar to \cite{Bai2014b}. However for LA, this model underestimates the coverage. 

\paragraph{Generalized LOS ball model}
Similar to \cite{Singh2015}, $p_l$ is computed as the average LOS fractional area in a ball of radius $R_\mathrm{B}$ from the region under consideration. Since we consider only outdoor deployments for users and BSs, the fractional area is computed as the ratio of LOS area in ball of radius $R_\mathrm{B}$ centered at a user location and the total outdoor area in that ball  averaged over several such user locations. The choice of $R_\mathrm{B}$ is flexible in this model, but it should be large enough to make sure that with high probability, the serving link and dominant interfering base stations fall within the ball of radius $R_\mathrm{B}$ centered at the user. Generally, mmWave networks are envisioned to be dense with inter-site distance less than or equal to 200m: we choose $R_\mathrm{B} = 200$m. The corresponding $p_l= 0.3027$ for Austin and $p_l = 0.2419$ for LA. This model accurately fits the \ac{SINR} coverage obtained using both the urban regions under consideration, which is surprising considering the simplicity of the model. The observations on this model until now have suggested that the choice of the ball radius between 150-300m gives a good fit for dense random deployment of BSs (with cell radius typically lesser than $R_\mathrm{B}$) in Manhattan, Chicago, LA and Austin regions considered in \cite{Singh2015} and this paper.  Further empirical studies, including possible joint optimization of  $R_\mathrm{B}$ and $p_l$ to optimize the SINR curve fit, would be useful.

All the blockage models mentioned in this comparison section neglect the correlation of two links being blocked by the same obstruction. The Poisson line model\cite{Baccelli2015} can handle correlations, but is difficult to validate because it assumes a very specific street and user geometry quite different than all the other models (or most real cities). The LOS ball and the random shape theory models are simple to incorporate in the  analysis, wherein the above observations imply that an appropriate choice of the blockage parameters potentially reflects real world blockage scenarios. The 3GPP-like urban micro-cellular model is more complex to incorporate in analysis, and it was observed that fitting the empirical LOS probability function does not guarantee a good fit to the coverage estimates, in fact it in most cases has a much poorer fit that the LOS ball or the random shape theory blocking model.  

We conclude by noting that the analytical approach developed starting in Sect. \ref{sec:analysis} depends on the blocking model only through the use of a generic $P_{\rm LOS}(d)$ function so an arbitrary blocking model can be used. The analytical results, however, can be obtained in simple forms, when certain LOS probability models, e.g. the LOS ball model, are applied in the derivation.

\section{Novel Modeling Aspects: Large Antenna Arrays}
\label{sec:antennas}

The use of large -- in terms of the number of elements, not the physical size -- antenna arrays at the base station and mobile users is a key feature of \ac{mmWave} cellular systems. The ways these antennas are used at mmWave differs from lower frequencies owing to hardware limitations on MIMO transceiver architectures. In this section, we discuss how \ac{mmWave} single-user/multi-user MIMO transmission techniques differ from their counterparts at lower frequencies. Understanding these large antenna array aspects is essential for proper modeling and analysis of \ac{mmWave} cellular systems.

\subsection{Hardware constraints and the need for different transceiver architectures} \label{subsec:hardware}

\begin{figure}[t] 
	\centering
	\subfloat[Fully-digital architecture]{
		\includegraphics[height=170pt,width=.3\columnwidth]{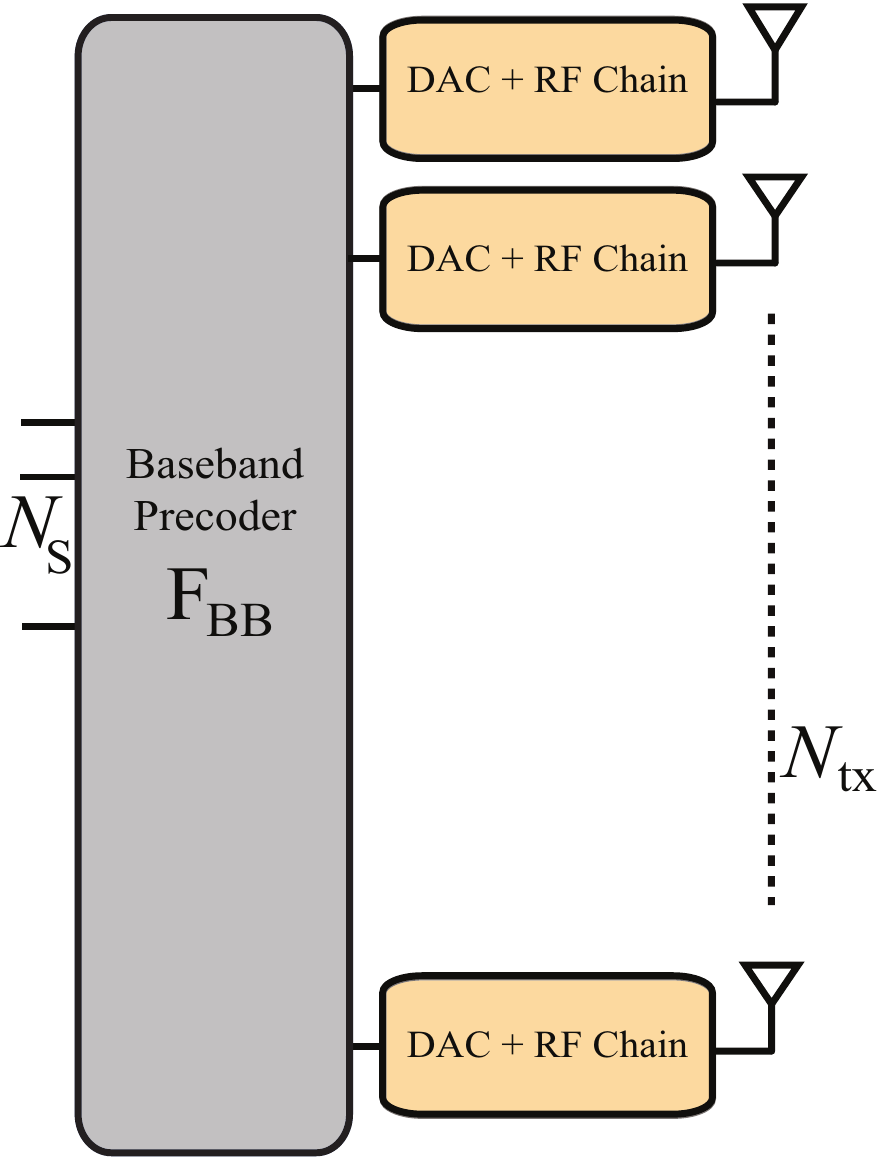}
		\label{fig:Digital_Arch}}
	\subfloat[Analog-only architecture]{
		\includegraphics[height=170pt,width=.27\columnwidth]{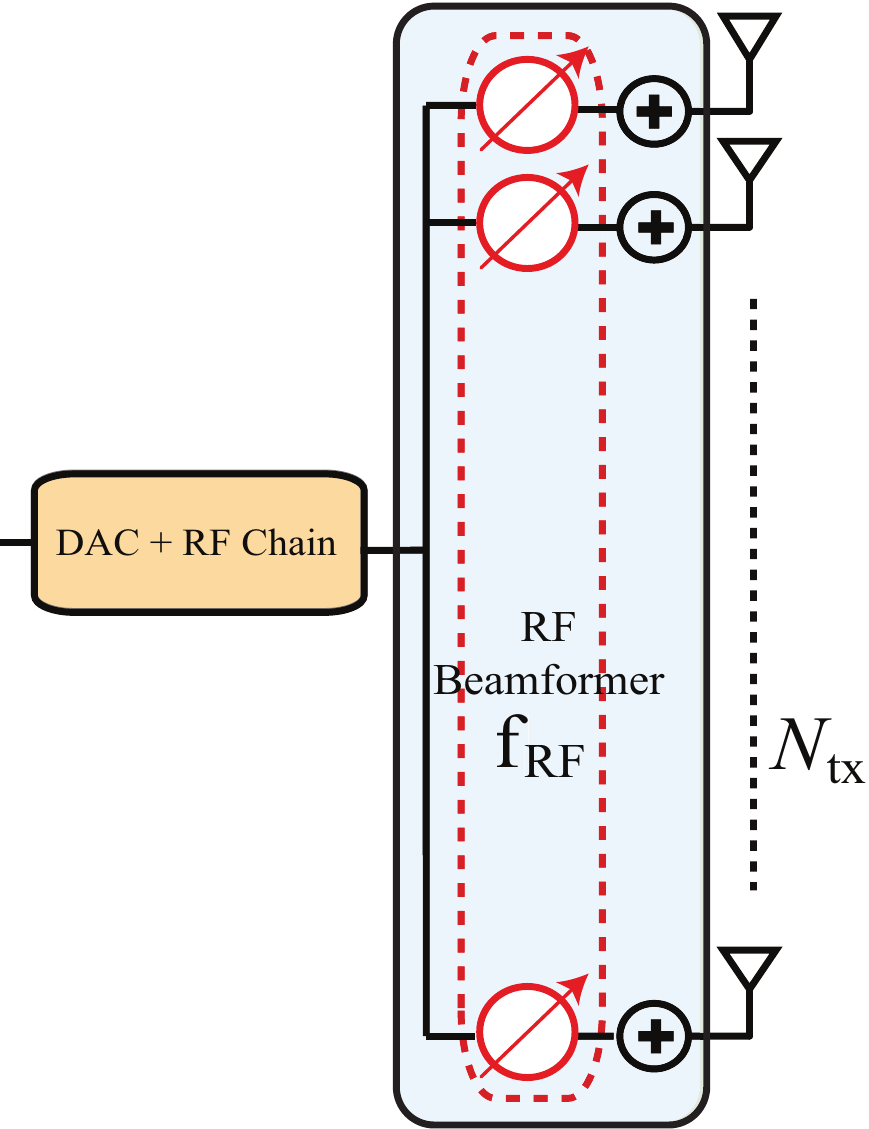}
		\label{fig:Analog_Arch}}
	\subfloat[Hybrid analog/digital architecture]{
		\includegraphics[height=170pt,width=.4\columnwidth]{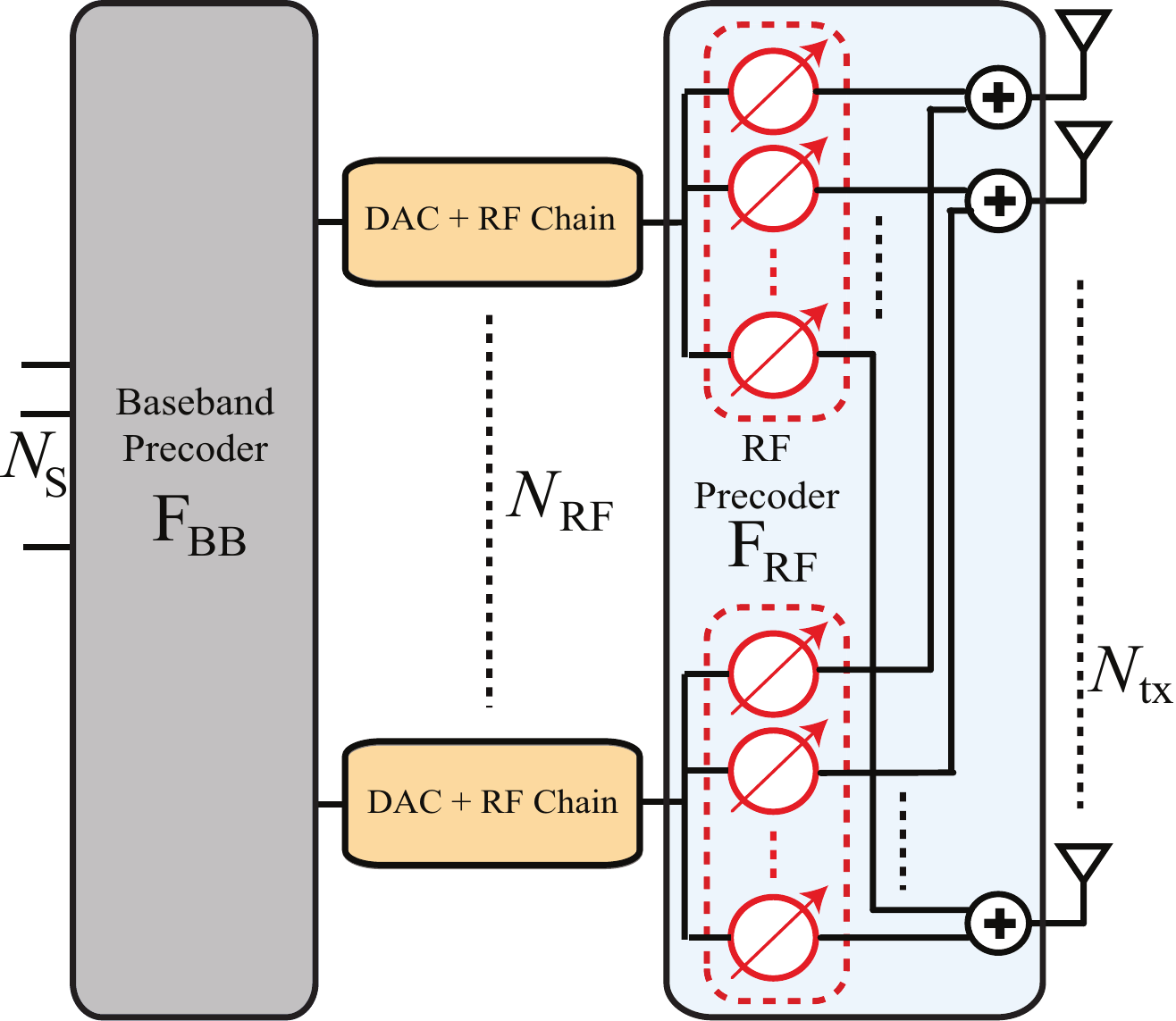}
		\label{fig:Hybrid_Arch}}
	\caption{This figure shows a tranmsitter having $\Ntx$ antennas with a fully-digital, analog-only, or hybrid analog/digital architecture. In the hybrid architecture,  $N_\mathrm{RF} \ll \Ntx$ RF chains are deployed.}
	\label{fig:Trans_Arch}
\end{figure}
 
Initial \ac{mmWave} research and prototypes suggest array sizes of $32 - 256$ antennas at the base station and $4-16$ antennas at the mobile users \cite{Roh2014,Khan2011,Hong2014,Han2015}. Realizing these numbers of antennas in a small package is feasible thanks to the recent developments in antenna circuit design \cite{Boers2014,Chen2010,Emami2011,Hong2014,Sun2013,Biglarbegian2011}. The large arrays, though, can not be used at mmWave in the same way they are used at lower frequencies due to the high power consumption of the mixed-signal components.  

In conventional cellular systems, precoding and combining is performed at baseband using digital signal processing. This allows better control over the precoding/combining matrices, which in turn facilitates the implementation of sophisticated single user, multiple user, and multi-cell precoding algorithms. Performing such baseband precoding/combining processing assumes that the transceiver dedicates an RF chain per antenna as shown in Fig. \ref{fig:Trans_Arch}(a). This fully-digital processing is hard to realize at \ac{mmWave} frequencies with wide bandwidths and large antenna arrays. This is mainly due to the high cost and power consumption of mixed-signal components, like high-resolution analog-to-digital converters (ADCs) \cite{Singh2009,Murmann2015}. For example, it is presently infeasible for mmWave receivers to have 32-256 full-resolution ADCs, so traditional MIMO transceiver architectures that allocate an RF chain for each antenna are very difficult to realize. Different transceiver architectures that comply to these hardware constraints have therefore been proposed \cite{ElAyach2014,Mo15,Alkhateeb2014d,Han2015,Mendez-Rial2016,Brady2013}. We now overview some key candidate transceiver architectures for \ac{mmWave} wireless systems.

\paragraph{Analog beamforming} An immediate solution to overcome the limitation on the number of RF chains is to perform beamforming entirely in the RF domain using analog processing. Analog beamforming is normally implemented using networks of phase shifters as shown in Fig. \ref{fig:Trans_Arch}(b) \cite{Wang2009,Xia2008b}. The weights of these phase shifters are tuned to shape and steer the transmit and receive beams along the dominant propagation directions. Mathematically, if the transmitter wants to transmit a symbol $s$, with the $\Ntx \times 1$ beamforming vector $\bff_\mathrm{RF}$, then the transmitted vector $\bx$ can be written as 
\beq
\bx=\bff_\mathrm{RF} s,
\eeq  
where the entries of the RF beamforming vector are subject to a constant modulus constraint due to the implementation with phase shifters. Therefore, these entries can be expressed as $\left(\bff_\mathrm{RF}\right)_n=e^{j \theta_n}, n=1, 2, ..., \Ntx$. Depending on the channel and the antenna array geometry, these phases $\left\{\theta_n\right\}_{n=1}^{\Ntx}$ are designed normally to maximize the beamforming gain at the receiver. To avoid the overhead of explicitly estimating the large \ac{mmWave} channel, analog beamforming weights can be directly trained using beam training \cite{Wang2009}. One common approach for beam training is to use a codebook of beam patterns at different resolutions, and iteratively find the the best beamforming vector codeword from this codebook \cite{11ad,Wang2009,Hur2013}.  Despite its simplicity, analog beamforming is subject to hardware constraints such as the phase shifter quantization, which make analog beamforming/combining solutions limited to single-stream transmission and difficult to extend to multi-stream or multi-user MIMO communication. Analog beamforming is available already in Wireless HD and IEEE 802.11ad products, therefore it is seen as commercially viable in the near-term. Much of the analysis in Section \ref{sec:analysis} assumes analog beamforming. 

\paragraph{Hybrid precoding} Hybrid analog/digital architectures provide a flexible compromise between hardware complexity and system performance \cite{Zhang2005a,ElAyach2014,Alkhateeb2014b,Alkhateeb2015a,Han2015,Ni2016,Yu2016,Sohrabi2015}. In hybrid architectures, the precoding/combining is divided between the analog and digital domains as illustrated in Fig. \ref{fig:Trans_Arch}(c). This allows the use of a number of RF chains $\Nrf$ much less than the number of antennas, i.e. $\Nrf \ll \Ntx$. One key advantage of hybrid precoding is that it permits the transmitter and receiver to communicate via several independent data streams, and hence achieve spatial multiplexing gains \cite{Alkhateeb2014d}. Consider a BS transmitting $N_\mathrm{S}$ data streams to a mobile user, and both of them employing hybrid architectures with $\Nrf$ RF chains. Let the $N_{\rm RF} \times N_{\rm S}$ matrix $\bF_{\rm BB}$, and the $\Ntx \times \Nrf$ matrix $\bF_{\rm RF}$ denote the baseband and RF precoders at the BS, and the $N_{\rm RF} \times N_{\rm S}$ matrix $\bW_{\rm BB}$, and the $\Nrx \times \Nrf$ matrix $\bW_{\rm RF}$ represent the baseband and RF combiners at the mobile user. Then, the received signal after processing can be written as 
\beq
\by=\bW^*_{\rm BB} \bW^*_{\rm RF} \bH \bF_{\rm RF} \bF_{\rm BB} \bs + \bW^*_{\rm BB} \bW^*_{\rm RF} \bn,
\eeq
where the RF precoders/combiners are subject to a similar implementation constraints as those discussed in the analog beamforming section. 

Despite the much smaller number of RF chains compared to the number of antennas, hybrid architectures were shown to achieve near-optimal performance compared to fully-digital transceivers in \cite{ElAyach2014,Alkhateeb2014b,Alkhateeb2015a,Han2015,Ni2016,Yu2016,Sohrabi2015}. To further reduce the power consumption, \cite{Mendez-Rial2016,Alkhateeb2016a} proposed to replace the phase shifter networks in the hybrid architectures with a network of switches. The RF precoding matrices can also be realized using lens antennas, which compute the spatial Fourier transform, and can work as analog beamforming vectors with a DFT structure \cite{Brady2013}. As the power consumption in the full-resolution ADCs may be a challenge at mmWave, \cite{Singh2009,Mo15} proposed using low-resolution ADCs. Hybrid architectures with few-bit ADC receivers have also been recently investigated in \cite{Mo2016}. Extending the system analysis in Section \ref{sec:analysis} to include all the facets of hybrid precoding or other architectures is largely a topic for future work. 

\subsection{Spatial channel modeling} \label{subsec:channel-models}

Measurements of outdoor \ac{mmWave} channels show that they normally have a small number of dominant scattering clusters \cite{Rappaport2013a,Akdeniz2013a,Sam16}. Therefore, geometric channel models with a few clusters are commonly adopted to describe \ac{mmWave} channels for system capacity analysis or precoder design \cite{Akdeniz2013a,ElAyach2014,HeathJr2015}. Most studies assume a channel which is non-selective in both time and frequency for simplicity, although there is some recent work on designing precoders and combiners for frequency selective \ac{mmWave} channels\cite{Alkhateeb2015a}. Let the $\Nrx \times \Ntx$ matrix $\bH_\il$ denote the downlink channel from the $\il^\text{th}$ BS at $X_\il$ to a typical user at origin. Then, $\bH_\il$ can be written as
\begin{equation}
\label{eq:channel}
\bH_\il = \frac{1}{\sqrt{\ell(|X_\il|)}}\sum_{p=1}^{\eta_\il} h_{\il,p} \ \ba_\MS\left(\theta_{\il, p}\right) \ba^*_\BS\left(\phi_{\il, p}\right),
\end{equation}
where $h_{\il,p}$ is the small-scale fading of the $p$th path, $\ell(|X_\il|)$ is the path loss, $\eta_\il$ is the total number of paths between the BS and user, wherein each path is a representative of a cluster of paths due to a scatterer in the environment. The angles $\theta_{\il, p}$ and $\phi_{\il, p}$ denote the $p$th path spatial angles of arrival and departure (AoA/AoD) at the user and the BS. Finally, $\ba_\MS\left(\theta_{\il,p}\right)$ and $\ba_\BS\left(\phi_{\il, p}\right)$ are the array response vectors at the MS and BS, respectively. The spatial angles are a function of the physical AoA/AoD as well as the array geometry. For a uniform linear array (ULA) with $N$ antennas, where $N\in\{\Ntx,\Nrx\}$, inter-antenna spacing $\mathrm{d}$ and steered at some physical AoA/AoD given by $\varphi$, the corresponding spatial angle is $\theta = 2\pi \mathrm{d} \sin(\varphi)/\lambda_c$ and the array response vector is given by 
\begin{equation}
\ba(\theta) = \begin{bmatrix}
1 & \exp(-j\theta) & \exp(-2j\theta) & \ldots & \exp(-j(N-1)\theta)
\end{bmatrix}^*.
\end{equation}
The distribution of the AoAs/AoDs can be modeled using the empirically observed power angular spectrum\cite{SamRap14,Rappaport2015}. 

For uniform arrays, a useful representation of the channel in \eqref{eq:channel} can be obtained by characterizing the channel response at the spatial quantized angles $0,2\pi/N\ldots,2\pi(N-1)/N$. This is particularly useful for network level analysis with hybrid or analog precoders/combiners using phase shifters or lenses and a large number of antennas at the BSs and MSs, as it gives rise to the ON/OFF nature of interference\cite{Alkhateeb2014b,KulGhoAnd16}. The reason is that each array response vector becomes now equivalent to a column of the $N$-point DFT matrix. This channel characterization, which is called the virtual channel representation \cite{Sayeed2002}, is defined as
\begin{equation}
\bH_\il = \bA_\mathrm{R} \widetilde{\bH_\il} \bA^*_\mathrm{T} = \sum_{k=1}^{\Ntx}\sum\limits_{l=1}^{\Nrx}[\widetilde{\bH_\il}]_{k,l} \ba_\MS\left(\theta_{k}\right) \ba^*_\BS\left(\phi_{l}\right)
\end{equation}
where $\bA_\mathrm{R}$ and $\bA_\mathrm{T}$ contain the array response vectors for the receiver and transmitter with spatial AoAs (AoDs) taken over a uniform grid of size $\Nrx$ ($\Ntx$), and $\tilde{\bH_\il}$ is a matrix with each entry representing the channel gain corresponding to a different combination of the permissible AoAs/AoDs. Exploiting the sparseness of the \ac{mmWave} channel in the spatial domain, most of the terms in the double summation will be zero and the above representation can be equivalently represented as a single summation over the distinct paths between the BS and user, as given in (\ref{eq:channel}) but with quantized spatial AoA/AoD.

\subsection{Single stream analog beamforming} \label{subsec:single-stream}
\begin{figure}[t]
	\centering
	\includegraphics[width=.4\columnwidth]{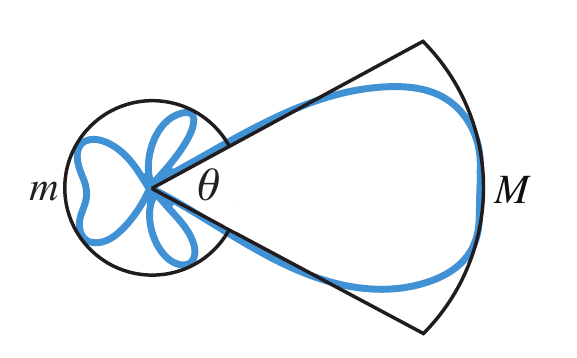}	
	\caption{Approximated sectored-pattern antenna model with main-lobe gain $G_{\rm BS}$, side-lobe gain $g_{\rm BS}$, and main-lobe beamwidth $\Theta_{\rm BS}$.}
	\label{fig:Beam_Pattern}
\end{figure}

In single stream beamforming, the BS and mobile user use the antenna arrays to transmit/receive one data stream. Let $\bff$ and $\bw$ denote the beamforming/combining vectors, the receive $\snr$ can be expressed as
\beq
\snr =\frac{|\bw^* \bH \bff|^2}{\sigma^2}.
\eeq

The design objective for the beamforming/combining vectors is usually to maximize this $\mathsf{SNR}$. When the channel is dominated with a LOS path or when the number of scatterers is small, it becomes reasonable to design the beamforming vectors to maximize the beamforming gain in a certain desired direction $\theta_d$, which is called beamsteering. One way to do that is by adjusting the beamforming weights to match the array response vector in the desired direction, i.e., to set  $\bff=\ba\left(\theta_d\right)$. This results in a beampattern with a main lobe in the desired direction. Other beam designs that trade-off main lobe and side lobe levels are also possible. For analytical tractability, it is common to approximate the actual array beam pattern by a step function with a constant main-lobe over the beamwidth and a constant side-lobe otherwise, shown in  \figref{fig:Beam_Pattern}. Such a model has been adopted in \cite{BaiHea14,Singh2015,Renzo2015,Alkhateeb2016} for tractable coverage and rate analysis of \ac{mmWave} cellular networks.

Thanks to its digital processing layer, hybrid architectures offer more degrees of freedom in beamforming design than purely analog beamforming. This can be used, for example, to realize beam patterns with better characteristics \cite{Alkhateeb2014}. For steering the beam in the azimuthal as well as the vertical directions, it is desirable to have a uniform planar array (UPA). Most industry papers assume a uniform planar array for single stream beamforming \cite{Roh2014,Ghosh14}. Existing analysis of \ac{mmWave} cellular networks has been focused on deployments of base stations and UEs on a 2-D plane\cite{BaiHea14,Singh2015}. In this case, the step beam pattern can be modeled with an antenna gain corresponding to the entire 2-D UPA whereas the 3 dB beamwidth corresponds to only the number of antenna elements of the UPA in the azimuth direction. Omni-directional antenna arrays give rise to an {\em image} beam in a non-desirable direction. This back lobe gain is equal to the gain in the desired direction. Therefore, using antenna elements which themselves have a non-omnidirectional pattern makes sense. To provide omni-directional coverage with directional antenna elements, each  access point may need to employ several antenna arrays with each one serving a different sector \cite{Akdeniz2013a,Ran14,Ghosh14,Kim13}. Dense networks are desirable at \ac{mmWave} and one cheap (but possibly suboptimal) way of densifying is having multiple sectors per site that reuse time-frequency resources, as the operators do not need to lease more sites or spectrum. Thus, unlike in Sub-6GHz networks, using the same time-frequency resources across all sectors of an access point could be feasible at \ac{mmWave}\cite{Ghosh14}.

\subsection{SU-MIMO} \label{subsec:SU_MIMO}

To improve the spectral efficiency in single-user MIMO systems, spatial multiplexing -- where multiple streams are simultaneously transmitted -- is an obvious solution. In conventional single user MIMO systems with fully-digital transceivers and perfect channel knowledge, channel capacity is achieved with singular value decomposition (SVD) precoding/combining and a water-filling power allocation \cite{Telatar1999,Goldsmith2003}.  Mathematically, let $\bH=\bU \boldsymbol{\Sigma} \bV^*$ denote the singular value decomposition (SVD) of the channel matrix $\bH$, then set the transmitter precoding matrix as $\bF=\bV \boldsymbol{\Gamma}$, with $\boldsymbol{\Gamma}$ being a diagonal water-filling power allocation matrix, and the receiver combining matrix as $\bW=\bU $.  LTE systems operating in closed loop spatial multiplexing (transmission mode 4) can be viewed as performing a crudely quantized approximation of this SVD-based procedure motivated by information theory \cite{LTEBook}.  It also does not work particularly well due to excessive quantization of the channel state information.   At \ac{mmWave}, the hardware constraints on the entries and dimensions of the precoding and combining matrices makes using an approximation of SVD precoding even more dubious. This motivates research to develop new precoding solutions for SU-MIMO \ac{mmWave} systems. 

Exploiting the sparsity of \ac{mmWave} channels, low-complexity hybrid precoding algorithms were proposed in \cite{ElAyach2014} to approximate the spectral efficiency achieved with SVD and fully-digital precoding. With some approximations, the hybrid precoding design problem was formulated as 
\beq \label{eq:Hybrid_Opt}
\left\{\Frf^\star, \Fbb^\star\right\}=\argmin_{\substack{\Frf \in \mathcal{A} \\ \left\|\Frf \Fbb\right\|_{\rm F}^2=\Nrf}}  \left\|\bF_\mathrm{opt}-\Frf \Fbb\right\|_\mathrm{F}^2,
\eeq
where the first constraint is due to the hardware constraints on the RF precoding matrix, which limits it to a certain set of precoding matrices $\cA$, and the second constraint is a power constraint. If the \ac{mmWave} channel has $\eta_\il$ paths with known angles of departure at the transmitter, then \cite{ElAyach2014} develops a matching pursuit variant to greedily design the RF and baseband precoding matrices. Following \cite{ElAyach2014}, the work in \cite{Ni2015,Yu2016,Chen2015,Mendez-Rial2015a}  used matrix decomposition, alternative minimization, and other techniques to design the hybrid precoders adopting the same optimization problem in \eqref{eq:Hybrid_Opt}. In terms of modeling, the solution in \cite{ElAyach2014} can be interpreted as a number of $\Nrf$ beam patterns, that can be approximated as that in \figref{fig:Beam_Pattern}, representing the column of the RF precoding matrix $\Frf$, with additional processing done in the baseband using the $\Fbb$. Other hybrid precoding designs that do not directly rely on the approximation in \eqref{eq:Hybrid_Opt} have been developed in \cite{Alkhateeb2015a,Sohrabi2015,Kim2013} with the same of objective of maximizing the system spectral efficiency. The solutions in \cite{ElAyach2014,Ni2015,Yu2016,Chen2015,Mendez-Rial2015a,Alkhateeb2015a,Sohrabi2015,Kim2013} showed that hybrid precoding can generally achieve very good spectral efficiencies compared to the fully-digital SVD solution in \ac{mmWave} systems, specially when the number of RF chains is close to the number of dominant channel paths. 

\subsection{MU-MIMO} \label{subsec:MU_MIMO}

\begin{figure}[t]
	\centering
	\includegraphics[width=.8\columnwidth]{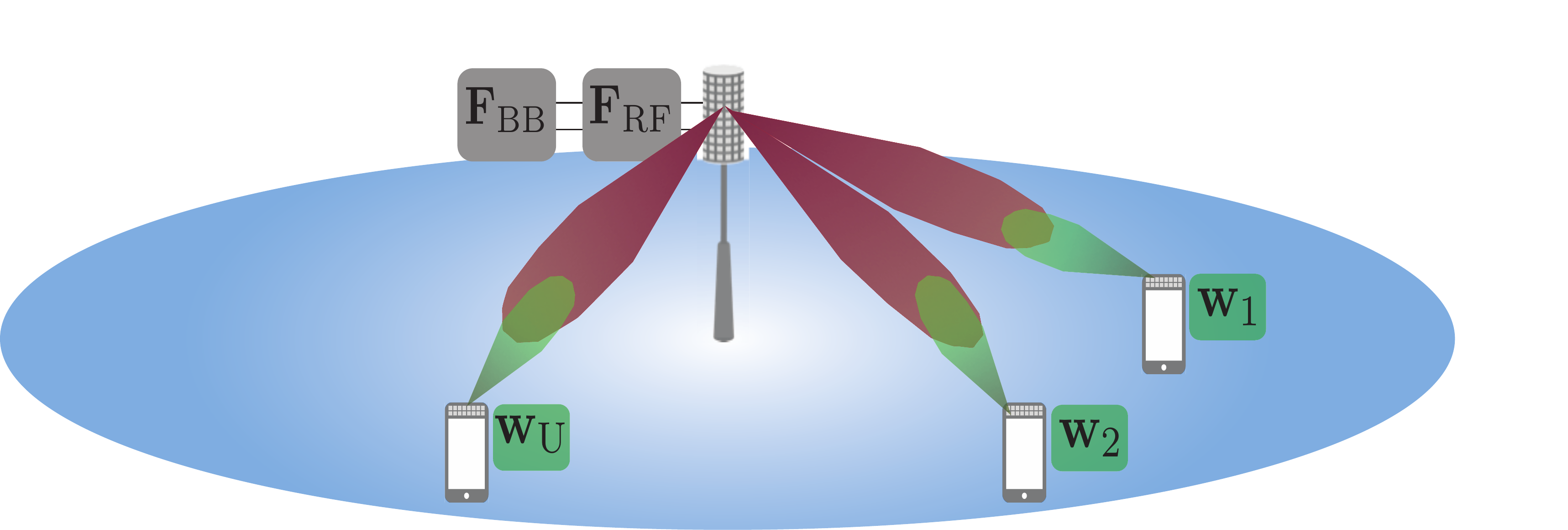}	
	\caption{A multi-user \ac{mmWave} downlink system model, in which a BS uses hybrid analog/digital precoding and a large antenna array to serve $U$ Mobile users.}
	\label{fig:MU_Hybrid}
\end{figure}
 
The antenna arrays can also be used to used to support multi-user MIMO, where users share the same time/frequency resources. To enable efficient multi-user precoding processing in mmWave systems, \cite{Alkhateeb2014b} proposed a two stage hybrid precoding technique. The first stage assigns a different analog beam to each user to maximize the received signal power, as illustrated in \figref{fig:MU_Hybrid}. Considering the effective channels, further baseband processing is performed to cancel the inter-user interference. This simple precoding strategy was shown to achieve very close results to the unconstrained digital solutions, despite its requirement of low training and feedback overhead. Consider the multi-user hybrid precoding system model in \figref{fig:MU_Hybrid}, with a BS employing hybrid analog/digital architecture and serving $U$ users that use analog-only combining. Then for single-path channels in a single cell setup, the $\mathsf{SINR}_u$ of user $u$ can be lower bounded by \cite{Alkhateeb2014b}
\begin{equation} \label{eq:SINR_MU_Hybrud}
\sinr_u \geq \snr_u  \ G\left(U, \Ntx, \eta_\il\right), 
\end{equation} 
where $G\left(U, \Ntx, \eta_\il\right)$, is a constant that depends only on $U, \Ntx, \eta_\il$, and represents the signal power penalty resulted from canceling the multi-user interference. Note that $\snr_u$ is the $\snr$ of user $u$ without inter-user interference, i.e., when the BS serves only this user. Similar expressions can be derived for the multi-path case by using the virtual channel approximation in Section \ref{subsec:channel-models}. One advantage of the discussed multi-user hybrid precoding technique is its relative analytical tractability in the stochastic geometry framework, as the distribution of the $\sinr_u$ in \eqref{eq:SINR_MU_Hybrud} can be easily characterized \cite{KulGhoAnd16}. Similar multi-user mmWave beamforming algorithms have been proposed based on lens antenna arrays \cite{Brady2013,Gao2016}, where the DFT properties of the lens antennas are exploited to dedicate orthogonal directions to different users. Multi-user mmWave combining has also been studied for the uplink system model with hybrid architectures \cite{Li2016}.

\section{Downlink SINR and Rate Distribution}
\label{sec:analysis}

Having considered the ways that \ac{mmWave} cellular systems diverge from conventional ones, we now turn our attention to techniques to analyze their performance.  

\subsection{Performance metrics}

We focus on two fundamental performance metrics:

\textbf{The $\sinr$ and coverage probability.}  The post processing $\sinr$ after the receiver combining operations in both the downlink and uplink is the fundamental metric to understanding how \ac{mmWave} cellular systems perform.  We focus on the downlink $\sinr$, which is a complicated random variable depending on many constituent random variables including (i) the distance separating the desired transmitter and receiver, (ii) the relative distances from all interfering transmissions, (iii) the random channel effects in both the desired and interfering links, including blocking, fading, and possibly shadowing, and (iv) the beam patterns appear randomly oriented, particularly for the interferers, and (v) the thermal noise.  

The $\sinr$ is most commonly characterized as the \emph{coverage probability} defined as $\mathcal{S}(T) = \PP[\sinr > T]$ relative to an $\sinr$ threshold $T$.  It is precisely the CCDF of $\sinr$ and describes the fraction of users that will achieve $\sinr > T$ in the network (averaged over time and space).  Unlike in Sub-6GHz networks, the $\snr$, which is a special case of $\sinr$ with the interference being negligible, is also a useful metric. In many cases \ac{mmWave} cellular systems will be noise-limited (or equivalently, power-limited), due to the path loss and blocking effects discussed earlier, in conjunction with the large bandwidth which brings in much more noise power.   In such cases, the $\snr$ distribution can be used as an approximation of the $\sinr$, allowing much simplification.  This is in stark contrast to Sub-6GHz cellular networks, which are often interference-limited, meaning $\sir \approx \sinr$ instead.

\textbf{Data rate, characterized by the rate coverage probability $\mathcal{R}(\tau) = \PP[R > \tau]$.}   This metric builds upon $\sinr$ and is the most important metric from a performance standpoint, since it most directly affects the perceived experience.  Here $R$ is the (random) data rate per active user, in units of either bits per second (bps) or bps/Hz if normalized by the bandwidth.  However note that this is not simply the system spectral efficiency, which is usually interpreted to be the aggregate rate (not the per user rate $R$ here) divided by the bandwidth.  Rate is also a random variable depending on two complicated underlying random variables: (i) the $\sinr$ through the usual $\log(1+\sinr)$ relation assuming Gaussian signaling and (ii) the fraction of resources that a given user receives over a fairly short time-frame (fractions of a second). 

To analyze these two metrics, we now describe a tractable model for downlink mmWave cellular network, based on the physical characteristics of \ac{mmWave} systems as described in the preceding sections.

\subsection{Baseline {mmWave} cellular system model}\label{sec:analysis:system_model}

The model is based on the traditional analytical framework for Sub-6GHz cellular networks \cite{AndBac11,AndGup16}, but incorporates blockage effects and directional beamforming. The main aspects of the baseline model are now enumerated. 

\begin{enumerate}
\item Base station locations.   We assume the BSs are all outdoors, but still independently distributed according to a homogeneous \ac{PPP} $\Phi=\{X_\il:\il\ge0\}$ of density $\lambda$ on a plane, where $X_\il$ is the location of the $\il$-th base station.   The impact of indoor \ac{mmWave} base stations is ignored, due to the large penetration losses.  Conceptually, this means we ignore the locations of buildings as far as determining BS locations (or equivalently assume they are mounted on the buildings when they happen to be ``dropped'' there).  We discuss this model's applicability further below.

\item User locations.  The users are also assumed to be outdoors, and form an independent \ac{PPP} $\Phi_u$ on the same plane, with density $\lambda_u$. Each user associates with the base station that has the smallest path loss.  Analysis is conducted for a user at the origin for mathematical simplicity.  Because of the stationarity of the PPP, this user can be considered ``typical'' and thus has the same average performance as a user at any location.  The serving base station for this user is denoted as $X_0$.  In this section, we assume the user density is sufficiently high that all base stations are transmitting constantly, which is pessimistic for $\sinr$ \cite{DhiAnd13}.   An extension to consider user loads can be found in the subsequent discussion on rate distributions. 

\item Blocking.  Recall that both BSs and users are assumed to be outdoors.  For this scenario, \ac{mmWave} base stations can be divided into two sub-processes: the \ac{LOS} base stations (unblocked) and the \ac{NLOS} base stations (blocked). The probability that a link of length $d$ is LOS is given by $P_\mathrm{LOS}(d)$ as in Section \ref{sec:blocking}. The events that any two distinct links in the network are LOS are assumed to be independent.  We leave $P_\mathrm{LOS}(d)$ as a general expression in our analysis, i.e. it can follow any of the models discussed in Section \ref{sec:blocking} that result in independent blocking for all links. Therefore, the \ac{LOS} and \ac{NLOS} base stations form two independent non-homogeneous \ac{PPP}s $\Phi_\mathrm{L}$ and $\Phi_\mathrm{N}$, to which different path loss laws will be applied. Note that the non-homogeneity of the LOS and NLOS base stations processes are due to the distance-dependence of the LOS probability function $P_\mathrm{LOS}(d)$.

\item Beamforming:  Analog beamforming is applied at both base stations and mobile stations. The extensions to hybrid beamforming will be discussed in Section \ref{sec:implications}. The typical user and its associated base station are assumed to have perfect channel knowledge, and adjust their steering orientation to achieve the maximum directionality gain. The steering angles of the interfering base stations are uniformly distributed in space. We approximate the actual array pattern by the sectored model as shown in Fig. \ref{fig:Beam_Pattern}. Let $G_\il$ be the total directivity gain (of both transmitter and receiver beamforming) in the link from the typical user to base station $X_\il$. Then, for the interfering link, i.e. for $\il>0$, the directivity gain $G_\il$ is a (discrete) random variable with the probability distribution
	as
	$G_\il=a_k$
  with probability $b_k$ $(k\in\{1,2,3,4\})$; $a_k$ and $b_k$ are constants defined in Table \ref{table:constant}; $c_\mathrm{BS}=\frac{\theta_\mathrm{BS}}{2\pi}$, and $c_\mathrm{MS}=\frac{\theta_\mathrm{MS}}{2\pi}$; and for $s\in\{\mathrm{MS},\mathrm{BS}\}$, $M_s$, $m_s$, and $\theta_s$ are the main lobe gain, side lobe gain, and main lobe beamwidth for the base stations and mobile stations (as plotted in Fig. \ref{fig:Beam_Pattern}). For the desired signal link, perfect beam pointing is assumed with $G_0=M_\mathrm{BS}M_\mathrm{MS}$.

\item Path loss model: Different path loss exponents are applied to the cases of \ac{LOS} and \ac{NLOS} links. Given a link has length $d$, its path loss $\ell(d)$ is 
    \begin{equation}
    \label{eq:pathloss}
    \ell(d) = \left\{
                \begin{array}{ll}
                  C_{\mathrm{L}} d^{-\alpha_{\mathrm{L}}} ~~ w.p. ~ P_{\rm LOS}(d)\\
                C_{\mathrm{N}} d^{-\alpha_{\mathrm{N}}} ~~ w.p. ~1-P_{\rm LOS}(d), 
                \end{array}
                \right.
    \end{equation}
    where $\alpha_{\mathrm{N}}$ are the \ac{LOS} and \ac{NLOS} path loss exponents and $C_\mathrm{L}$, $C_\mathrm{N}$ are the intercepts of the LOS and NLOS path loss formulas. The intercepts $C_\mathrm{L}$ and $C_\mathrm{N}$ are the same for LOS and NLOS links, when the same closed-in reference distance $d_\mathrm{ref}$ is used, e.g. $d_\mathrm{ref}=1$ meter in \cite{Rappaport2015}.  Typical values for \ac{mmWave} path loss exponents can be found in measurement results, e.g. in \cite{Rappaport2013,Rappaport2013a,Rappaport2015}, and for simplicity we use $\alpha_{\mathrm{L}} = 2$ and $\alpha_{\mathrm{N}} = 4$ as default values.
    
	\item Small-scaling fading: Measurements shows that small-scale fading has a relatively minor impact in \ac{mmWave} cellular systems.  The Rayleigh fading model for the sub-6GHz band, which is predicated on a large amount of local scattering, does not apply for \ac{mmWave} bands, especially when directional beamforming is applied \cite{Rappaport2013a}.  Therefore, we assume independent Nakagami fading for each link, which is more general but still tractable. Different parameters of Nakagami fading $\nu_\mathrm{L}$ and $\nu_\mathrm{N}$ are assumed for \ac{LOS} and \ac{NLOS} links. Let $h_\il$ be the small-scale fading in signal power in the $\il$-th link. Then under the Nakagami fading assumption, $h_\il$ is a normalized Gamma random variable. For simplicity, we assume $\nu_\mathrm{L}$ and $\nu_\mathrm{N}$ are positive integers, and ignore the frequency selectivity in fading. Shadowing is ignored in our baseline model, but can be incorporated using the approach in \cite{Singh2015} at some cost in tractability.\\

\end{enumerate}

\begin{table}
	\begin{center}
		\caption{Probability mass function of $G_\il$ ($\il>0$)}\label{table:constant}
			\small{\begin{tabular}{ c | c | c |c |c}
					\hline
					\mbox{k} & 1&2&3&4 \\ \hline
					\mbox{$a_k$} &$M_\mathrm{MS}M_\mathrm{BS}$& $M_\mathrm{MS}m_\mathrm{BS}$  & $m_\mathrm{MS}M_\mathrm{BS}$&$m_\mathrm{MS}m_\mathrm{BS}$\\ \hline
					\mbox{$b_k$}&$c_\mathrm{MS}c_\mathrm{BS}$ &$c_\mathrm{MS}(1-c_\mathrm{BS})$  &$(1-c_\mathrm{MS})c_\mathrm{BS}$ &$(1-c_\mathrm{MS})(1-c_\mathrm{BS})$\\ \hline
			\end{tabular}}
		\end{center}
\end{table}

Based on this model, the downlink SINR can be expressed as
\begin{align}\label{eqn:sinr}
\mbox{SINR}=\frac{h_0G_0 \ell\left(\left|X_0\right|\right)}{\sigma^2+\sum_{\il>0:X_\il\in\Phi}h_\il G_\il \ell\left(\left|X_\il\right|\right)},
\end{align}
where $\sigma^2$ is the noise power and $|X_\il|$ is the norm of the location $X_\il$, which denotes the distance of the link from $X_\il$ to the typical user at the origin.

As many readers will be aware, significant progress has been recently made in the characterization of $\sinr$ distributions of similar form to \eqref{eqn:sinr}.  The key underlying toolset descends from stochastic geometry \cite{BacNOW,HaenggiBook,HaeAnd09}, which relies on the base stations following a random spatial distribution like we have here.    Although the PPP is an idealized distribution that assumes independent BS locations, it has been found to describe important SINR trends observed under actual BS distributions \cite{AndBac11}, and is typically accurate to within a 2-3 dB fixed SINR gap (and is pessimistic versus real BS locations) \cite{GuoHae15,Hae14}.  Since there are not any actual \ac{mmWave} cellular deployments at the time of writing to compare against, any model -- tractable or not -- is speculative, but given that the PPP results in similar $\sinr$ curves to a very large class of BS locations (including grids), it seems quite reasonable for a baseline model.

A contemporary entry-level tutorial cellular analysis using stochastic geometry for both rate and $\sinr$ for Sub-6GHz systems can be found in the recent reference \cite{AndGup16}, which includes a summary of key tools and definitions and a step-by-step description of how to get the SINR distribution for downlink, uplink and multi-tier downlink cellular networks following.  Thus, we do not repeat this here, but instead extend and generalize these now well-known results to the \ac{mmWave} cellular case.

\subsection{SINR downlink coverage probability}
\label{sec:sinrderivation}

We now derive the $\sinr$ coverage probability for \ac{mmWave} cellular systems.  The new technical difficulties compared with the Sub-6GHz analysis are: (i) the small-scale fading is modeled as Nakagami, not Rayleigh distributed; (ii) the base station locations are seen as a superposition of inhomogeneous \ac{PPP}s representing LOS and NLOS base stations from a typical user in the network; and (iii) the directivity gain from beamforming introduces additional randomness in the interference, compared with the omni-antenna case.

To overcome the first difficulty on Nakagami fading,  Alzer's Lemma\cite{Alzer1997} on the CCDF of a gamma random variable with integer parameter can be applied. This relates the CDF of a gamma random variable into an weighted sum of the CDFs of exponential random variables. The approximation is shown to be generally tight in numerical simulations with different system parameters, and was used in prior analysis of Sub-6GHz MIMO networks in \cite{Huang2007}.

\begin{lemma}[Alzer's Lemma \cite{Alzer1997}]
	\label{lem:inequ1} Let $h$ be a normalized gamma random variable with parameter $\nu$. For a constant $\gamma>0$, the probability $\mathbb{P}(h<\gamma)$ can be tightly upper bounded by
	$$
	\mathbb{P}(h>\gamma)\le1-\left[1-\mathrm{e}^{-\eta \gamma}\right]^\nu=\sum_{n=1}^{\nu}(-1)^{n+1}{\nu\choose{n}}\mathrm{e}^{-\eta n\gamma},
	$$
	where $\eta=\nu(\nu!)^{-\frac{1}{\nu}}$, and the equality holds when $\nu=1$.
\end{lemma}

Then, we can compute the SINR coverage probability as the Laplace functional of the interference as:
\begin{align}
\mathcal{S}(T)&=\mathbb{P}(\mathrm{SINR}>T)\nonumber\\&=\mathbb{P}\left(h_0>\frac{T(\sigma^2+I)}{G_0 \ell\left(\left|X_0\right|\right)}\right)\nonumber\\
&\stackrel{(a)}\approx\sum_{n=1}^{\nu}(-1)^{n+1}{\nu\choose{n}}\mathbb{E}\left[\mathrm{e}^{-\frac{\eta n T(\sigma^2+I)}{G_0 \ell\left(\left|X_0\right|\right)}}\right]\nonumber\\
&\stackrel{(b)}=\sum_{n=1}^{\nu}(-1)^{n+1}{\nu\choose{n}}\mathbb{E}\left[\mathrm{e}^{-n T \mu\sigma^2}\right]\mathbb{E}\left[\mathrm{e}^{-n T \mu I}\right]\nonumber\\
&\stackrel{(c)}=\sum_{n=1}^{\nu}(-1)^{n+1}{\nu\choose{n}}\mathbb{E}\left[\mathrm{e}^{-n T \mu\sigma^2}\right]\mathcal{L}_I(n T\mu)\label{eqn:trick1}
\end{align}
where $I=\sum_{\il>0:X_\il\in\Phi}h_\il G_\il \ell\left(\left|X_\il\right|\right)$ is the total interference; $\sigma$ is the normalized noise power by the transmit power ; step (a) follows from Lemma \ref{lem:inequ1}; in step (b), we denote $\mu=\frac{\eta}{G_0 \ell\left(\left|X_0\right|\right)}$, and in (c) we denote the Laplace functional of the interference $I$ as $\mathcal{L}_I(s)=\mathbb{E}\left[\mathrm{e}^{-sI}\right]$.

For the second difficulty on base station inhomogeneity, we consider the LOS and NLOS base stations as two independent tiers of BSs as seen from the typical user at origin, as we ignore the correlations between the LOS probabilities between nearby links. The non-homogeneous PPP representing LOS BSs has density equal to $\lambda P_\mathrm{LOS}(d)$, where $d$ is the distance from origin. Similarly, the NLOS BS process is PPP with density $\lambda (1-P_\mathrm{LOS}(d))$.

Based on the above discussion, we illustrate the method to compute the Laplace function of the interference. Given that the desired signal link is LOS and has a length of $|X_0|=d$, based on the minimum path loss association rule, all the LOS interfering base stations are farther than distance $d$, and all NLOS interfering base stations are farther than distance $d^{\alpha_\mathrm{L}/\alpha_\mathrm{N}}$ from the typical user. The Laplace transform of the interference $\mathcal{L}_I(t)=\mathbb{E}\left[\mathrm{e}^{-t\left(I_\mathrm{L}+I_\mathrm{N}\right)}\right]$, where $t>0$, $I_\mathrm{L}=\sum_{\il>0:X_\il\in\Phi_\mathrm{L}}h_\il G_\il \ell\left(\left|X_\il\right|\right)$ represents the LOS interference, and $I_\mathrm{N}=\sum_{\il>0:X_\il\in\Phi_\mathrm{N}}h_\il G_\il \ell\left(\left|X_\il\right|\right)$ the NLOS interference. Given that the typical user associates with a LOS BS at distance $d$, this can be simplified as follows. 
\begin{align}
&\mathcal{L}_I(t)\stackrel{(a)}=\mathbb{E}\left[\mathrm{e}^{-tI_\mathrm{L}}\right]\mathbb{E}\left[\mathrm{e}^{-tI_\mathrm{N}}\right]\nonumber\\
&\nonumber\stackrel{(c)}=\exp\left(-2\pi\int^{\infty}_{d}\left(1-\mathbb{E}_{h,G}\left[\mathrm{e}^{-C_\mathrm{L}r^{-\alpha_\mathrm{L}}hGt}\right]\right)\lambda P_\mathrm{LOS}(r)r\mathrm{d}r\right)\\
&\hspace{1cm}\exp\left(-2\pi\int^{\infty}_{d^{\frac{\alpha_\mathrm{L}}{\alpha_\mathrm{N}}}}\left(1-\mathbb{E}_{h,G}\left[\mathrm{e}^{-C_\mathrm{N}r^{-\alpha_\mathrm{N}}hGt}\right]\right)\lambda (1-P_\mathrm{LOS}(r))r\mathrm{d}r\right)\label{eqn:trick2}
\end{align}
where step (a) follows from the independence between $\Phi_\mathrm{L}$ and $\Phi_\mathrm{N}$; and (b) follows from the probability generating functional of a PPP\cite{HaenggiBook}. Here, $h$ and $G$ are dummy random variables for the small-scale fading and directivity gain in interference channels. The Laplace transform for interference can be similarly derived given that the user associates with a NLOS base station at distance $d$. Here, all NLOS interferers are farther than $d$ from the user at origin and all LOS interferers are farther than $d^{\alpha_\mathrm{N}/\alpha_\mathrm{L}}$.

For the third difficulty on antenna gain, note that the randomness in the directivity gain is incorporated in the random variable $G$ in (\ref{eqn:trick2}). As the actual antenna pattern can be intractable to incorporate, the sectored antenna approximation has been proposed to simplify the computation. In the approximation, the directivity gain $G$ in the interference channel is modeled as a discrete random variable with a simply distribution, as shown in Table~\ref{table:constant}. Therefore, for $\mathrm{s}\in\{\mathrm{L},\mathrm{N}\}$, the term $\mathbb{E}_{h,G}\left[\mathrm{e}^{-C_\mathrm{s} r^{-\alpha_\mathrm{s}}hGt}\right]$ in (\ref{eqn:trick2}) can be computed as
\begin{align}
\mathbb{E}_{h,G}\left[\mathrm{e}^{-C_\mathrm{s} r^{-\alpha_\mathrm{s}}hGt}\right]&\stackrel{(a)}=\sum_{k=1}^{4}b_k\mathbb{E}_{h}\left[\mathrm{e}^{-C_\mathrm{s} r^{-\alpha_\mathrm{s}}h a_k t}\right]\\
&\stackrel{(b)}=\sum_{k=1}^{4}b_k\left(1+C_\mathrm{s} r^{-\alpha_\mathrm{s}}a_kt\right)^{-\nu_\mathrm{s}}
\end{align}
where in (a) $a_k$ and $b_k$ are the constants given in Table \ref{table:constant}, and step (b) follows from computing the moment generating function of the gamma distributed random variable $h$.

The Laplace functional in (\ref{eqn:trick2}) can be applied to compute the SINR coverage probability as shown in  (\ref{eqn:trick1}). The remaining steps are deconditioning the LOS/NLOS status and the length of the desired signal link, whose distributions are given in the following lemma.

\begin{lemma}[Association probability]\label{lem:neardis}
The probability $A_\mathrm{L}$ that the typical user is associated with a LOS base station is
\begin{align}
A_\mathrm{L}=\int_{0}^{\infty}\mathrm{e}^{-2\pi\lambda\int_{0}^{x^{\alpha_\mathrm{L}/\alpha_\mathrm{N}}}(1-P_\mathrm{LOS}(t))t\mathrm{d}t}g_\mathrm{L}(x)\mathrm{d}x,
\end{align}
where $g_\mathrm{L}(x)=\pi\lambda xP_\mathrm{LOS}(x)\mathrm{e}^{-2\pi\lambda\int_{0}^{x}rP_\mathrm{LOS}(r)\mathrm{d}r}$, and
the probability that the user is associated with a \ac{NLOS} base station is $A_\mathrm{N}=1-A_\mathrm{L}$. Given that a user is associated with a \ac{LOS} base station, the probability density function of the distance to its serving base station is
\begin{align}
{f}_\mathrm{L}(x)=\frac{g_\mathrm{L}(x)}{A_\mathrm{L}}\mathrm{e}^{-2\pi\lambda\int_{0}^{x^{\alpha_\mathrm{L}/\alpha_\mathrm{N}}}(1-P_\mathrm{LOS}(t))t\mathrm{d}t},
\end{align}
when $x>0$. Given the user is served by a \ac{NLOS} base station, the probability density function of the distance to its serving base station is
\begin{align}
{f}_\mathrm{N}(x)=\frac{g_\mathrm{N}(x)}{A_\mathrm{N}}\mathrm{e}^{-2\pi\lambda\int_{0}^{x^{\alpha_\mathrm{N}/\alpha_\mathrm{L}}}P_\mathrm{LOS}(t)t\mathrm{d}t},
\end{align}
where $x>0$, and $g_\mathrm{N}(x)=2\pi\lambda x(1-P_\mathrm{LOS}(x))\mathrm{e}^{-2\pi\lambda\int_{0}^{x}r(1-P_\mathrm{LOS}(r))\mathrm{d}r}$.
\end{lemma}

Now we present the main SINR coverage result in the following theorem.  The detailed proof of Theorem~\ref{thm:sinrcov} and Lemma~\ref{lem:neardis} can be found in \cite{BaiHea14}.

\begin{theorem}[SINR coverage results] \label{thm:sinrcov}

The \ac{SINR} coverage probability $\mathcal{S}(T)$  can be computed as
\begin{align}
\mathcal{S}(T)=A_\mathrm{L}\mathcal{S}_{\mathrm{L}}(T)+A_\mathrm{N}\mathcal{S}_{\mathrm{N}}(T),
\end{align}
where for $\mathrm{s}\in\{\mathrm{L},\mathrm{N}\}$, $\mathcal{S}_{\mathrm{s}}(T)$ is the conditional coverage probability given that the user is associated with a base station in $\Phi_\mathrm{s}$. Further, $\mathcal{S}_\mathrm{s}(T)$  can be evaluated as
\begin{align}
\mathcal{S}_{\mathrm{L}}(T)&\approx\sum_{n=1}^{\nu_\mathrm{L}}(-1)^{n+1}{\nu_\mathrm{L}\choose{n}}\int_{0}^{\infty}\mathrm{e}^{-\frac{n \eta_\mathrm{L} x^{\alpha_\mathrm{L}}T\sigma^2}{C_\mathrm{L}M_\mathrm{MS}M_\mathrm{BS}}-Q_n(T,x)-V_n(T,x)}{f}_\mathrm{L}(x)\mathrm{d}x,\label{eqn:pc1}
\end{align}
and
\begin{align}
\mathcal{S}_{\mathrm{N}}(T)&\approx\sum_{n=1}^{\nu_\mathrm{N}}(-1)^{n+1}{\nu_\mathrm{N}\choose{n}}\int_{0}^{\infty}\mathrm{e}^{-\frac{n \eta_\mathrm{N} x^{\alpha_\mathrm{N}}T\sigma^2}{C_\mathrm{N} M_\mathrm{MS}M_\mathrm{BS}}-W_n(T,x)-Z_n(T,x)}{f}_\mathrm{N}(x)\mathrm{d}x,\label{eqn:pc2}
\end{align}
where
\begin{align}
Q_{n}(T,x)=2\pi\lambda\sum_{k=1}^{4} b_k \int_{x}^{\infty}F\left(\nu_\mathrm{L},\frac{n\eta_\mathrm{L}\bar{a}_kTx^{\alpha_\mathrm{L}}}{\nu_\mathrm{L}t^{\alpha_\mathrm{L} }}\right)P_\mathrm{LOS}(t)t\mathrm{d}t,
\end{align}
\begin{align}
V_{n}(T,x)=&2\pi\lambda\sum_{k=1}^{4} b_k \int_{x^{\alpha_\mathrm{L}/\alpha_\mathrm{N}}}^{\infty}F\left(\nu_\mathrm{N},\frac{nC_\mathrm{N}\eta_\mathrm{L}\bar{a}_kTx^{\alpha_\mathrm{L}}}{C_\mathrm{L}\nu_\mathrm{N}t^{\alpha_\mathrm{N}}}\right)(1-P_\mathrm{LOS}(t))t\mathrm{d}t,
\end{align}
\begin{align}
W_{n}(T,x)=&2\pi\lambda\sum_{k=1}^{4} b_k \int_{x^{\alpha_\mathrm{N}/\alpha_\mathrm{L}}}^{\infty}F\left(\nu_\mathrm{L},\frac{nC_\mathrm{L}\eta_\mathrm{N}\bar{a}_kTx^{\alpha_\mathrm{N}}}{C_\mathrm{N} \nu_\mathrm{L} t^{\alpha_\mathrm{L}}}\right)P_\mathrm{LOS}(t)t\mathrm{d}t,
\end{align}
\begin{align}
Z_{n}(T,x)=&2\pi\lambda\sum_{k=1}^{4} b_k \int_{x}^{\infty}F\left(\nu_\mathrm{N},\frac{n\eta_\mathrm{N}\bar{a}_kTx^{\alpha_\mathrm{N}}}{\nu_\mathrm{N}t^{\alpha_\mathrm{N}}}\right)(1-P_\mathrm{LOS}(t))t\mathrm{d}t,
\end{align}
and $F(\nu,x)=1-1/(1+x)^\nu$. For $s\in\{\mathrm{L},\mathrm{N}\}$, $\eta_s=\nu_s(\nu_{s}!)^{-\frac{1}{\nu_s}}$, $\nu_{s}$ are the parameters of the Nakagami small-scale fading; for $k\in\{1,2,3,4\}$, $\bar{a}_k=\frac{a_k}{M_\mathrm{BS}M_\mathrm{MS}}$, $a_k$ and $b_k$ are defined in Table \ref{table:constant}.
\end{theorem}
\vspace{0.2in}

Note that in a noise-limited network, the SINR distribution can be replaced by the SNR distribution that has a much simpler analytical expression to compute \cite{Singh2015,Renzo2015}. For example, when the interference is ignored, the expression in Theorem \ref{thm:sinrcov} is largely simplified, as $W_{n}(\cdot)$, $Q_{n}(\cdot)$, $V_{n}(\cdot)$, and $Z_{n}(\cdot)$ all become zero. In an interference-limited scenario, substituting the noise power $\sigma^2=0$ gives us the SIR distribution.

We will discuss key extensions of the SINR result in Sect. \ref{sec:extensions} and further implications of these results in Sect. \ref{sec:implications}.  We now turn our attention to characterizing the rate distribution for the baseline model.

\subsection{Rate coverage probability}
The per user rate in \ac{bps} depends largely on the user perceived $\sinr$ and the amount of time-frequency resources it can use. Treating interference as noise, the achievable per user rate in \ac{bps}/Hz is close to the point-to-point link capacity given by $\log_2 (1+\sinr)$, also called the spectral efficiency. Conventionally, this has been considered to be the metric of interest in evaluating the performance of wireless networks and is still largely used today for analytical purposes given the mathematical challenge involved in characterizing the distribution of amount of per user resources. Once the $\sinr$ coverage is known, computing the distribution of the spectral efficiency is straightforward. However, cellular networks today are becoming increasingly heterogeneous and thus, user association and offloading have been hot topics of research. Further, it is likely that \ac{mmWave} networks will coexist with traditional Sub-6GHz networks. Thus, the problem of offloading from one frequency band to another is still relevant for \ac{mmWave} cellular networks. For studying such problems, incorporating the impact of load on the rate characterization is essential\cite{And14}. Wireless backhauling and dynamic resource allocation are some other intriguing research directions for \ac{mmWave} cellular\cite{Rangan2014,Singh2015,GuptaKul2016}. Incorporating the impact of load on the rate characterization is essential for studying such problems as well.

A common assumption in the literature for analytical tractability has been to assume round robin scheduling, which is also equivalent to random user selection in each time-frequency resource block. In this case, the per user rate in \ac{bps} can be modeled as
\begin{equation}
\label{eq: rate}
\rate  =\frac{\bandwidth}{\load}\log_2(1+\sinr),
\end{equation}
where $\bandwidth$ is the total bandwidth and $\load$ is the total number of users connected to the base station serving the user whose performance is being evaluated. Rate coverage is defined as $\ratecov = \mathbb{P}(\rate>\tau)$, where $\tau$ is the rate threshold in \ac{bps}. Recall that association cell of a BS at $X\in\mathbb{R}^2$ is a collection of user locations in space that would associate with $X$ based on instantaneous channel conditions. In general the $\sinr$ and load distribution of serving BS are correlated since larger the association cell implies more load and longer link distances (that is smaller $\sinr$). Thus, knowing the joint distribution of association areas and SINR is necessary to find the load distribution. As of now, even characterizing the marginal distribution of the Poisson-Voronoi (PV) tesselations is an open problem. Voronoi tesselations are association cells in Sub-6GHz networks wherein a user connects to the nearest base station. As was shown in \cite{Singh2015}, blockage effects lead to very irregular association cells in mmWave cellular networks, of which PV tesselations is a special case. Given that the association scheme is stationary or translation invariant (refer \cite{SinBacAnd13} for a formal definition), the mean association area of a typical cell can be characterized as $1/\BSdensity$, which is the same as the mean association area of a typical cell in a PV tesselation. Note that this is different than the mean association area of the cell containing the user at origin, which is a size-biased version of the area of typical cell\cite{SinBacAnd13}. In the case of mmWave path loss model in \eqref{eq:pathloss}, the minimum path loss association rule is stationary since given the stationary point processes for base stations and user locations, the association of a user to a base station depends only on independent distance dependent random variables which are invariant under a translation. Thus, the mean number of users in a typical cell is equal to $\UEdensity/\BSdensity$\cite{Singh2015}. 

The above observations lead to the following approximations for tractability.
\begin{enumerate}
\item Association area distribution: The distribution of association area of a typical BS in a \ac{mmWave} cellular network is assumed to be same as the approximate distribution of a typical PV cell with the same mean area as proposed in \cite{Ferenc2007}.
\item User point process: The point process for UE locations is a homogeneous \ac{PPP} with intensity $\UEdensity$.
\item Independence: The load distribution of the BS serving the typical user at origin is independent of the load of other BSs in the network, and all these are independent of the user perceived $\sinr$. 
\end{enumerate} 

The approximation for volume of a typical PV cell proposed in \cite{Ferenc2007} was used to derive and validate the load distribution of the serving as well as the other BSs in the network in \cite{Singh2013,YuKim13}. Such an approximation was subsequently verified numerically in \cite{SinAnd14,Singh2014,Wang14, DhiRate14} for Sub-6GHz networks and in \cite{Singh2015,KulGhoAnd16} for \ac{mmWave} networks. We provide the formulas for load distribution here, and interested readers can find the proof in Appendix B of \cite{Singh2013}.

\begin{lemma}
The \ac{PMF} of the number of users $\load_0$ associated with the BS at $X_0$ serving the user at origin is given as
\begin{equation}
\label{eq:loadtagged}
\mathbb{P}(\load_0 = n) = \Upsilon(n) =  \frac{3.5^{3.5}}{(n-1)!}\frac{\Gamma(n+3.5)}{\Gamma(3.5)}\left(\UEdensity/\BSdensity\right)^{n-1} \left(3.5+\UEdensity/\BSdensity\right)^{-n-3.5},
\end{equation}
for $n\geq 1$ and $\mathbb{P}(\load_0 = 0)=0$. The corresponding mean is $1+1.28\UEdensity/\BSdensity$.

For a typical BS located at $X_l$, the load distribution is given as
\begin{equation}
\label{eq:loaduntagged}
\mathbb{P}(\load_l = n) = \frac{3.5^{3.5}}{n!}\frac{\Gamma(n+3.5)}{\Gamma(3.5)}\left(\UEdensity/\BSdensity\right)^{n} \left(3.5+\UEdensity/\BSdensity\right)^{-n-3.5},
\end{equation}
for $n\geq 0$. The corresponding mean is $\UEdensity/\BSdensity$.
\end{lemma}

Based on the assumptions given above, the rate coverage can be found from the following theorem.
\begin{theorem}
\label{thm:ratecov}
The rate coverage of a typical user in a \ac{mmWave} cellular network for a rate threshold $\tau$ is given by
\begin{equation}
\label{eq:ratecov}
\ratecov(\tau) = \sum_{n\geq 1} \Upsilon(n) \sinrcov\left(2^{\tau n/\bandwidth} - 1\right),
\end{equation}
where $\sinrcov(\cdot)$ is the $\sinr$ coverage derived in the Theorem~
\ref{thm:sinrcov}.
\end{theorem}

Although the rate coverage expression is an infinite summation, it can be computed as a finite summation up to $n_\mathrm{max}$ terms without much loss in accuracy\cite{Singh2013,Singh2015}. A rule of thumb is to choose $n_\mathrm{max}$ as a multiple of  $\UEdensity/\BSdensity$, where the multiplicative constant can be found from numerical investigations. A faster but less accurate mean load approximation can also be done by substituting the random variable representing load by the corresponding mean given by $1+1.28\UEdensity/\BSdensity$.

\section{Implications of Models and Analysis}
\label{sec:implications}

In this section we consider the design and deployment implications of the baseline model analysis.

\subsection{When will \ac{mmWave} systems be noise/power-limited?}

An ongoing major challenge for Sub-6GHz cellular networks has been the management of interference from neighboring cells using the same time-frequency resources.  Such networks, especially in urban areas, are typically interference-limited, meaning that $\sinr\approx\sir$ and so increasing transmission power does not increase $\sinr$, on a network-wide basis.  Another way to view the interference-limited behavior is that the network density is sufficiently high such that further densifying the network does not substantially improve the $\sinr$ distribution, because the increase in $\snr$ is counter-acted by the increase in interference, as shown in \cite{AndBac11, DhiGan12}.  For mmWave networks, the behavior is somewhat different.

First, the received $\snr$ is nominally very low in mmWave due to the small per antenna area, the large bandwidth and blocking/penetration effects.  As we have discussed, this is compensated for using large antenna arrays that achieve highly directional transmission and reception.  Improving the $\snr$ through beamforming, as opposed to increasing it via higher transmitter power or base station density, should not increase the average interference, because the average transmit power is unchanged.  Meanwhile, the interfering signals experience blocking and typically have misaligned beams.  These factors seem to indicate that mmWave systems are much more likely to be noise-limited than their Sub-6GHz counterparts. This has many important implications on the system design, as will be discussed in the subsequent sections.

In this section, we explore what circumstances effect the phase transition between noise and interference domination of the $\sinr$ denominator, to better understand when noise or interference limited behaviors will be observed. Important system parameters to consider include: the base station and user density, antenna gains and beamwidths, operational bandwidth, blockage model,  the LOS and NLOS path loss exponents, and the choice of MIMO technique.  For example, our baseline analysis shows, as discussed in \cite{BaiHea14}, that the SINR coverage probability exhibits a non-monotonic trend with base station density. This implies that the network eventually transitions from noise to interference limited behavior when there are enough interfering BSs.  This of course also holds for Sub-6GHz systems, but occurs at much lower densities.  Because many parameters simultaneously effect whether the mmWave system is described better by the $\sir$ or $\snr$, it is not possible to give crisp threshold values where such transitions occur. In this section, we first identify specific scenarios where \ac{mmWave} networks have been previously observed to follow noise-limited behavior, and then we present some numerical results to illustrate the dependence of these trends on the aforementioned system parameters. 

In the simulations, we compare the interference-to-noise ratio (INR, or $I/N$), and plot the probability $\mathbb{P}(I/N>1)$ as a function of the average inter-site distance between neighboring base stations.  We consider two carrier frequencies: 73 GHz in E-band and 28 GHz in LMDS band. The 73 GHz system will likely have a larger bandwidth which increases noise power, but due to the smaller wavelength also more directionality. In addition, we use the Austin and Los Angeles city, for which LOS functions have been fitted using real data in Section \ref{sec:blocking}, as examples for moderately-dense and ultra-dense building environments.

\begin{figure}[t]
	\centering
	\subfloat[Comparison of INR at 28 GHz and 73 GHz.]{
		\includegraphics[height=170pt,width=.48\columnwidth]{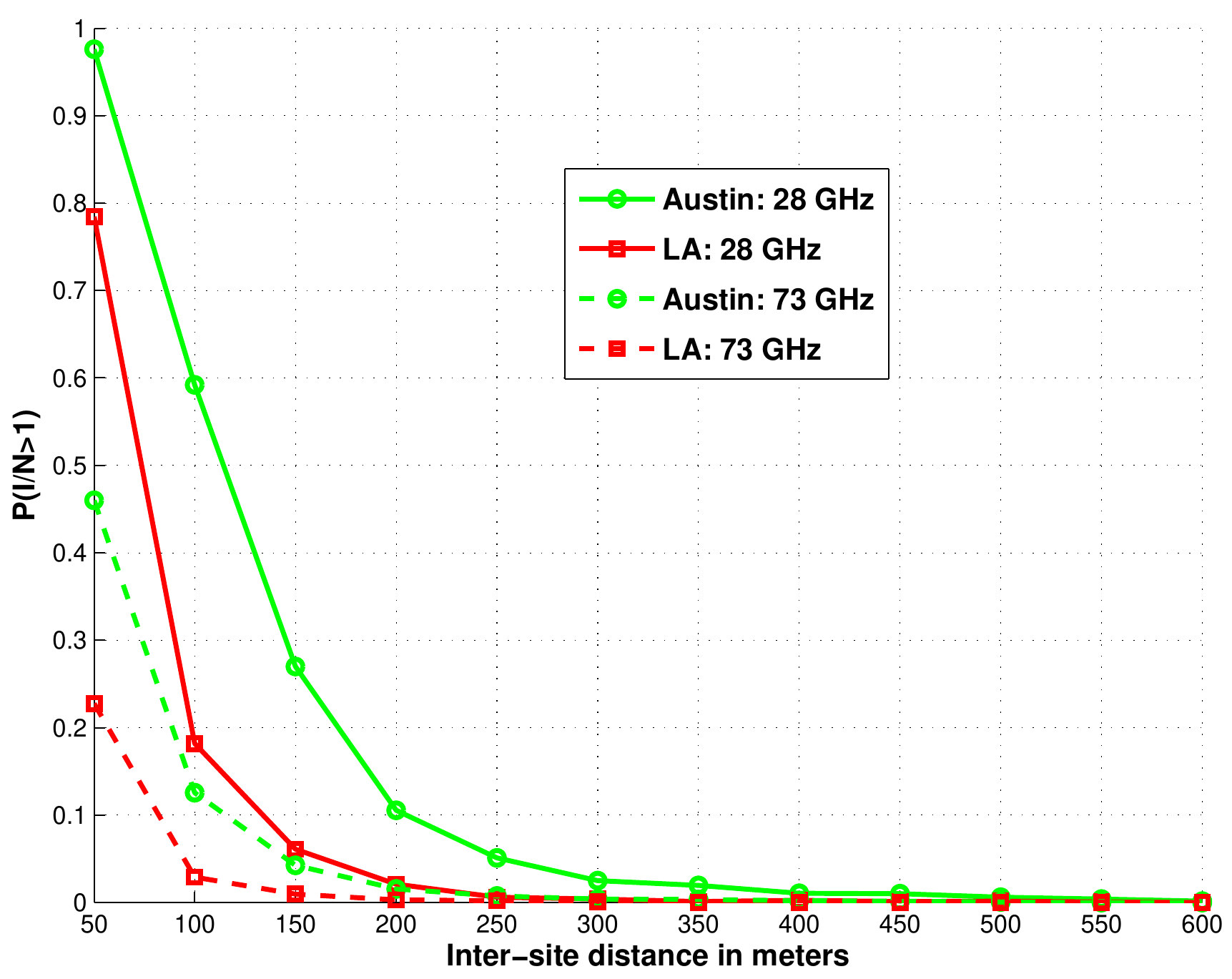}
	}
	\subfloat[Impact of different parameters on INR]{
		\includegraphics[height=170pt,width=.48\columnwidth]{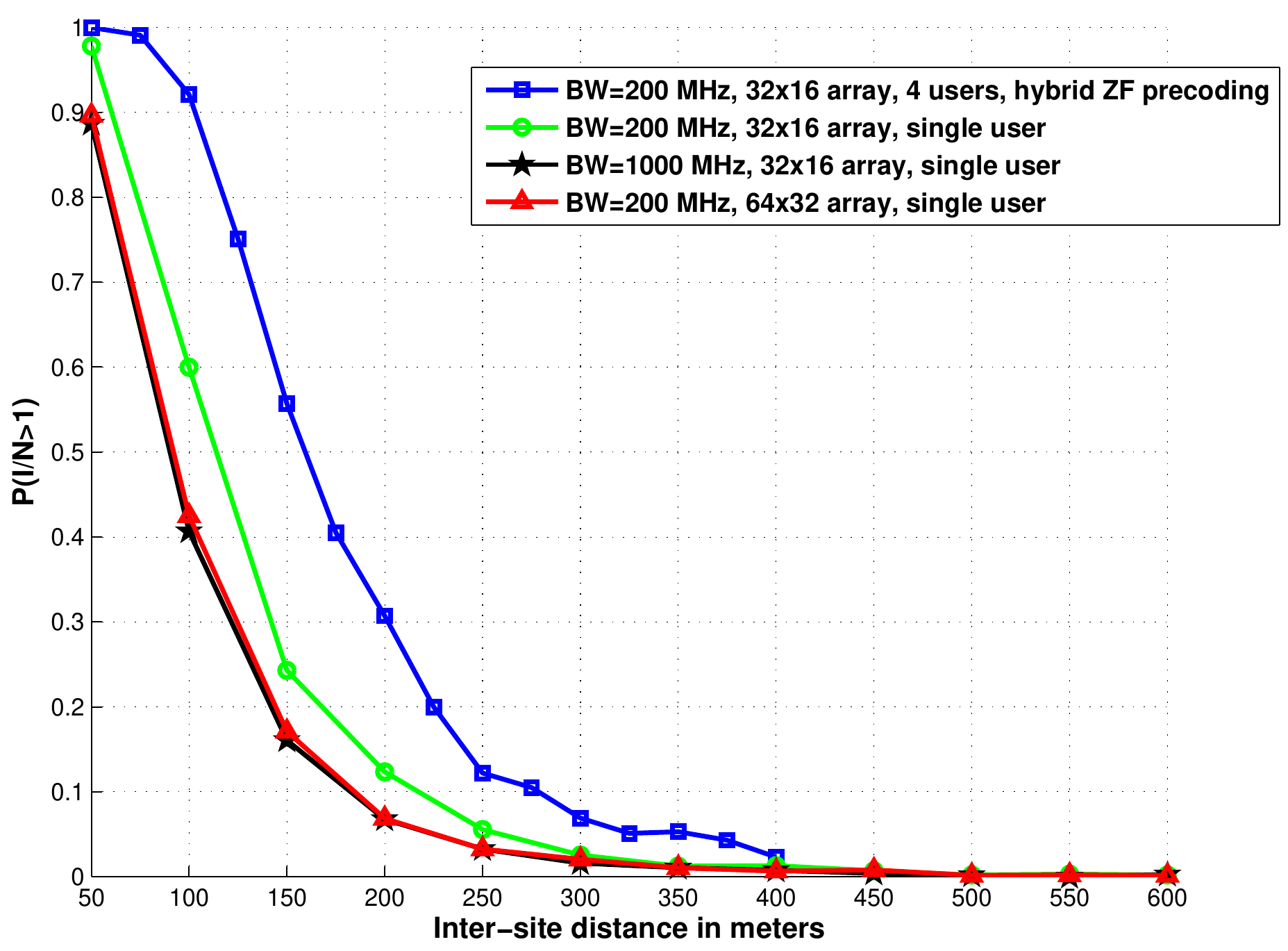}
	}
	\caption{Comparison of INR with different system parameters. We assume 200 MHz bandwidth, 32-antenna uniform linear array (ULA) at base stations, and 16-antenna ULA for mobiles for 28 GHz systems;  whereas for 73 GHz we assume 1000 MHz bandwidth, 64-antenna uniform linear array (ULA) at base stations, and 32-antenna ULA for the mobiles. The path loss exponents are $\alpha_\mathrm{L}=2$, and $\alpha_\mathrm{N}=4$. Random shape theory blockage model is used with parameters corresponding to Austin and LA regions in Section~\ref{sec:blockage:summary}. MU-MIMO simulations use the hybrid precoding algorithm in \cite{Alkhateeb2014b} with 4 RF chains and zero-forcing for digital precoding.  In (a), we plot $\mathbb{P}(I/N>1)$ typically 73 GHz systems are much more noise-limited than 28 GHz systems due to availability of larger bandwidth.  In (b), we show that INR increases with decreasing bandwidth, smaller antenna arrays, and use of MU-MIMO in the 28 GHz system. }\label{fig:INR}
\end{figure}

As stochastic geometry analysis in \cite{Singh2015,Renzo2015} indicated, Fig. \ref{fig:INR}(a) shows that with a larger bandwidth and more directionality (which reduces the effect of nearby interferers), the 73 GHz system tends to be noise-limited even when inter-site distance (ISD) is as little as 100 meters, corresponding to a base station density as high as to $\lambda = 200$ BSs per square km. Similar to \cite{BaiHea14}, with the smaller 200 MHz bandwidth, a larger impact from interference can be observed in the 28 GHz band.  Moreover, compared with the performance in Austin, the impact of the interference becomes smaller in Los Angeles due to the denser building blockages. In Fig. \ref{fig:INR}(b), we simulate the distribution of $\mathbb{P}(I/N>1)$ for a 28 GHz system in Austin with different system parameters.  Numerical results indicate that the impact of interference increases with a smaller bandwidth, and a large beamwidth in the beamforming (resulted from applying fewer antennas in the ULA), and a larger number of users in the MU MIMO. 

The results indicate that it is difficult to provide a general and crisp answer whether \ac{mmWave} cellular networks are noise-limited or noise-limited, but several observations can be made.
\begin{enumerate}
\item Higher carrier frequencies, which typically will allow for considerably larger bandwidths and higher directionality, as well as possibly having more sensitivity to blockage (especially penetration into buildings), will be significantly more noise-limited.  For example, to have $I \approx N$ with probability $0.1$ requires about a factor of 2 smaller ISD at 73 GHz vs. 28 GHz for both Austin and LA.  In other words, four times the BS density is required for 73 GHz to experience similar interference (relative to noise) as at 28 GHz.
\item More blockages make systems more noise-limited, not just by blocking the desired signal, but also by blocking (in probability) the stronger interfering signals.  Thus, the denser the buildings and obstacles, the more noise-limited the system is. 
\item  The primary concern for initial mmWave cellular deployments should be achieving sufficient $\snr$.  In large bandwidth and highly directional mmWave networks, interference will only become a significant factor once mmWave cellular has become successful, with extremely dense deployments and heavily loaded cells\footnote{Note that our results assume the worst-case interference, i.e. neighboring BSs are always transmitting, which is very pessimistic from an $\sinr$ viewpoint particularly for a high-rate small-cell mmWave system with few active users per cell \cite{DhiAnd13}.}. However, if initial deployment itself requires very dense network deployment (inter-site distance roughly less than 100 m) to fill in coverage holes, which might happen due to use of smaller antenna arrays or high blockage effects, then interference effects need not be negligible.  
\end{enumerate}

We now discuss several related implications, in view of the system generally being noise-limited.

\subsection{Initial access in \ac{mmWave} systems is challenging}
Initial access is a very important challenging problem for \ac{mmWave} systems, given the low-SNR before beamforming design and the need for large antenna gains to close the link budget  \cite{jeong2015random,Desai2014,BaratiNt.2015}.  The initial access procedure consists of at least two stages, one for the downlink referred to as \emph{cell search} whereby the UE acquires the BS signal, and another for the uplink which we refer to as \emph{random access} whereby the BS acquires the UE signal and learns from the UE which beamforming direction it can receive data on.  Without successfully completing this procedure, it is not even possible to communicate.

The initial access procedure in LTE is  sophisticated \cite{shen2012neighboring}, but in \ac{mmWave} it would be more difficult due to the need for beam alignment, since beams must be searched at  both the BS and UE side  to align them at both ends \cite{barati2015directional}, as our baseline model assumes.  Since it is not possible to beamform user data in one angular direction over one frequency band and do a beam search in another direction over a different frequency band, this leads to the constraint that the initial access procedure must be performed over the entire band because of the necessary use of analog beamforming.  This raises the overhead cost of initial access  and accentuates many difficult tradeoffs.   The more beams that are tested, the better the beam alignment will be and the higher directionality can be achieved, but at the cost of spending a lot of slots testing useless beam combinations at both ends.  The more often the procedure is performed, the more robustly beams can be aligned and new users (or moving users) can be quickly attain connectivity.  But this eats into the time that can be spent transmitting data.  

To explore the impact of initial beam training/association on the coverage and rate trends of mmWave cellular systems, \cite{Alkhateeb2016}  developed a tractable model based on beam sweeping and downlink control pilot reuse. A new metric, called  the effective reliable rate, was defined to capture the  
\ac{SINR} penalty that results from possible beam misalignment errors and the resources overhead that is needed for the beam training process. 
The effective reliable rate 
was derived for two special cases: with near-orthogonal control pilots and with full pilot reuse.  The results showed that unless the employed beams are very wide or the system coherence block length is very small, exhaustive search with full pilot reuse is nearly as good as perfect beam alignment.  Exploring and optimizing initial access further is a key research area for the coming years.

\subsection{The promise of self-backhauling}
Self-backhauling is attractive for mmWave, since it  requires dense deployments and  high-speed backhaul (on the order of Gbps burst rates), which can be difficult to achieve with a wired network. 
Because mmWave networks typically do not have strong interference, and the backhaul links will be directional (and usually not interfering with the access links), self-backhauling is a scalable solution, wherein a fraction of base stations with wired backhaul provide in-band wireless backhaul to the remaining base stations. 

An analytical model and analysis for rate in self-backhauled \ac{mmWave} cellular networks was developed in \cite{Singh2015}. It was observed that increasing the fraction of base stations with wired backhaul improves the peak rates in the network. However, the per user rate saturates if the density of BSs is increased keeping the density of wired backhaul base stations constant. The saturation density was found to be proportional to the density of base stations with wired backhaul. Owing to the subdued interference effects at \ac{mmWave}, it could be even feasible for access and in-band backhaul links to operate on the same time resources\cite{PiKha11,Rangan2014}. This has motivated the investigation of whether dynamic time division duplexing is feasible in self-backhauled \ac{mmWave} networks\cite{Rangan15,Rois15}. 

\subsection{Spectrum license sharing among cellular operators is possible}
There is an early proposal by FCC to use the  28 GHz (27.5-28.35 GHz), 37 GHz (37-38.6 GHz) and 39 GHz (38.6-40 GHz) bands  for cellular services \cite{FCCNOIProposal}. There is also large amount of unlicensed \ac{mmWave} spectrum (57GHz-71GHz) which will be available for lower-power operations similar to WiGig. Although the \ac{mmWave} bands have a relatively large amount of available spectrum,  effective utilization of the spectrum is always important.  Because it will be particularly difficult to deploy a network with truly nationwide coverage at mmWave, exclusively licensing the spectrum to a single entity seems particularly inefficient in terms of spectrum utilization and may even by unnecessary.  

One way to increase the spectrum utilization 
is to use 
authorized shared access \cite{Matinmikko2013} and licensed shared access \cite{Khun-Jush2013}. These  frameworks allow spectrum sharing by a limited number of parties by letting members define sharing rules that can protect them from  interference from each other. Another way to resolve the transmission conflicts between multiple licensees  is by the use of a central database, possibly owned by a third party, which keeps track of transmissions of each licensee to ensure fair and reliable services for each licensee \cite{FCCNOIProposal}.  Along with these coordinated sharing of spectrum, it is also possible that operators can simply share their spectrum licenses without any explicit coordination \cite{Gupta2016} and still achieve higher rates when compared to the rate achieved when  exclusive licensing is used, owing to the low  inter-operator interference in \ac{mmWave}. Moreover, it was shown in \cite{GupAlk16} that static coordination can further improve the overall network performance  while providing a way to differentiate the spectrum access between the different operators. 
 
To evaluate various types of spectrum sharing, one important aspect is  to correctly model a multi-operator system with possible correlation among their locations. It is anticipated for \ac{mmWave} systems (as with the current cellular systems)  that the entity owing the BS site and the cellular operator using that BS may be two independent entities. The site owners can lease the locations to multiple operators which will results in BSs of multiple operators co-located at a single location \cite{And5G}. Our recent work \cite{Gupta2016} presents two possible  cases of a multi-operator system. In the first case, the BS locations of each operator are modeled by a PPP and  assumed to be independent to BSs locations of other operators.  In the second case, consisting of a single PPP representing locations of the sites where BSs of all operators are co-located.  These can be seen as two extreme cases with a real mmWave deployment (and its corresponding SINR and rate performance) somewhere in between.  We found that uncoordinated sharing of spectrum licenses is possible in both the cases as long as the antenna beams are relatively narrow, on the order of about 30 degrees or less. 

\begin{figure}
\centering
\includegraphics[width=0.6\textwidth,trim=27 18 0 0,clip]{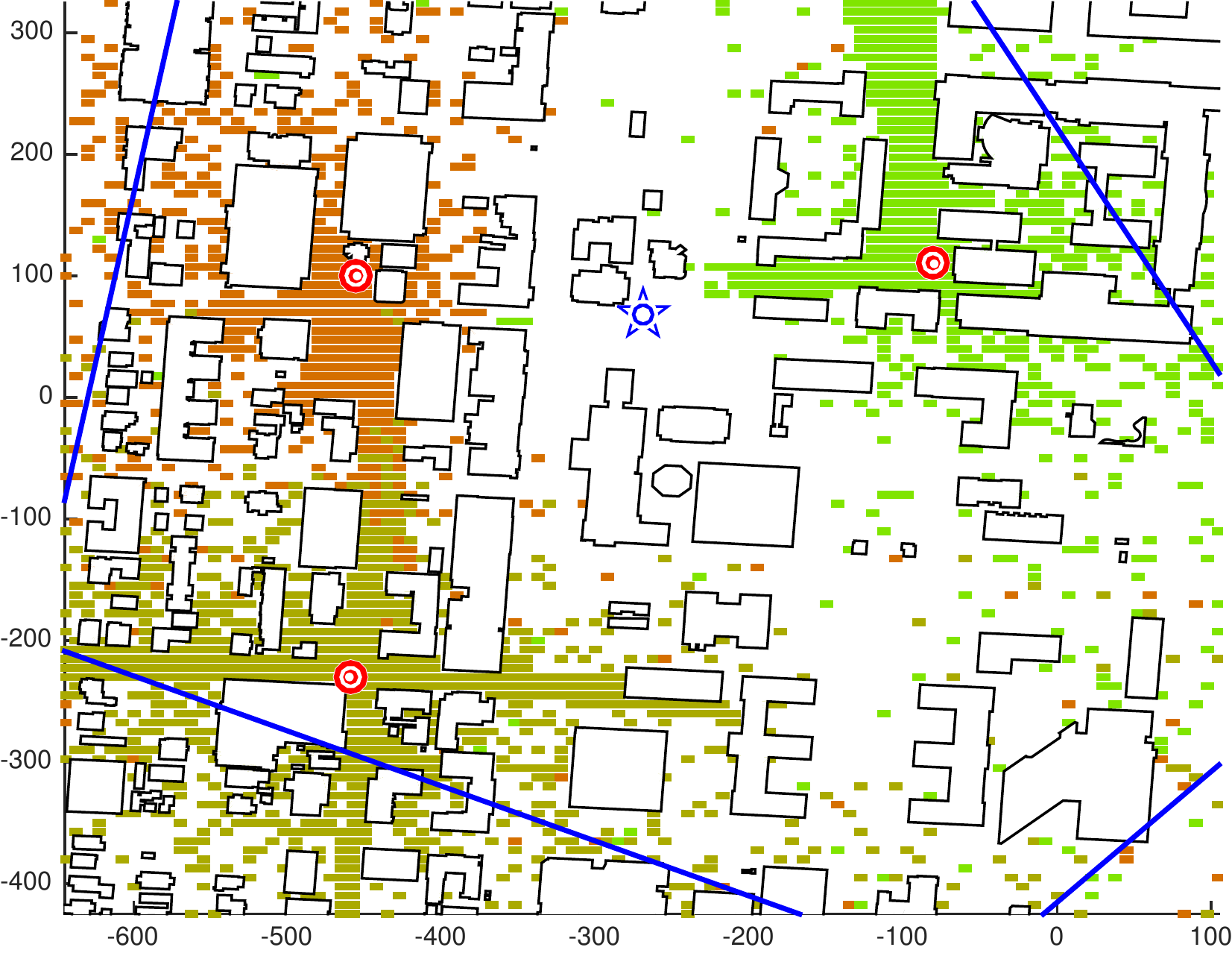}
\caption{Association cells of \ac{mmWave} (the three circles) and a Sub-6GHz BSs (star) overlaid in a portion of Austin, Texas. The shaded areas show the three association cells of the \ac{mmWave} BSs with the macrocell boundaries showed by the solid lines.}
\label{fig:ltemmwaveassoc}
\end{figure}

\subsection{Ultra-densification in \ac{mmWave} networks}
Due to high penetration losses and sensitivity to blockages, the coverage areas of \ac{mmWave} BSs are generally small and irregular, as shown in Fig. \ref{fig:ltemmwaveassoc}.   This issue can be overcome by densification of the network. When  the BS density is low, BS densification decreases the distance of the serving BS and  increases the probability of serving BSs being LOS BS. This results in higher serving power to a typical user. The interference power, however, is not affected significantly as the interferers are still far enough to be LOS to the typical user. Therefore, densification generally helps \ac{mmWave} systems and improve its SINR and rate coverage. 
After a certain threshold, though there will be enough interferers in the LOS region causing the SINR degrade significantly. It was shown in \cite{BaiHea14} that if LOS path loss exponent $\alpha_\mathrm{L}$is below 2, the SINR coverage will finally become zero as BS density $\lambda \rightarrow \infty$. If $\alpha_\mathrm{L}>2$, the SINR coverage converges to a nonzero value at infinite densification. 
Therefore, there is an optimal density (termed as ``critical density'') after which the performance of \ac{mmWave} systems starts degrading \cite{BaiHea14}.

\begin{figure} [!ht]
	\centerline{ 
		\includegraphics[width=0.6\columnwidth]{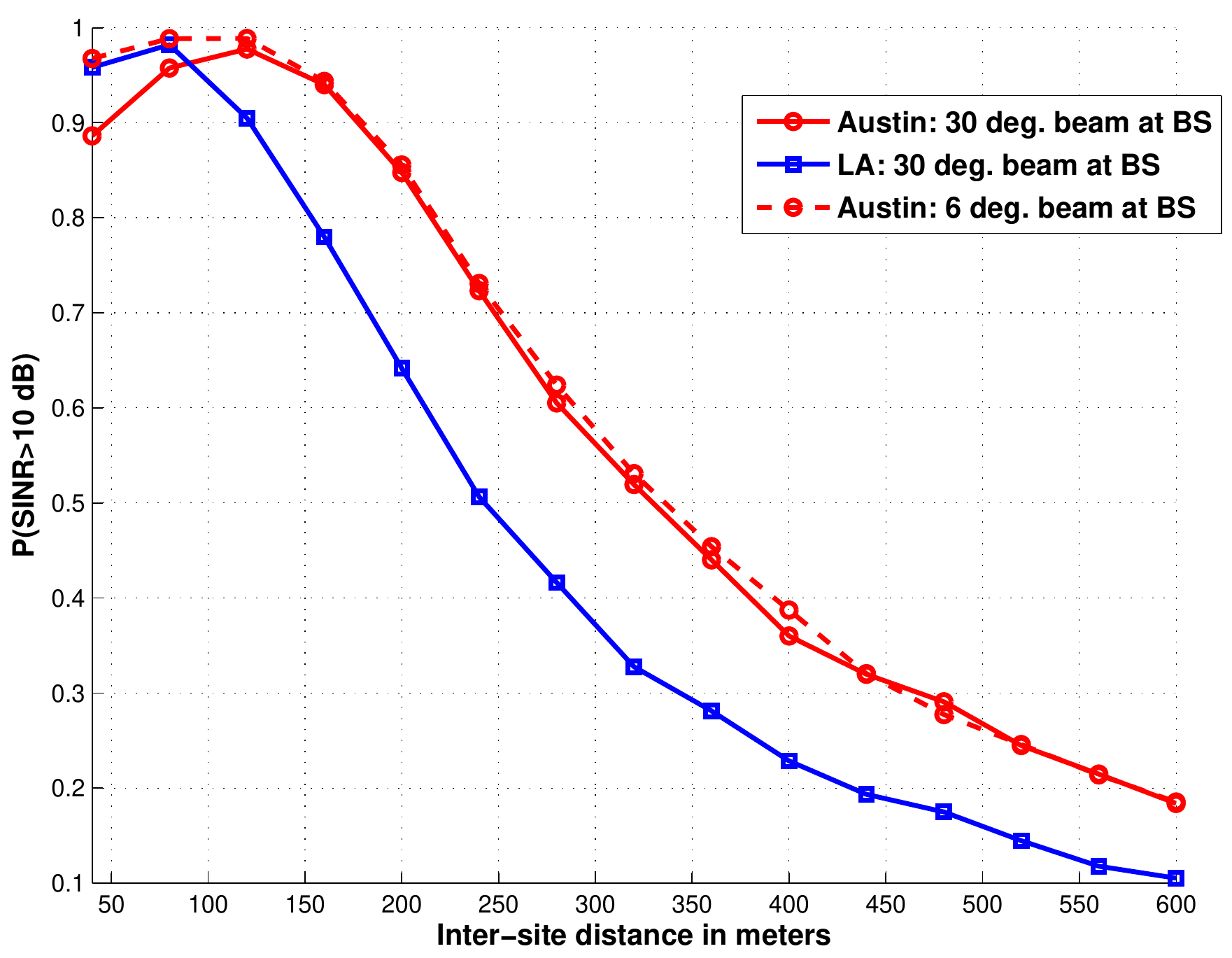}
	}
	\caption{SINR coverage probability targeting 10 dB with different inter-site distances. We simulate the SINR in 28 GHz cellular systems. The bandwidth is 200 MHz. For base station beamforming, the main lobe antenna gain is 20 dB; for mobile station, the main lobe gain is 10 dB, and the beamwidth is $45^\circ$.}\label{fig:SINR_step}
\end{figure}

It was also shown in \cite{Alkhateeb2016} that this critical density is normally in the order of the LOS range, and gets smaller as narrower beams are employed. To illustrate that, we plot in Fig. \ref{fig:SINR_step} the probability $\mathbb{P}(\mathrm{SINR}>10 dB)$ as a function of inter-site distances between base stations in a 28 GHz system. Numerical results indicate that the critical density is proportional to the average LOS range that is 43 meters for Los Angeles, and 85 meters for Austin, according to the data fitting for the random shape theory model in Section \ref{sec:blocking}; and that the critical density decreases with a smaller beam width, as the interference is reduced.

This behavior is similar to that for Sub-6GHz systems under dual-slope pathloss models \cite{ZhaAnd15}, wherein a ``close-in'' path loss
exponent  is used for distances less than a deterministic corner distance $R_c$,
and then changes to larger path loss exponent outside $R_c$. These two regions in the dual slope path loss model are analogous  to the LOS and NLOS regions in the \ac{mmWave} context. The main difference between the \ac{mmWave} and the dual-slope case is that the variable  which defines the LOS region around a user is not deterministic, but a random variable with mean value equal to $R_\mathrm{B}$. However, this difference does not distinguish the general behavior of \ac{mmWave} and Sub-6GHz systems in the ultra-dense regime.

\section{Extensions to the Baseline Model}
\label{sec:extensions}

The SINR and rate coverage results of the previous section can be extended in many 
ways, given the complexity, scope, and large design space of \ac{mmWave} cellular systems.  Crucial extensions include the uplink, as well as more realistic and heterogeneous topologies which would at a minimum include spatial overlap with Sub-6GHz macrocells (that can be used for fallback coverage and control signaling) as well as handling indoor coverage.  Further, given the number of ways that the inherently large antenna arrays in \ac{mmWave} systems can be used (as discussed in Sect. \ref{sec:antennas}), we will also overview how to extend these baseline results to other multi-antenna transmission and reception techniques.

\subsection{Uplink} 
It is natural to extend the downlink SINR analysis to the uplink. Prior analysis in Sub-6GHz cellular networks shows different distributions of interferers between downlink and uplink \cite{Novlan2013,Singh2014}. For example, when the base stations are assumed to be distributed as a \ac{PPP}, then the scheduled users in all cells do not form a \ac{PPP}, due to the Voronoi cell structure \cite{Novlan2013,Singh2014}.  \ac{mmWave} cellular networks will inherit such difference in the network topology between downlink and uplink. More importantly, based on recent study on electromagnetic field exposure \cite{Colombi2015}, the uplink transmit power in \ac{mmWave} networks is expected to be even smaller than that of Sub-6GHz system. Consequently, the uplink tends to be even more noise-limited than the downlink. Therefore, a somewhat trivial but possibly useful extension for obtaining uplink $\sinr$ coverage is to use $\sinr\approx\snr$ and then use the downlink coverage probability for uplink with transmit power replaced with that of the mobile users \cite{Singh2015}. Power control could be possibly neglected for \ac{mmWave} networks considering that these networks are power-limited, so mobiles will typically be transmitting at close to full power (and this will not have much effect on other users' SINRs).

One promising approach to compute the uplink interference distribution is to extend the approach in \cite{Singh2014} that models uplink interferers as a non-homogeneous PPP in Sub-6GHz heterogeneous networks. For example, in \cite{Bai2015ee}, the density functions for LOS other-cell user process $\lambda_\mathrm{u,L}(r)$and NLOS user process $\lambda_\mathrm{u,N}(r)$ are computed as
\begin{align*}
\lambda_\mathrm{u,L}(r)&=\lambda_\mathrm{b}p_\mathrm{LOS}(r)Q(r^{\alpha_\mathrm{L}}/C_\mathrm{L}),\\
\lambda_\mathrm{u,N}(r)&=\lambda_\mathrm{b}\left(1-p_\mathrm{LOS}(r)\right)Q(r^{\alpha_\mathrm{N}}/C_\mathrm{N}),
\end{align*}
where $\lambda_\mathrm{b}$ is the density of base stations, $r$ is the distance to the serving base station of the typical user, and 
\begin{align*}
Q(y)=1-\exp\left(-2\pi\lambda_{\mathrm{b}}\left(\int^{(yC_\mathrm{N})^{1/\alpha_N}}_{0}s\left(1-p_\mathrm{LOS}(s)\right)\mathrm{d}s+\int^{(yC_\mathrm{L})^{1/\alpha_L}}_{0}sp_\mathrm{LOS}(s)\mathrm{d}s\right)\right)
\end{align*}
is the probability that a user's path loss to its associated base station is smaller than $y^{-1}$. A similar approach was used in \cite{GuptaKul2016} to study uplink SINR and rate coverage in \ac{mmWave} cellular networks. 

\subsection{Joint coverage with sub-6GHz  systems}
It seems self-evident that \ac{mmWave} systems cannot be deployed stand alone and still achieve a high level of coverage in an urban area, much less nationwide coverage without a tremendous amount of infrastructure. Rather, a \ac{mmWave} network will generally be overlaid on an LTE-like network to provide high-rate hotspots, with the \ac{mmWave} base stations being used whenever possible to offload from the more congested Sub-6GHz network.  LTE macrocells can be used in multiple ways to assist \ac{mmWave} networks. 

The first way is to help with control signaling.  As discussed above, initial access (and thus all control signaling) poses a particularly troublesome ``chicken and egg'' problem for mmWave cellular.  It seems very likely that some control signaling will happen over the legacy Sub-6GHz spectrum, similar to how some of the variants of unlicensed LTE are using the licensed spectrum for the control plane.  Second, the Sub-6GHz BSs may also be used to provide dual connectivity to both systems \cite{Ghosh14} in which mobile users can be simultaneously connected to both LTE (or future Sub-6GHz system) and \ac{mmWave} BSs, and even receive data from both BSs when possible. In \cite{Singh2015}, an offloading technique was proposed where the data services are mainly provided from \ac{mmWave} BSs (whenever available), but when the \ac{mmWave} link quality drops below a certain threshold, the user reverts to an LTE macrocell.  In this sense, \ac{mmWave} can be viewed as an additional carrier aggregation technique for existing LTE systems.

Recent work \cite{Singh2015,Ishii2012,Li2013a} has analytically modeled the coexistence of \ac{mmWave} and LTE systems to derive  performance metrics and design insights.  In \cite{Singh2015,Li2013a}, the locations of macrocells and \ac{mmWave} BSs  were modeled as independent PPPs with different densities and the coexistence is modeled by superposition of these two PPPs. A clustered point process such as Neyman-Scott process can also be used to model the co-existence of LTE and \ac{mmWave} where Sub-6GHz  BS will be located at the center of the clusters while \ac{mmWave} BSs will be spread around the center.

Technically, a coexisting Sub-6GHz and \ac{mmWave} cellular network is a type of \ac{HetNet} where the tiers do not interfere with each other, and also have a vast disparity in terms of bandwidth and other parameters. By the law of total probability, the rate coverage in such a \ac{HetNet} is given by
\begin{equation}
\ratecov(\tau) = \mathcal{A}_{\text{mmW}}\mathcal{R}_{\text{mmW}}(\tau) +\mathcal{A}_{\text{UHF}} \mathcal{R}_{\text{UHF}}(\tau),
\end{equation}
where $\mathcal{A}_{\text{mmW}}$ and $\mathcal{A}_{\text{UHF}}$ are the association probabilities, and $\mathcal{R}_{\text{mmW}} $ and $\mathcal{R}_{\text{UHF}}$ are rate coverages conditioned on association to the respective tiers. If the BSs in both tiers are modeled as \ac{PPP}, $\mathcal{R}_{(.)}$ can be computed from Theorem~\ref{thm:ratecov} by replacing $\BSdensity$ in $\sinrcov(.)$ and $\kappa(.)$ expressions with the density of BSs in the respective tier.

It was shown in \cite{Shaer16} that incorporating beamforming gains during cell association based on downlink/uplink received signal power significantly boosts the probability of connecting to a large bandwidth \ac{mmWave} BS which is beneficial from a rate standpoint. Due to the bandwidth disparity in the two tiers, it may be desirable to connect to a \ac{mmWave} BS offering much lower $\snr$ than a \ac{UHF} BS. However, there is an optimum value of bias towards \ac{mmWave} BSs after which the rate again starts to decrease due to weak $\snr$ over the \ac{mmWave} link. This insight is in line with that in \cite{Singh2015}, wherein it was shown that users should be offloaded to \ac{UHF} networks only if communication over \ac{mmWave} link is infeasible. Thus, studying robust modulation and coding schemes that can decode low $\snr$ signals is an intriguing avenue of physical layer research for \ac{mmWave} cellular networks.

In \cite{Shaer16} the BS locations of \ac{mmWave} and Sub-6GHz base stations were modeled as independent PPPs for tractable analysis. In reality, however, some of the mmW and Sub-6GHz base stations may be co-located or their locations may be correlated depending on hotspot traffic areas. 
In future work, examinations that consider the correlations in both base stations locations and load are essential to understand the gain of such multi-band joint-coverage system.
 
\subsection{Outdoor-to-indoor coverage}
The framework in Section \ref{sec:analysis} mainly focuses on the performance analysis in outdoor \ac{mmWave} cellular networks. Due to the huge penetration losses from outer walls of buildings, indoor users -- if not near a window -- will unlikely be served by outdoor \ac{mmWave} base stations. Therefore, users inside buildings need to be served either by indoor base stations or by lower carrier frequency systems.  This is a very serious limitation of mmWave cellular systems, given the prevalence of indoor cellular data usage: over 80\% of all current cellular communications are with indoor users according to \cite{NokiaIndoor14,CiscoIndoor2012}, which does not even include WiFi offloading.    One silver lining, as discussed in \cite{Taranetz2014},  is that when indoor small cells are deployed, the indoor SINR coverage and rate performance improves with a larger penetration loss through the outer walls, due to the reduction of the interference from outdoor base stations.  Thus, in many cases the \ac{mmWave} system may serve effectively as a wireless backhaul network to a separate indoor wireless data network (which could be \ac{mmWave} or WiFi) \cite{Taori2015}. 

Extending our baseline outdoor only model to various indoor scenarios is important given the prevalence of indoor use, but technically challenging.  Challenges including that the outdoor model does not incorporate (i) the first-order reflections (from walls) that can have comparable strengths with the direct paths in certain indoor scenarios \cite{Xu2002}, and (ii) the partition losses, e.g. from inner walls, that contribute to a large fraction of the overall indoor path loss \cite{Anderson2004}.  In \cite{Venugopal2016,GeoLoz2015}, given the semi-specular nature of the \ac{mmWave} signal propagation, the first-order reflections were taken account by considering the walls as mirrors, and modeling the images of transmitters approximately. To simplify the analysis, the locations of the image transmitters were approximated as an independent process of the original transmitters with the same density. Numerical results show that such approximation brings in minor losses in accuracy when computing the SINR distributions in certain indoor scenarios.  Meanwhile, the partition loss has been characterized for the indoor performance analysis in \cite{Zhang2015c}, where the inner walls of a building were modeled as a Poisson line model. Based on the proposed model, the distribution of the interference at a typical indoor user was derived, assuming the partition losses from the walls dominate the path loss, i.e., ignoring the free-space path loss.

\subsection{MIMO techniques beyond analog beamforming}

Analog beamforming is the default approach for communication in current commercial systems like IEEE 802.11ad and WirelessHD. The large antenna arrays at \ac{mmWave} beyond analog beamforming, allow more sophisticated forms of MIMO respecting the hardware constraints in Section \ref{sec:antennas}.

There has not been much work on extension of the baseline model for analog beamforming to other MIMO techniques.  Recent work in \cite{KulGhoAnd16} uses the virtual channel representation to develop an analytical model for MU-MIMO based on the hybrid precoding algorithm proposed in \cite{Alkhateeb2014b}. Perfect channel state information and the narrowband channel model in \eqref{eq:channel} was assumed. The analysis enabled comparison with hybrid precoding based SU-MIMO and single-user analog beamforming, and also highlighted several design insights. For example, (a) the optimum base station density in terms of $\sinr$ coverage decreases with increasing degree of multiuser transmission, (b) the cell edge rates suffer with round robin scheduling which highlights the importance of user selection while employing MU-MIMO, (c) a denser single-user beamforming network provides higher cell edge rates than a less dense MU-MIMO network that consumes same power per unit area. 

Several extensions of the baseline model are possible. We briefly mention some key extensions here.

\begin{enumerate}
\item Incorporating a channel with rank greater than one: 
The impact of multiple spatial paths from the transmitter to receiver may affect the interference statistics. Also, enabling multi-stream transmission highly relies on the rank of the channel. 
\item Incorporating the impact of different MIMO transceivers in network level analysis: 
Impact of analog beamforming, hybrid beamforming, low resolution ADCs at receivers, CAP-MIMO on network-level performance could be evaluated and compared. This can provide insight on which MIMO architectures are suitable in what scenarios depending on network parameters like available base station density, number of antennas, number of RF chains, power consumption constraints, etc. The virtual channel representation could provide simple means to incorporate the impact of different precoders and combiners. 
\item Imperfect channel state information: 
Most existing stochastic geometry analysis of \ac{mmWave} networks assumes perfect channel state information at the transmitter apart from a few exceptions like \cite{Renzo2015,Alkhateeb2016}.  Although developing analytical models with perfect channel information is a good first step, studying the impact of imperfect channel information and channel aging is important. Given the importance of narrow beams to enable outdoor mmWave cellular, wrong estimates of angles of departure or arrival may significantly drop the post processing SINR. 
\item Massive MIMO at mmWave:
Another extension is to analyze the performance of \ac{mmWave} massive MIMO networks, a natural extension of massive MIMO networks at lower frequencies \cite{Swindlehurst2014}. Due to the aforementioned difference in propagation channels and hardware constraints, prior stochastic geometry models for Sub-6GHz massive MIMO networks, e.g. in \cite{Bjornson2016,Bai2015a}, do not directly apply to the \ac{mmWave} bands. In \cite{Bai2015ee}, key features in \ac{mmWave} channels, including the blockage effects and channel sparsity (in terms of multi-paths), and certain differences in hardware, e.g. the beamforming at mobile stations, were incorporated to model \ac{mmWave} massive MIMO cellular networks. Based on the model, the asymptotic SINR and rate performance were analyzed, when the number of base station antennas goes to infinity. Numerical results based on the analysis show that mmWave massive MIMO requires a high base station density to achieve good SINR coverage; the SINR distribution in the asymptotic case is a good approximation for the cases with more than 256 antennas, in certain dense mmWave networks. For future work, it would be valuable to incorporate \ac{mmWave}-compatible channel training into the system analysis, e.g. the compressed sensing based approach in the angular domain \cite{Ramasamy2013,Alkhateeb2015,Mendez-Rial2016,Choi2015b} which could potentially reduce training time \cite{Swindlehurst2014}. 
\end{enumerate}

\section{Conclusions}

The upcoming standardization and development of mmWave cellular systems is one of the largest leaps forward in wireless communications in the last two decades. Going to mmWave though introduces novel design challenges and research questions.  This paper described the two most important physical challenges -- susceptibility to blocking and the need for strong directionality -- and has provided a baseline mathematical model and analysis for these systems accounting for these factors.  

There are many open questions and key extensions remaining, some of which we overview in Sects. \ref{sec:implications} and \ref{sec:extensions}.  For example, the crucial topic of outdoor-to-indoor coverage is essentially neglected in work to date.  How to do load balancing and offloading in light of directionality and blocking, both between mmWave cells and between mmWave and Sub-6GHz cells, is not well understood.  The support of mobility is also not discussed here, which will require considerable effort (and system overhead) to keep the beams aligned in both the downlink and uplink directions.  Handoffs also will no longer be based mostly on distance, but blocking (including by one's own body) may often be the dominant factor, making the need for handoff more difficult to predict. In short, we expect this new paradigm for cellular communication to challenge wireless engineers for some time. We expect that the models developed in this paper will continue to be improved and extended to help aid the understanding and design of these mmWave cellular systems.

\bibliographystyle{IEEEtran}
\bibliography{mmCellularBib}

\begin{thebibliography}{100}
\providecommand{\url}[1]{#1}
\csname url@samestyle\endcsname
\providecommand{\newblock}{\relax}
\providecommand{\bibinfo}[2]{#2}
\providecommand{\BIBentrySTDinterwordspacing}{\spaceskip=0pt\relax}
\providecommand{\BIBentryALTinterwordstretchfactor}{4}
\providecommand{\BIBentryALTinterwordspacing}{\spaceskip=\fontdimen2\font plus
\BIBentryALTinterwordstretchfactor\fontdimen3\font minus
  \fontdimen4\font\relax}
\providecommand{\BIBforeignlanguage}[2]{{%
\expandafter\ifx\csname l@#1\endcsname\relax
\typeout{** WARNING: IEEEtran.bst: No hyphenation pattern has been}%
\typeout{** loaded for the language `#1'. Using the pattern for}%
\typeout{** the default language instead.}%
\else
\language=\csname l@#1\endcsname
\fi
#2}}
\providecommand{\BIBdecl}{\relax}
\BIBdecl

\bibitem{PiKha11}
Z.~Pi and F.~Khan, ``An introduction to millimeter-wave mobile broadband
  systems,'' \emph{IEEE Commun. Mag.}, vol.~49, no.~6, pp. 101 --107, June
  2011.

\bibitem{Fel56}
R.~G. Fellers, ``Millimeter waves and their applications,'' \emph{Electrical
  Engineering}, vol.~75, no.~10, pp. 914--17, Oct. 1956.

\bibitem{Rappaport2013a}
T.~Rappaport, S.~Sun, R.~Mayzus, H.~Zhao, Y.~Azar, K.~Wang, G.~Wong, J.~Schulz,
  M.~Samimi, and F.~Gutierrez, ``Millimeter wave mobile communications for {5G}
  cellular: It will work!'' \emph{IEEE Access}, vol.~1, pp. 335--349, May 2013.

\bibitem{Walke1985}
B.~Walke and R.~Briechle, ``A local cellular radio network for digital voice
  and data transmission at {60} {GHz},'' in \emph{Proc. International Cellular
  \& Mobile Comm.}\hskip 1em plus 0.5em minus 0.4em\relax London: Online, Nov.
  1985, pp. 215--225.

\bibitem{Currie1987}
N.~C. Currie and C.~E. Brown, \emph{Principles and applications of
  millimeter-wave radar}.\hskip 1em plus 0.5em minus 0.4em\relax Artech House,
  1987.

\bibitem{Russell1997}
M.~E. Russell, A.~Crain, A.~Curran, R.~A. Campbell, C.~A. Drubin, and W.~F.
  Miccioli, ``Millimeter-wave radar sensor for automotive intelligent cruise
  control {(ICC)},'' \emph{{IEEE} Trans. Microwave Theory Tech.}, vol.~45,
  no.~12, pp. 2444--2453, Dec 1997.

\bibitem{Clark1998}
S.~Clark and H.~Durrant-Whyte, ``Autonomous land vehicle navigation using
  millimeter wave radar,'' in \emph{Proc. IEEE International Conference on
  Robotics and Automation}, vol.~4, May 1998, pp. 3697--3702 vol.4.

\bibitem{Woodward2002}
R.~M. Woodward, B.~E. Cole, V.~P. Wallace, R.~J. Pye, D.~D. Arnone, E.~H.
  Linfield, and M.~Pepper, ``Terahertz pulse imaging in reflection geometry of
  human skin cancer and skin tissue,'' \emph{Phys. Med. Biol.}, vol.~47,
  no.~21, p. 3853, 2002.

\bibitem{Taylor2011}
Z.~D. Taylor, R.~S. Singh, D.~B. Bennett, P.~Tewari, C.~P. Kealey, N.~Bajwa,
  M.~O. Culjat, A.~Stojadinovic, H.~Lee, J.~P. Hubschman, E.~R. Brown, and
  W.~S. Grundfest, ``{THz} medical imaging: in vivo hydration sensing,''
  \emph{IEEE Trans. THz Sci. Technol.}, vol.~1, no.~1, pp. 201--219, Sept 2011.

\bibitem{Kato2001}
A.~Kato, K.~Sato, and M.~Fujise, ``{ITS} wireless transmission technology:
  Technologies of millimeter-wave inter-vehicle communications: Propagation
  characteristics,'' \emph{Journal of the Communications Research Laboratory},
  vol.~48, no.~4, pp. 99--110, 2001.

\bibitem{Meinel1995}
H.~H. Meinel, ``Commercial applications of millimeterwaves: History, present
  status, and future trends,'' \emph{IEEE Trans. Microw. Theory Techn.},
  vol.~43, no.~7, pp. 1639 -- 1653, 1995.

\bibitem{Tsugawa2005}
S.~Tsugawa, ``Issues and recent trends in vehicle safety communication
  systems,'' \emph{IATSS Research}, vol.~29, pp. 7--15, 2005.

\bibitem{VaShiBan:Millimeter-Wave-Vehicular:16}
V.~Va, T.~Shimizu, G.~Bansal, and R.~W. {Heath Jr.}, \emph{Millimeter Wave
  Vehicular Communications: A Survey}.\hskip 1em plus 0.5em minus 0.4em\relax
  NOW: Foundations and Trends in Networking, 2016.

\bibitem{ISO-21216-CALM-MM}
{ISO Standard}, ``Intelligent transportation systems--communication access for
  land mobiles ({CALM})--millimetre wave air interface,'' Feb. 2012, iSO/FDIS
  21216.

\bibitem{Kenney2011}
J.~B. Kenney, ``Dedicated short-range communications ({DSRC}) standards in the
  {U}nited {S}tates,'' \emph{Proceedings of the IEEE}, vol.~99, no.~7, pp.
  1162--1182, July 2011.

\bibitem{Park2007}
C.~Park and T.~S. Rappaport, ``Short-range wireless communications for
  next-generation networks: {UWB}, 60 {GHz} millimeter-wave {WPAN}, and
  {ZigBee},'' \emph{{IEEE} Wireless Commun. Mag.}, vol.~14, no.~4, pp. 70--78,
  Aug. 2007.

\bibitem{DanHea07}
R.~C. Daniels and R.~W. Heath, ``60 {GHz} wireless communications: emerging
  requirements and design recommendations,'' \emph{IEEE Veh. Technol. Mag.},
  vol.~2, no.~3, pp. 41--50, Sept. 2007.

\bibitem{11ad}
\BIBentryALTinterwordspacing
{IEEE 802.11ad}, ``{IEEE} 802.11ad standard draft {D0.1}.'' [Online].
  Available: \url{www.ieee802.org/11/Reports/tgad update.htm}
\BIBentrySTDinterwordspacing

\bibitem{Baykas2011}
T.~Baykas, C.-S. Sum, Z.~Lan, J.~Wang, M.~Rahman, H.~Harada, and S.~Kato,
  ``{IEEE} 802.15.3c: the first {IEEE} wireless standard for data rates over 1
  {Gb/s},'' \emph{{IEEE} Commun. Mag.}, vol.~49, no.~7, pp. 114--121, July
  2011.

\bibitem{WirelessHDSrandard2007}
{WirelessHD Standard}, ``{WirelessHD} specification version 1.0 overview,''
  \emph{WirelessHD}, Oct. 2007.

\bibitem{Rappaport2014book}
T.~S. Rappaport, R.~W. {Heath, Jr.}, R.~C. Daniels, and J.~N. Murdock,
  \emph{Millimeter Wave Wireless Communications}.\hskip 1em plus 0.5em minus
  0.4em\relax Prentice Hall, 2014.

\bibitem{Tharek1988}
A.~R. Tharek and J.~P. McGeehan, ``Propagation and bit error rate measurements
  within buildings in the millimeter wave band about 60 {GHz},'' in \emph{Proc.
  European Conference on Electrotechnics}, June 1988, pp. 318--321.

\bibitem{Davies1991}
R.~Davies, M.~Bensebti, M.~A. Beach, and J.~P. McGeehan, ``Wireless propagation
  measurements in indoor multipath environments at 1.7 {GHz} and 60 {GHz} for
  small cell systems,'' in \emph{Proc. IEEE VTC}, May 1991, pp. 589--593.

\bibitem{Manabe1994}
T.~Manabe, K.~Taira, K.~Sato, T.~Ihara, Y.~Kasashima, and K.~Yamaki,
  ``Multipath measurement at 60 {GHz} for indoor wireless communication
  systems,'' in \emph{Proc. IEEE VTC}, June 1994, pp. 905--909 vol.2.

\bibitem{Smulders1995}
P.~F.~M. Smulders and A.~G. Wagemans, ``Frequency-domain measurement of the
  millimeter wave indoor radio channel,'' \emph{{IEEE} Trans. Instrum. Meas.},
  vol.~44, no.~6, pp. 1017--1022, Dec 1995.

\bibitem{Smulders1997}
P.~Smulders and L.~Correia, ``Characterisation of propagation in 60 {GHz} radio
  channels,'' \emph{Electronics \& Communication Engineering Journal}, vol.~9,
  no.~2, pp. 73--80, 1997.

\bibitem{Manabe1996}
T.~Manabe, Y.~Miura, and T.~Ihara, ``Effects of antenna directivity and
  polarization on indoor multipath propagation characteristics at 60 {GHz},''
  \emph{{IEEE} J. Sel. Areas Commun.}, vol.~14, no.~3, pp. 441--448, Apr 1996.

\bibitem{Xu2002}
H.~Xu, V.~Kukshya, and T.~Rappaport, ``Spatial and temporal characteristics of
  60-{GHz} indoor channels,'' \emph{{IEEE} J. Sel. Areas Commun.}, vol.~20,
  no.~3, pp. 620--630, Apr. 2002.

\bibitem{Moraitis2004a}
N.~Moraitis and P.~Constantinou, ``Indoor channel measurements and
  characterization at 60 {GHz} for wireless local area network applications,''
  \emph{IEEE Trans. Antennas Propag.}, vol.~52, no.~12, pp. 3180--3189, Dec
  2004.

\bibitem{Anderson2004}
C.~Anderson and T.~Rappaport, ``{In-building wideband partition loss
  measurements at 2.5 and 60 GHz},'' \emph{IEEE Trans. Wireless Commun.},
  vol.~3, no.~3, pp. 922--928, May 2004.

\bibitem{Collonge2004}
S.~Collonge, G.~Zaharia, and G.~E. Zein, ``Influence of the human activity on
  wide-band characteristics of the 60 {GHz} indoor radio channel,''
  \emph{{IEEE} Trans. Wireless Commun.}, vol.~3, no.~6, pp. 2396--2406, Nov.
  2004.

\bibitem{Zwick2005a}
T.~Zwick, T.~J. Beukema, and H.~Nam, ``Wideband channel sounder with
  measurements and model for the 60 {GHz} indoor radio channel,'' \emph{{IEEE}
  Trans. Veh. Technol.}, vol.~54, no.~4, pp. 1266--1277, July 2005.

\bibitem{Geng2009}
S.~Geng, J.~Kivinen, X.~Zhao, and P.~Vainikainen, ``Millimeter-wave propagation
  channel characterization for short-range wireless communications,''
  \emph{{IEEE} Trans. Veh. Technol.}, vol.~58, no.~1, pp. 3--13, May 2009.

\bibitem{WyneMolisch2011TWC}
S.~Wyne, K.~Haneda, S.~Ranvier, F.~Tufvesson, and A.~Molisch, ``Beamforming
  effects on measured mm-wave channel characteristics,'' \emph{IEEE Trans.
  Wireless Commun.}, vol.~10, no.~11, pp. 3553--3559, Sept. 2011.

\bibitem{Ben-Dor2011}
E.~Ben-Dor, T.~S. Rappaport, Y.~Qiao, and S.~J. Lauffenburger, ``Millimeter
  wave 60 {GHz} outdoor and vehicle {AOA} propagation measurements using a
  broadband channel sounder,'' in \emph{Proc. IEEE GLOBECOM}, Dec 2011, pp.
  1--6.

\bibitem{murdock201238}
J.~Murdock, E.~Ben-Dor, Y.~Qiao, J.~Tamir, and T.~Rappaport, ``A 38 {GHz}
  cellular outage study for an urban outdoor campus environment,'' in
  \emph{Proc. IEEE WCNC}, Apr. 2012, pp. 3085--3090.

\bibitem{rappaport2012cellular}
T.~Rappaport, Y.~Qiao, J.~Tamir, J.~Murdock, and E.~Ben-Dor, ``Cellular
  broadband millimeter wave propagation and angle of arrival for adaptive beam
  steering systems,'' in \emph{Proc. {IEEE} Radio Wireless Symp.}, Jan. 2012,
  pp. 151--154.

\bibitem{Rappaport2015}
T.~Rappaport, G.~Maccartney, M.~Samimi, and S.~Sun, ``Wideband millimeter-wave
  propagation measurements and channel models for future wireless communication
  system design,'' \emph{IEEE Trans. Commun.}, vol.~63, no.~9, pp. 3029--3056,
  Sept. 2015.

\bibitem{Weiler14}
R.~J. Weiler, M.~Peter, W.~Keusgen, H.~Shimodaira, K.~T. Gia, and K.~Sakaguchi,
  ``Outdoor millimeter-wave access for heterogeneous networks: Path loss and
  system performance,'' in \emph{Proc. IEEE PIMRC}, Sept. 2014, pp. 2189--2193.

\bibitem{haneda20165g}
K.~Haneda, L.~Tian, Y.~Zheng, H.~Asplund, J.~Li, Y.~Wang, D.~Steer, C.~Li,
  T.~Balercia, S.~Lee \emph{et~al.}, ``{5G} {3GPP}-like channel models for
  outdoor urban microcellular and macrocellular environments,'' \emph{arXiv
  preprint arXiv:1602.07533}, 2016.

\bibitem{MacCartney2015}
G.~R. MacCartney, M.~K. Samimi, and T.~S. Rappaport, ``Exploiting
  directionality for millimeter-wave wireless system improvement,'' in
  \emph{Proc. {IEEE ICC}}, June 2015, pp. 2416--2422.

\bibitem{Va2015}
V.~Va, J.~Choi, and R.~W. Heath~Jr, ``The impact of beamwidth on temporal
  channel variation in vehicular channels and its implications,''
  \emph{submitted to IEEE Veh. Technol., arXiv preprint arXiv:1511.02937},
  2015.

\bibitem{Goldsmith2005}
A.~Goldsmith, \emph{Wireless Communications}.\hskip 1em plus 0.5em minus
  0.4em\relax Cambridge University Press, 2005.

\bibitem{NSN13}
S.~Larew, T.~Thomas, and A.~Ghosh, ``Air interface design and ray tracing study
  for 5{G} millimeter wave communications,'' in \emph{Proc. IEEE GLOBECOM B4G
  Workshop}, Dec. 2013, pp. 117 -- 122.

\bibitem{Cud14}
M.~Cudak, T.~Kovarik, T.~A. Thomas, A.~Ghosh, Y.~Kishiyama, and T.~Nakamura,
  ``Experimental mm{W}ave 5{G} cellular systems,'' in \emph{Proc. IEEE GLOBECOM
  Workshops}, Dec. 2014, pp. 377--381.

\bibitem{Ghosh14}
A.~Ghosh, T.~A. Thomas, M.~C. Cudak, R.~Ratasuk, P.~Moorut, F.~W. Vook, T.~S.
  Rappaport, G.~MacCartney, S.~Sun, and S.~Nie, ``Millimeter-wave enhanced
  local area systems: A high data rate approach for future wireless networks,''
  \emph{{IEEE} J. Sel. Areas Commun.}, vol.~32, no.~6, pp. 1152--1163, June
  2014.

\bibitem{Roh2014}
W.~Roh, J.-Y. Seol, J.~Park, B.~Lee, J.~Lee, Y.~Kim, J.~Cho, K.~Cheun, and
  F.~Aryanfar, ``Millimeter wave beamforming as an enabling technology for {5G}
  cellular communications: theoretical feasibility and prototype results,''
  \emph{{IEEE} Commun. Mag.}, vol.~52, no.~2, pp. 106--113, Feb. 2014.

\bibitem{Hong2014}
W.~Hong, K.-H. Baek, Y.~Lee, Y.~Kim, and S.-T. Ko, ``Study and prototyping of
  practically large-scale {mmWave} antenna systems for {5G} cellular devices,''
  \emph{{IEEE} Commun. Mag.}, vol.~52, no.~9, pp. 63--69, Sept. 2014.

\bibitem{Wu2015}
T.~Wu, T.~S. Rappaport, and C.~M. Collins, ``The human body and millimeter-wave
  wireless communication systems: Interactions and implications,'' in
  \emph{Proc. {IEEE ICC}}, June 2015, pp. 2423--2429.

\bibitem{Wu2015a}
------, ``Safe for generations to come: Considerations of safety for millimeter
  waves in wireless communications,'' \emph{IEEE Microw. Mag.}, vol.~16, no.~2,
  pp. 65--84, March 2015.

\bibitem{Akoum2012}
S.~Akoum, E.~O. Ayach, and R.~W. Heath~Jr., ``Coverage and capacity in {mmWave}
  cellular systems,'' in \emph{Proc. ASILOMAR}, Nov. 2012, pp. 688--692.

\bibitem{BaiHea14}
T.~Bai and {R. W. Heath}{ Jr.}, ``Coverage and rate analysis for millimeter
  wave cellular networks,'' \emph{{IEEE} Trans. Wireless Commun.}, vol.~14,
  no.~2, pp. 1100--1114, Oct. 2014.

\bibitem{Rangan2014}
S.~Rangan, T.~Rappaport, and E.~Erkip, ``Millimeter-wave cellular wireless
  networks: Potentials and challenges,'' \emph{Proc. {IEEE}}, vol. 102, no.~3,
  pp. 366--385, Mar. 2014.

\bibitem{Akdeniz2013a}
M.~Akdeniz, Y.~Liu, M.~Samimi, S.~Sun, S.~Rangan, T.~Rappaport, and E.~Erkip,
  ``Millimeter wave channel modeling and cellular capacity evaluation,''
  \emph{IEEE J. Sel. Areas Commun.}, vol.~32, no.~6, pp. 1164--1179, June 2014.

\bibitem{Kim13}
T.~Kim, J.~Park, J.-Y. Seol, S.~Jeong, J.~Cho, and W.~Roh, ``Tens of {G}bps
  support with {mmWave} beamforming systems for next generation
  communications,'' in \emph{Proc. IEEE GLOBECOM}, Dec. 2013, pp. 3685--3690.

\bibitem{Abouelseoud13}
M.~Abouelseoud and G.~Charlton, ``System level performance of millimeter-wave
  access link for outdoor coverage,'' in \emph{Proc. IEEE WCNC}, Apr. 2013, pp.
  4146--4151.

\bibitem{AndBac11}
J.~G. Andrews, F.~Baccelli, and R.~K. Ganti, ``A tractable approach to coverage
  and rate in cellular networks,'' \emph{IEEE Trans. Commun.}, vol.~59, no.~11,
  pp. 3122--3134, Nov. 2011.

\bibitem{HaeAnd09}
M.~Haenggi, J.~G. Andrews, F.~Baccelli, O.~Dousse, and M.~Franceschetti,
  ``Stochastic geometry and random graphs for the analysis and design of
  wireless networks,'' \emph{IEEE J. Sel. Areas Commun.}, vol.~27, no.~7, pp.
  1029--46, Sept. 2009.

\bibitem{BacNOW}
F.~Baccelli and B.~Blaszczyszyn, \emph{Stochastic Geometry and Wireless
  Networks}.\hskip 1em plus 0.5em minus 0.4em\relax NOW: Foundations and Trends
  in Networking, 2010.

\bibitem{HaenggiBook}
M.~Haenggi, \emph{Stochastic Geometry for Wireless Networks}.\hskip 1em plus
  0.5em minus 0.4em\relax Cambridge University Publishers, 2012.

\bibitem{Hae14}
------, ``The mean interference-to-signal ratio and its key role in cellular
  and amorphous networks,'' \emph{IEEE Wireless Commun. Lett.}, vol.~3, no.~6,
  pp. 597--600, Dec. 2014.

\bibitem{GuoHae15}
A.~Guo and M.~Haenggi, ``Asymptotic deployment gain: A simple approach to
  characterize the {SINR} distribution in general cellular networks,''
  \emph{IEEE Trans. Commun.}, vol.~63, no.~3, pp. 962--76, Mar. 2015.

\bibitem{Bai2014b}
T.~Bai, R.~Vaze, and R.~Heath, ``Analysis of blockage effects on urban cellular
  networks,'' \emph{IEEE Trans. Wireless Commun.}, vol.~13, no.~9, pp.
  5070--5083, Sept. 2014.

\bibitem{Singh2015}
S.~Singh, M.~Kulkarni, A.~Ghosh, and J.~Andrews, ``Tractable model for rate in
  self-backhauled millimeter wave cellular networks,'' \emph{IEEE J. Sel. Areas
  Commun.}, vol.~33, no.~10, pp. 2196--2211, Oct. 2015.

\bibitem{3GPPTR36.8142010}
{3GPP TR 36.814}, ``Further advancements for {E-UTRA} physical layer aspects
  ({R}elease 9),''
  \url{http://www.3gpp.org/ftp/Specs/archive/36series/36.814/36814-900.zip},
  Mar. 2010.

\bibitem{3GPP3D}
{3GPP TR 36.873}, ``{Technical Specification Group Radio Access Network; Study
  on 3D Channel Model for LTE (Release 12)},'' Sept. 2014.

\bibitem{5GChannel}
\BIBentryALTinterwordspacing
``{5G} channel model for bands up to 100 {GHz} (white paper),'' Dec. 2015.
  [Online]. Available: \url{http://www.5gworkshops.com/5GCM.html}
\BIBentrySTDinterwordspacing

\bibitem{ZhaRyu15}
Z.~Zhang, J.~Ryu, S.~Subramanian, and A.~Sampath, ``Coverage and channel
  characteristics of millimeter wave band using ray tracing,'' in \emph{Proc.
  IEEE ICC}, June 2015, pp. 1380--1385.

\bibitem{Langen1994}
B.~Langen, G.~Lober, and W.~Herzig, ``Reflection and transmission behaviour of
  building materials at 60 {GHz},'' in \emph{Proc. of IEEE PIMRC}, vol.~2, Sep.
  1994, pp. 505--509.

\bibitem{Lu2012}
J.~S. Lu, D.~Steinbach, P.~Cabrol, and P.~Pietraski, ``Modeling human blockers
  in millimeter wave radio links,'' \emph{ZTE Communications}, vol.~10, no.~4,
  pp. 23--28, Dec. 2012.

\bibitem{Rajagopal2012}
S.~Rajagopal, S.~Abu-Surra, and M.~Malmirchegini, ``Channel feasibility for
  outdoor non-line-of-sight {mmWave} mobile communication,'' in \emph{Proc.
  IEEE VTC}, Sept. 2012, pp. 1--6.

\bibitem{Ding2015}
M.~Ding, P.~Wang, D.~Lopez-Perez, G.~Mao, and Z.~Lin, ``Performance impact of
  {LoS} and {NLoS} transmissions in dense cellular networks,'' \emph{IEEE
  Trans. Wireless Commun.}, vol.~15, no.~3, pp. 2365--2380, Nov. 2015.

\bibitem{Bai2014c}
T.~Bai and R.~W. Heath~Jr., ``Analysis of self-body blocking effects in
  millimeter wave cellular networks,'' in \emph{Proc. ASILOMAR}, Nov. 2014, pp.
  1921--1925.

\bibitem{ZhaAnd15}
X.~Zhang and J.~G. Andrews, ``Downlink cellular network analysis with
  multi-slope path loss models,'' \emph{IEEE Trans. Commun.}, vol.~63, no.~5,
  pp. 1881--94, May 2015.

\bibitem{WINNERII}
\BIBentryALTinterwordspacing
``Winner {II} channel models,'' 2007. [Online]. Available:
  \url{http://www.cept.org/files/1050/documents/winner2%20-%20final%20report.pdf}
\BIBentrySTDinterwordspacing

\bibitem{Sun2015}
S.~Sun, T.~A. Thomas, T.~S. Rappaport, H.~Nguyen, I.~Z. Kov{\'{a}}cs, and
  I.~Rodriguez, ``Path loss, shadow fading, and line-of-sight probability
  models for {5G} urban macro-cellular scenarios,'' in \emph{Proc. IEEE
  GLOBECOM Workshops}, Dec. 2015, pp. 1--7.

\bibitem{Xia93}
H.~Xia, H.~L. Bertoni, L.~R. Maciel, A.~Lindsay-Stewart, and R.~Rowe, ``Radio
  propagation characteristics for line-of-sight microcellular and personal
  communications,'' \emph{IEEE Trans. Antennas Propag.}, vol.~41, no.~10, pp.
  1439--1447, Oct. 1993.

\bibitem{Jarvelainen2016}
J.~Jarvelainen, S.~Nguyen, K.~Haneda, R.~Naderpour, and U.~Virk, ``Evaluation
  of millimeter-wave line-of-sight probability with point cloud data,''
  \emph{IEEE Wireless Commun. Lett.}, vol.~PP, no.~99, pp. 1--1, 2016.

\bibitem{Cowan1989}
\BIBentryALTinterwordspacing
R.~Cowan, ``\BIBforeignlanguage{English}{Objects arranged randomly in space: An
  accessible theory},'' \emph{\BIBforeignlanguage{English}{Advances in Applied
  Probability}}, vol.~21, no.~3, pp. 543--569, 1989. [Online]. Available:
  \url{http://www.jstor.org/stable/1427635}
\BIBentrySTDinterwordspacing

\bibitem{bai2013e}
T.~Bai and R.~W. Heath~Jr., ``Coverage in dense millimeter wave cellular
  networks,'' in \emph{Proc. ASILOMAR}, Nov. 2013, pp. 2062--2066.

\bibitem{KulBlockageCodes15}
\BIBentryALTinterwordspacing
M.~N. Kulkarni. {MATLAB} codes for converting building location data from shape
  files to {MAT} files. [Online]. Available: \url{https://goo.gl/Ie39k7}
\BIBentrySTDinterwordspacing

\bibitem{Baccelli2015}
F.~Baccelli and X.~Zhang, ``A correlated shadowing model for urban wireless
  networks,'' in \emph{Proc. {IEEE INFOCOM}}, Apr. 2015, pp. 801--809.

\bibitem{Rappaport2015a}
T.~Rappaport and S.~Deng, ``73 {GHz} wideband millimeter-wave foliage and
  ground reflection measurements and models,'' in \emph{Proc. IEEE ICC
  Workshops}, June 2015, pp. 1238--1243.

\bibitem{ThomasVook14}
T.~A. Thomas and F.~W. Vook, ``System level modeling and performance of an
  outdoor {mmWave} local area access system,'' in \emph{Proc. PIMRC}, Sept.
  2014, pp. 108--112.

\bibitem{Venugopal2016}
K.~Venugopal and R.~W. Heath, ``Millimeter wave networked wearables in dense
  indoor environments,'' \emph{IEEE Access}, vol.~4, pp. 1205--1221, Mar. 2016.

\bibitem{Austinbuildings13}
\BIBentryALTinterwordspacing
(2016, Mar.) Building footprints 2013. City of Austin. [Online]. Available:
  \url{https://goo.gl/7bc3Gl}
\BIBentrySTDinterwordspacing

\bibitem{LAbuildings08}
\BIBentryALTinterwordspacing
(2012, Nov.) Building outlines from {LAR-IAC2} 2008. City of Los Angleles.
  [Online]. Available: \url{http://goo.gl/Svgyie}
\BIBentrySTDinterwordspacing

\bibitem{Khan2011}
F.~Khan and Z.~Pi, ``{mmWave} mobile broadband {(MMB)}: Unleashing the
  3-300{GHz} spectrum,'' in \emph{Proc. IEEE Sarnoff Symposium}, May 2011, pp.
  1--6.

\bibitem{Han2015}
S.~Han, C.-L. I, Z.~Xu, and C.~Rowell, ``Large-scale antenna systems with
  hybrid analog and digital beamforming for millimeter wave {5G},''
  \emph{{IEEE} Commun. Mag.}, vol.~53, no.~1, pp. 186--194, Jan. 2015.

\bibitem{Boers2014}
M.~Boers, B.~Afshar, I.~Vassiliou, S.~Sarkar, S.~Nicolson, E.~Adabi,
  B.~Perumana, T.~Chalvatzis, S.~Kavvadias, P.~Sen, W.~Chan, A.-T. Yu,
  A.~Parsa, M.~Nariman, S.~Yoon, A.~Besoli, C.~Kyriazidou, G.~Zochios,
  J.~Castaneda, T.~Sowlati, M.~Rofougaran, and A.~Rofougaran, ``A {16TX/16RX}
  60 {GHz} 802.11ad chipset with single coaxial interface and polarization
  diversity,'' \emph{{IEEE} J. Solid-State Circuits}, vol.~49, no.~12, pp.
  3031--3045, Dec 2014.

\bibitem{Chen2010}
X.-P. Chen, K.~Wu, L.~Han, and F.~He, ``Low-cost high gain planar antenna array
  for {60-GHz} band applications,'' \emph{IEEE Trans. Antennas Propag.},
  vol.~58, no.~6, pp. 2126--2129, June 2010.

\bibitem{Emami2011}
S.~Emami, R.~Wiser, E.~Ali, M.~Forbes, M.~Gordon, X.~Guan, S.~Lo, P.~McElwee,
  J.~Parker, J.~Tani, J.~Gilbert, and C.~Doan, ``A {60GHz} {CMOS} phased-array
  transceiver pair for {multi-Gb/s} wireless communications,'' in \emph{Proc.
  IEEE ISSCC}, Feb. 2011, pp. 164--166.

\bibitem{Sun2013}
H.~Sun, Y.-X. Guo, and Z.~Wang, ``60-{GHz} circularly polarized u-slot patch
  antenna array on {LTCC},'' \emph{IEEE Trans. Antennas Propag.}, vol.~61,
  no.~1, pp. 430--435, Jan. 2013.

\bibitem{Biglarbegian2011}
B.~Biglarbegian, M.~Fakharzadeh, D.~Busuioc, M.~Nezhad-Ahmadi, and
  S.~Safavi-Naeini, ``Optimized microstrip antenna arrays for emerging
  millimeter-wave wireless applications,'' \emph{IEEE Trans. Antennas Propag.},
  vol.~59, no.~5, pp. 1742--1747, May 2011.

\bibitem{Singh2009}
J.~Singh, S.~Ponnuru, and U.~Madhow, ``Multi-gigabit communication: the {ADC}
  bottleneck,'' in \emph{Proc. IEEE ICUWB}, Vancouver, BC, Sept. 2009, pp.
  22--27.

\bibitem{Murmann2015}
\BIBentryALTinterwordspacing
B.~Murmann, ``{ADC} performance survey 1997-2015,'' 2015. [Online]. Available:
  \url{http://www.stanford.edu/~murmann/adcsurvey.html}
\BIBentrySTDinterwordspacing

\bibitem{ElAyach2014}
O.~El~Ayach, S.~Rajagopal, S.~Abu-Surra, Z.~Pi, and R.~Heath, ``Spatially
  sparse precoding in millimeter wave {MIMO} systems,'' \emph{{IEEE} Trans.
  Wireless Commun.}, vol.~13, no.~3, pp. 1499--1513, Mar. 2014.

\bibitem{Mo15}
J.~Mo and R.~W. Heath, ``Capacity analysis of one-bit quantized {MIMO} systems
  with transmitter channel state information,'' \emph{{IEEE} Trans. Signal
  Process.}, vol.~63, no.~20, pp. 5498--5512, Oct. 2015.

\bibitem{Alkhateeb2014d}
A.~Alkhateeb, J.~Mo, N.~Gonzalez-Prelcic, and R.~Heath, ``{MIMO} precoding and
  combining solutions for millimeter-wave systems,'' \emph{IEEE Commun. Mag.},
  vol.~52, no.~12, pp. 122--131, Dec. 2014.

\bibitem{Mendez-Rial2016}
R.~Mendez-Rial, C.~Rusu, N.~Gonzalez-Prelcic, A.~Alkhateeb, and R.~Heath,
  ``Hybrid {MIMO} architectures for millimeter wave communications: Phase
  shifters or switches?'' \emph{IEEE Access}, vol.~4, pp. 247--267, Jan. 2016.

\bibitem{Brady2013}
J.~Brady, N.~Behdad, and A.~Sayeed, ``Beamspace {MIMO} for millimeter-wave
  communications: System architecture, modeling, analysis, and measurements,''
  \emph{IEEE Trans. Antennas Propag.}, vol.~61, no.~7, pp. 3814--3827, July
  2013.

\bibitem{Wang2009}
J.~Wang, Z.~Lan, C.~Pyo, T.~Baykas, C.~Sum, M.~Rahman, J.~Gao, R.~Funada,
  F.~Kojima, H.~Harada, and S.~Kato, ``Beam codebook based beamforming protocol
  for multi-{Gbps} millimeter-wave {WPAN} systems,'' \emph{{IEEE} J. Sel. Areas
  Commun.}, vol.~27, no.~8, pp. 1390--1399, Oct. 2009.

\bibitem{Xia2008b}
P.~Xia, S.-K. Yong, J.~Oh, and C.~Ngo, ``Multi-stage iterative antenna training
  for millimeter wave communications,'' in \emph{Proc. IEEE GLOBECOM}, Nov.
  2008, pp. 1--6.

\bibitem{Hur2013}
S.~Hur, T.~Kim, D.~Love, J.~Krogmeier, T.~Thomas, and A.~Ghosh, ``Millimeter
  wave beamforming for wireless backhaul and access in small cell networks,''
  \emph{{IEEE} Trans. Commun.}, vol.~61, no.~10, pp. 4391--4403, Oct. 2013.

\bibitem{Zhang2005a}
X.~Zhang, A.~Molisch, and S.~Kung, ``Variable-phase-shift-based {RF}-baseband
  codesign for {MIMO} antenna selection,'' \emph{{IEEE} Trans. Signal
  Process.}, vol.~53, no.~11, pp. 4091--4103, Nov. 2005.

\bibitem{Alkhateeb2014b}
A.~Alkhateeb, G.~Leus, and R.~Heath, ``Limited feedback hybrid precoding for
  multi-user millimeter wave systems,'' \emph{{IEEE} Trans. Wireless Commun.},
  vol.~14, no.~11, pp. 6481--6494, Nov. 2015.

\bibitem{Alkhateeb2015a}
A.~Alkhateeb and R.~W. Heath~Jr, ``Frequency selective hybrid precoding for
  limited feedback millimeter wave systems,'' \emph{to appear in IEEE Trans.
  Commun., arXiv preprint arXiv:1510.00609}, 2015.

\bibitem{Ni2016}
W.~Ni and X.~Dong, ``Hybrid block diagonalization for massive multiuser {MIMO}
  systems,'' \emph{{IEEE} Trans. Commun.}, vol.~64, no.~1, pp. 201--211, Jan
  2016.

\bibitem{Yu2016}
X.~Yu, J.-C. Shen, J.~Zhang, and K.~Letaief, ``Alternating minimization
  algorithms for hybrid precoding in millimeter wave {MIMO} systems,''
  \emph{IEEE J. Sel. Topics Signal Process.}, vol.~10, no.~3, pp. 485--500,
  Apr. 2016.

\bibitem{Sohrabi2015}
F.~Sohrabi and W.~Yu, ``Hybrid digital and analog beamforming design for
  large-scale {MIMO} systems,'' in \emph{Proc. IEEE ICASSP, Brisbane,
  Australia}, Apr. 2015, pp. 2929 -- 2933.

\bibitem{Alkhateeb2016a}
A.~Alkhateeb, Y.~H. Nam, J.~Zhang, and R.~Heath, ``Massive {MIMO} combining
  with switches,'' \emph{IEEE Wireless Commun. Lett.}, vol.~PP, no.~99, pp.
  1--1, Jan. 2016.

\bibitem{Mo2016}
J.~Mo, A.~Alkhateeb, S.~Abu-Surra, and R.~W. Heath~Jr, ``Hybrid architectures
  with few-bit {ADC} receivers: Achievable rates and energy-rate tradeoffs,''
  \emph{submitted to Trans. Wireless Commun., arXiv preprint arXiv:1605.00668},
  2016.

\bibitem{Sam16}
M.~K. Samimi and T.~S. Rappaport, ``Local multipath model parameters for
  generating 5{G} millimeter-wave {3GPP}-like channel impulse response,''
  \emph{submitted to the 10th European Conference on Antennas and Propagation,
  arXiv preprint arXiv:1511.06941}, Apr. 2016.

\bibitem{HeathJr2015}
R.~W. Heath~Jr, N.~Gonzalez-Prelcic, S.~Rangan, W.~Roh, and A.~Sayeed, ``An
  overview of signal processing techniques for millimeter wave {MIMO}
  systems,'' \emph{IEEE J. Sel. Topics Signal Process.}, vol.~10, no.~3, pp.
  436--453, Apr. 2016.

\bibitem{SamRap14}
M.~K. Samimi and T.~S. Rappaport, ``Ultra-wideband statistical channel model
  for non line of sight millimeter wave urban channels,'' in \emph{Proc. IEEE
  GLOBECOM}, Dec. 2014, pp. 3483--3489.

\bibitem{KulGhoAnd16}
M.~N. Kulkarni, A.~Ghosh, and J.~G. Andrews, ``A comparison of {MIMO}
  techniques in downlink millimeter wave cellular networks with hybrid
  beamforming,'' \emph{to appear in IEEE Trans. Commun., arXiv preprint
  arXiv:1510.02845}, 2016.

\bibitem{Sayeed2002}
A.~Sayeed, ``Deconstructing multiantenna fading channels,'' \emph{{IEEE} Trans.
  Signal Process.}, vol.~50, no.~10, pp. 2563--2579, Oct. 2002.

\bibitem{Renzo2015}
M.~D. Renzo, ``Stochastic geometry modeling and analysis of multi-tier
  millimeter wave cellular networks,'' \emph{IEEE Trans. Wireless Commun.},
  vol.~14, no.~9, pp. 5038--5057, Sept. 2015.

\bibitem{Alkhateeb2016}
A.~Alkhateeb, Y.-H. Nam, M.~S. Rahman, J.~Zhang, and R.~W. Heath~Jr., ``Initial
  beam association in millimeter wave cellular systems: Analysis and design
  insights,'' \emph{submitted to IEEE Trans. Wireless Commun., arXiv preprint
  arXiv:1602.06598}, Feb. 2016.

\bibitem{Alkhateeb2014}
A.~Alkhateeb, O.~El~Ayach, G.~Leus, and R.~Heath, ``Channel estimation and
  hybrid precoding for millimeter wave cellular systems,'' \emph{{IEEE} J. Sel.
  Topics Signal Process.}, vol.~8, no.~5, pp. 831--846, Oct. 2014.

\bibitem{Ran14}
S.~Rangan, T.~S. Rappaport, and E.~Erkip, ``Millimeter wave cellular wireless
  networks: Potentials and challenges,'' \emph{Proc. {IEEE}}, vol. 102, no.~3,
  pp. 366--385, Mar. 2014.

\bibitem{Telatar1999}
\BIBentryALTinterwordspacing
E.~Telatar, ``Capacity of multi-antenna {G}aussian channels,'' \emph{European
  Transactions on Telecommunications}, vol.~10, no.~6, pp. 585--595, 1999.
  [Online]. Available: \url{http://dx.doi.org/10.1002/ett.4460100604}
\BIBentrySTDinterwordspacing

\bibitem{Goldsmith2003}
A.~Goldsmith, S.~Jafar, N.~Jindal, and S.~Vishwanath, ``Capacity limits of
  {MIMO} channels,'' \emph{{IEEE} J. Sel. Areas Commun.}, vol.~21, no.~5, pp.
  684--702, June 2003.

\bibitem{LTEBook}
A.~Ghosh, J.~Zhang, J.~G. Andrews, and R.~Muhamed, \emph{Fundamentals of
  LTE}.\hskip 1em plus 0.5em minus 0.4em\relax Prentice-Hall, 2010.

\bibitem{Ni2015}
W.~Ni, X.~Dong, and W.-S. Lu, ``Near-optimal hybrid processing for massive
  {MIMO} systems via matrix decomposition,'' \emph{arXiv preprint
  arXiv:1504.03777}, 2015.

\bibitem{Chen2015}
C.-E. Chen, ``An iterative hybrid transceiver design algorithm for millimeter
  wave {MIMO} systems,'' \emph{IEEE Wireless Commun. Lett.}, vol.~4, no.~3, pp.
  285--288, June 2015.

\bibitem{Mendez-Rial2015a}
R.~M{\'e}ndez-Rial, C.~Rusu, N.~Gonz{\'a}lez-Prelcic, and R.~W. Heath,
  ``Dictionary-free hybrid precoders and combiners for {mmWave} {MIMO}
  systems,'' in \emph{Proc. IEEE ISPAWC}, June 2015, pp. 151--155.

\bibitem{Kim2013}
C.~Kim, T.~Kim, and J.-Y. Seol, ``Multi-beam transmission diversity with hybrid
  beamforming for {MIMO}-{OFDM} systems,'' in \emph{Proc. IEEE GLOBECOM
  Workshops}, Atlanta, GA, Dec. 2013, pp. 61--65.

\bibitem{Gao2016}
X.~Gao, L.~Dai, Z.~Chen, Z.~Wang, and Z.~Zhang, ``Near-optimal beam selection
  for beamspace mm{W}ave massive {MIMO} systems,'' \emph{{IEEE} Commun. Lett.},
  vol.~PP, no.~99, pp. 1--1, 2016.

\bibitem{Li2016}
J.~Li, L.~Xiao, X.~Xu, and S.~Zhou, ``Robust and low complexity hybrid
  beamforming for uplink multiuser mm{W}ave {MIMO} systems,'' \emph{{IEEE}
  Commun. Lett.}, vol.~PP, no.~99, pp. 1--1, 2016.

\bibitem{AndGup16}
J.~G. Andrews, A.~K. Gupta, and H.~S. Dhillon, ``A primer on cellular network
  analysis using stochastic geometry,'' \emph{submitted to Commun. Surveys
  Tuts., arXiv preprint arXiv:1604.03183}, 2016.

\bibitem{DhiAnd13}
H.~S. Dhillon, R.~K. Ganti, and J.~G. Andrews, ``Load-aware modeling and
  analysis of heterogeneous cellular networks,'' \emph{IEEE Trans. Wireless
  Commun.}, vol.~12, no.~4, pp. 1666--1677, Apr. 2013.

\bibitem{Rappaport2013}
T.~Rappaport, F.~Gutierrez, E.~Ben-Dor, J.~Murdock, Y.~Qiao, and J.~Tamir,
  ``Broadband millimeter-wave propagation measurements and models using
  adaptive-beam antennas for outdoor urban cellular communications,''
  \emph{IEEE Trans. Antennas Propag.}, vol.~61, no.~4, pp. 1850--1859, Apr.
  2013.

\bibitem{Alzer1997}
H.~Alzer, ``\BIBforeignlanguage{English}{On some inequalities for the
  incomplete {Gamma} function},''
  \emph{\BIBforeignlanguage{English}{Mathematics of Computation}}, vol.~66, no.
  218, pp. 771--778, 1997.

\bibitem{Huang2007}
K.~Huang, R.~Heath~Jr., and J.~Andrews, ``Space division multiple access with a
  sum feedback rate constraint,'' \emph{IEEE Trans. Signal Process.}, vol.~55,
  no.~7, pp. 3879--3891, July 2007.

\bibitem{And14}
J.~G. Andrews, S.~Singh, Q.~Ye, X.~Lin, and H.~S. Dhillon, ``An overview of
  load balancing in {HetNets}: Old myths and open problems,'' \emph{IEEE
  Wireless Commun.}, vol.~52, no.~2, pp. 18--25, Apr. 2014.

\bibitem{GuptaKul2016}
A.~K. Gupta, M.~N. Kulkarni, E.~Visotsky, F.~Vook, A.~Ghosh, J.~G. Andrews, and
  R.~W. Heath, ``Rate analysis and feasibility of dynamic {TDD} in 5{G}
  cellular systems,'' in \emph{Proc. IEEE ICC}, May 2016.

\bibitem{SinBacAnd13}
S.~Singh, F.~Baccelli, and J.~G. Andrews, ``On association cells in random
  heterogeneous networks,'' \emph{IEEE Wireless Commun. Lett.}, vol.~3, no.~1,
  pp. 70--73, Feb. 2014.

\bibitem{Ferenc2007}
J.-S. Ferenc and Z.~N{\'e}da, ``On the size distribution of {P}oisson {V}oronoi
  cells,'' \emph{Physica A: Statistical Mechanics and its Applications}, vol.
  385, no.~2, pp. 518 -- 526, Nov. 2007.

\bibitem{Singh2013}
S.~Singh, H.~Dhillon, and J.~Andrews, ``Offloading in heterogeneous networks:
  Modeling, analysis, and design insights,'' \emph{IEEE Trans. Wireless
  Commun.}, vol.~12, no.~5, pp. 2484--2497, May 2013.

\bibitem{YuKim13}
S.~M. Yu and S.-L. Kim, ``Downlink capacity and base station density in
  cellular networks,'' in \emph{Proc. Workshop in Spatial Stochastic Models for
  Wireless Networks}, May 2013.

\bibitem{SinAnd14}
S.~Singh and J.~G. Andrews, ``Joint resource partitioning and offloading in
  heterogeneous cellular networks,'' \emph{IEEE Trans. Wireless Commun.},
  vol.~13, no.~2, pp. 888--901, Feb. 2014.

\bibitem{Singh2014}
S.~Singh, X.~Zhang, and J.~Andrews, ``Joint rate and {SINR} coverage analysis
  for decoupled uplink-downlink biased cell associations in {HetNets},''
  \emph{IEEE Trans. Wireless Commun.}, vol.~14, no.~10, pp. 5360--5373, Oct.
  2015.

\bibitem{Wang14}
H.~Wang, X.~Zhou, and M.~C. Reed, ``Coverage and throughput analysis with a
  non-uniform small cell deployment,'' \emph{IEEE Trans. Wireless Commun.},
  vol.~3, no.~4, pp. 2047--2059, Apr. 2014.

\bibitem{DhiRate14}
H.~S. Dhillon and J.~G. Andrews, ``Downlink rate distribution in heterogeneous
  cellular networks under generalized cell selection,'' \emph{IEEE Wireless
  Commun. Lett.}, vol.~3, no.~1, pp. 42--45, Feb. 2014.

\bibitem{DhiGan12}
H.~S. Dhillon, R.~K. Ganti, F.~Baccelli, and J.~G. Andrews, ``Modeling and
  analysis of {K}-tier downlink heterogeneous cellular networks,'' \emph{IEEE
  J. on Sel. Areas Commun.}, vol.~30, no.~3, pp. 550--560, Apr. 2012.

\bibitem{jeong2015random}
C.~Jeong, J.~Park, and H.~Yu, ``Random access in millimeter-wave beamforming
  cellular networks: Issues and approaches,'' \emph{IEEE Commun. Mag.},
  vol.~53, no.~1, pp. 180--185, Jan. 2015.

\bibitem{Desai2014}
V.~Desai, L.~Krzymien, P.~Sartori, W.~Xiao, A.~Soong, and A.~Alkhateeb,
  ``Initial beamforming for {mmWave} communications,'' in \emph{Proc.
  ASILOMAR}, Nov. 2014, pp. 1926--1930.

\bibitem{BaratiNt.2015}
C.~Barati~Nt., S.~Hosseini, S.~Rangan, P.~Liu, T.~Korakis, S.~Panwar, and
  T.~Rappaport, ``Directional cell discovery in millimeter wave cellular
  networks,'' \emph{{IEEE} Trans. Wireless Commun.}, vol.~14, no.~12, pp.
  6664--6678, July 2015.

\bibitem{shen2012neighboring}
Y.~Shen, T.~Luo, and M.~Z. Win, ``Neighboring cell search for {LTE} systems,''
  \emph{IEEE Trans. Wireless Commun.}, vol.~11, no.~3, pp. 0908--919, Mar.
  2012.

\bibitem{barati2015directional}
C.~N. Barati, S.~A. Hosseini, M.~Mezzavilla, S.~Rangan, T.~Korakis, S.~S.
  Panwar, and M.~Zorzi, ``Directional initial access for millimeter wave
  cellular systems,'' \emph{arXiv preprint arXiv:1511.06483}, 2015.

\bibitem{Rangan15}
R.~Ford, F.~Gomez-Cuba, M.~Mezzavilla, and S.~Rangan, ``Dynamic time-domain
  duplexing for self-backhauled millimeter wave cellular networks,'' in
  \emph{Proc. IEEE ICC Workshop}, June 2015, pp. 13--18.

\bibitem{Rois15}
J.~Garc{\'\i}a-Rois, F.~G{\'o}mez-Cuba, M.~R. Akdeniz, F.~J.
  Gonz{\'a}lez-Casta{\~n}o, J.~C. Burguillo, S.~Rangan, and B.~Lorenzo, ``On
  the analysis of scheduling in dynamic duplex multihop {mmWave} cellular
  systems,'' \emph{{IEEE} Trans. Wireless Commun.}, vol.~14, no.~11, pp.
  6028--6042, Nov. 2015.

\bibitem{FCCNOIProposal}
``Federal communications commission- notice of inquiry: {FCC} 14-154,'' Oct.
  2014.

\bibitem{Matinmikko2013}
M.~Matinmikko, M.~Palola, H.~Saarnisaari, M.~Heikkila, J.~Prokkola, T.~Kippola,
  T.~Hanninen, M.~Jokinen, and S.~Yrjola, ``Cognitive radio trial environment:
  First live authorized shared access-based spectrum sharing demonstration,''
  \emph{IEEE Veh. Technol. Mag.}, vol.~8, no.~3, pp. 30--37, Sept. 2013.

\bibitem{Khun-Jush2013}
J.~Khun-Jush, P.~Bender, B.~Deschamps, and M.~Gundlach, ``Licensed shared
  access as complementary approach to meet spectrum demands: Benefits for next
  generation cellular systems,'' in \emph{Proc. ETSI Workshop Reconfigurable
  Radio Syst., Cannes, France}, Dec 2013, pp. 1--7.

\bibitem{Gupta2016}
A.~K. Gupta, J.~G. Andrews, and R.~W. Heath~Jr, ``On the feasibility of sharing
  spectrum licenses in mm{W}ave cellular systems,'' \emph{submitted to IEEE
  Trans. Commun., arXiv preprint arXiv:1512.01290}, 2016.

\bibitem{GupAlk16}
A.~K. Gupta, A.~Alkhateeb, J.~G. Andrews, and R.~W. {Heath Jr}, ``Gains of
  restricted secondary licensing in millimeter wave cellular systems,''
  \emph{submitted to IEEE J. Sel. Areas Commun., arXiv preprint
  arXiv:1605.00205}, 2016.

\bibitem{And5G}
J.~G. Andrews, S.~Buzzi, W.~Choi, S.~Hanly, A.~Lozano, A.~Soong, and C.~Zhang,
  ``What will {5G} be?'' \emph{IEEE J. Sel. Areas Commun.}, vol.~32, no.~6, pp.
  1065--1082, June 2014.

\bibitem{Novlan2013}
T.~Novlan, H.~Dhillon, and J.~Andrews, ``Analytical modeling of uplink cellular
  networks,'' \emph{IEEE Trans. Wireless Commun.}, vol.~12, no.~6, pp.
  2669--2679, June 2013.

\bibitem{Colombi2015}
D.~Colombi, B.~Thors, and C.~T{\"o}rnevik, ``Implications of {EMF} exposure
  limits on output power levels for {5G} devices above 6 {GHz},'' \emph{IEEE
  Antennas Wireless Propag. Lett.}, vol.~14, pp. 1247--1249, Feb. 2015.

\bibitem{Bai2015ee}
T.~Bai and R.~W. Heath, ``Asymptotic {SINR} for millimeter wave massive {MIMO}
  cellular networks,'' in \emph{Proc. SPAWC}, June 2015, pp. 620--624.

\bibitem{Ishii2012}
H.~Ishii, Y.~Kishiyama, and H.~Takahashi, ``A novel architecture for
  {LTE-B}:{C-plane/U-plane} split and phantom cell concept,'' in \emph{Proc.
  IEEE GLOBECOM Workshops}, Dec. 2012, pp. 624--630.

\bibitem{Li2013a}
Q.~Li, H.~Niu, G.~Wu, and R.~Q. Hu, ``Anchor-booster based heterogeneous
  networks with {mmWave} capable booster cells,'' in \emph{Proc. IEEE GLOBECOM
  B4G Workshop}, Dec. 2013, pp. 93--98.

\bibitem{Shaer16}
H.~E. Shaer, M.~N. Kulkarni, F.~Boccardi, J.~G. Andrews, and M.~Dohler,
  ``Downlink and uplink cell association with traditional macrocells and
  millimeter wave small cells,'' \emph{submitted to IEEE Trans. Wireless
  Commun., arXiv preprint arXiv:1601.05281}, 2016.

\bibitem{NokiaIndoor14}
``Indoor deployment strategies,'' Nokia Networks, Tech. Rep., June 2014,
  available at: http://goo.gl/lcqbIr.

\bibitem{CiscoIndoor2012}
``Cisco service provider {W}i-{F}i: {A} platform for business innovation and
  revenue generation,'' Cisco, Tech. Rep., Nov. 2012, available at:
  http://goo.gl/GIewaH.

\bibitem{Taranetz2014}
M.~Taranetz, T.~Bai, R.~W. Heath, and M.~Rupp, ``Analysis of small cell
  partitioning in urban two-tier heterogeneous cellular networks,'' in
  \emph{Proc. IEEE ISWCS}, Aug. 2014, pp. 739--743.

\bibitem{Taori2015}
R.~Taori and A.~Sridharan, ``Point-to-multipoint in-band mmwave backhaul for
  {5G} networks,'' \emph{IEEE Commun. Mag.}, vol.~53, no.~1, pp. 195--201, Jan.
  2015.

\bibitem{GeoLoz2015}
G.~George and A.~Lozano, ``Impact of reflections in enclosed mmwave wearable
  networks,'' in \emph{Proc. of IEEE CAMSAP}, Dec. 2015, pp. 201--204.

\bibitem{Zhang2015c}
X.~Zhang, F.~Baccelli, and R.~W. Heath, ``An indoor correlated shadowing
  model,'' in \emph{Proc. IEEE GLOBECOM}, Dec. 2015, pp. 1--7.

\bibitem{Swindlehurst2014}
A.~Swindlehurst, E.~Ayanoglu, P.~Heydari, and F.~Capolino, ``Millimeter-wave
  massive {MIMO}: the next wireless revolution?'' \emph{IEEE Commun. Mag.},
  vol.~52, no.~9, pp. 56--62, Sept. 2014.

\bibitem{Bjornson2016}
E.~Bjornson, L.~Sanguinetti, and M.~Kountouris, ``Deploying dense networks for
  maximal energy efficiency: Small cells meet massive {MIMO},'' \emph{IEEE J.
  Sel. Areas Commun.}, vol.~PP, no.~99, pp. 1--1, Mar. 2016.

\bibitem{Bai2015a}
T.~Bai and R.~W. {Heath Jr.}, ``Analyzing uplink {SIR} and rate in massive
  {MIMO} systems using stochastic geometry,'' \emph{submitted to IEEE Trans.
  Commun., arXiv preprint arXiv:1510.02538}, 2015.

\bibitem{Ramasamy2013}
D.~Ramasamy, S.~Venkateswaran, and U.~Madhow, ``Compressive parameter
  estimation in {AWGN},'' \emph{IEEE Trans. Signal Process.}, vol.~62, no.~8,
  pp. 2012--2027, Apr. 2014.

\bibitem{Alkhateeb2015}
A.~Alkhateeb, G.~Leus, and R.~Heath, ``Compressed-sensing based multi-user
  millimeter wave systems: How many measurements are needed?'' in \emph{Proc.
  IEEE ICASSP, Brisbane, Australia}, Apr. 2015, pp. 2909 -- 2913.

\bibitem{Choi2015b}
J.~Choi, ``Beam selection in {mm-Wave} multiuser {MIMO} systems using
  compressive sensing,'' \emph{IEEE Trans. Commun.}, vol.~63, no.~8, pp.
  2936--2947, Aug 2015.

\end{thebibliography}
\end{document}